\documentclass[12pt]{article}
\usepackage{epsfig,amssymb,amsmath,psfrag}

\catcode`\@=11

\def\I {V}
\def \toy  {\rm (toy) }
\def \BES  { }

\def \be  {\begin{equation}}
\def \ee  {\end{equation}}
\def \ba  {\begin{eqnarray}}
\def \ea  {\end{eqnarray}}
\def \baa {\begin{eqnarray*}}
\def \eaa {\end{eqnarray*}}
\def \bb  {\begin {thebibliography} }
\def \eb  {\end{thebibliography}}
\def \lab #1 {\label{#1}}
\newcommand \ci [1] {\cite{#1}}

\newcommand\re[1]{(\ref{#1})}
\def \qqquad {\qquad\quad}
\def \qqqquad {\qquad\qquad}
\def \matrix #1 {\left(\begin{array}{cc} #1 \end{array}\right)}

\def \Im {\mathop{\rm Im}\nolimits}
\def \Re {\mathop{\rm Re}\nolimits}

\def \e  {\mathop{\rm e}\nolimits}
\newcommand\lr[1]{{\left({#1}\right)}}

\newcommand{\ft}[2]{{\textstyle\frac{#1}{#2}}}

\def\Xint#1{\mathchoice
   {\XXint\displaystyle\textstyle{#1}}%
   {\XXint\textstyle\scriptstyle{#1}}%
   {\XXint\scriptstyle\scriptscriptstyle{#1}}%
   {\XXint\scriptscriptstyle\scriptscriptstyle{#1}}%
   \!\int}
\def\XXint#1#2#3{{\setbox0=\hbox{$#1{#2#3}{\int}$}
     \vcenter{\hbox{$#2#3$}}\kern-.5\wd0}}

\def\dashint{\Xint-}

\def\numberbysection{\@addtoreset{equation}{section}
                     \def\theequation{\thesection.\arabic{equation}}}
\numberbysection

\begin{document}

\renewcommand{\thefootnote}{\fnsymbol{footnote}}

\begin{titlepage}
\begin{flushright}
\begin{tabular}{l}
LPT--Orsay--08--101 \\
Saclay/IPhT--T08/220
\end{tabular}
\end{flushright}

\vskip3cm

\begin{center}
 {\large \bf Nonperturbative scales in AdS/CFT}
\end{center}

\vspace{1cm}

\centerline{\sc B.~Basso${}^1$, G.P.~Korchemsky${}^2$\footnote{On leave from Laboratoire de Physique Th\'eorique, Universit\'e de Paris XI, 91405 Orsay C\'edex, France}}

\vspace{10mm}

\centerline{\it ${}^1$Laboratoire de Physique Th\'eorique\footnote{Unit\'e
                    Mixte de Recherche du CNRS  UMR 8627.},
                    Universit\'e de Paris XI}
\centerline{\it 91405 Orsay C\'edex, France}

\vspace{3mm}

\centerline{\it ${}^2$Institut de Physique Th\'eorique
\footnote{Unit\'e de Recherche Associ\'ee au CNRS URA 2306.},
                  CEA Saclay}
\centerline{\it 91191 Gif-sur-Yvette C\'edex, France}

\vspace{1cm}

\centerline{\bf Abstract}

\vspace{5mm}

The cusp anomalous dimension is a ubiquitous quantity in four-dimensional gauge theories,
ranging from QCD to maximally supersymmetric $\mathcal{N}=4$ Yang-Mills theory, and it
is  one of the best investigated observables in the AdS/CFT correspondence. In planar 
$\mathcal{N}=4$ SYM theory, its perturbative expansion at weak coupling has  a finite radius of  convergence while at  strong coupling it admits an expansion in inverse powers of the 't Hooft 
coupling which is given by a non-Borel summable asymptotic series. We study the cusp anomalous dimension in the transition regime from strong to weak coupling and argue that the transition 
is driven by nonperturbative,  exponentially suppressed corrections. To compute these corrections, 
we revisit the calculation of the cusp anomalous  dimension in planar $\mathcal{N}=4$ SYM theory and extend the previous analysis by taking into account nonperturbative effects. We demonstrate that the scale parameterizing nonperturbative corrections  
coincides with the mass gap of the two-dimensional bosonic O(6) sigma model embedded into the $\rm AdS_5\times S^5$ string theory. This result is in agreement with the prediction coming from the string
theory consideration.  

\end{titlepage}

\setcounter{footnote} 0

{\small \tableofcontents}

\newpage

\renewcommand{\thefootnote}{\arabic{footnote}}

\section{Introduction}

The AdS/CFT correspondence provides a powerful framework for studying maximally
supersymmetric ${\mathcal N}=4$ Yang-Mills theory (SYM) at strong
coupling~\cite{Mal97}. At present, one  of the best studied examples of the
conjectured gauge/string duality is the relationship between anomalous dimensions
of Wilson operators in planar ${\mathcal N}=4$  theory in the so-called $SL(2)$
sector and energy spectrum of folded strings spinning on  $\rm AdS_5\times
S^5$~\cite{GKP,FT}. The Wilson operators in this sector are given by single trace
operators built from $L$ copies of the same complex scalar field and $N$
light-cone components of the covariant derivatives. These quantum numbers
define, correspondingly,  the twist and the Lorentz spin of the Wilson operators
in $\mathcal{N}=4$ SYM theory (for a review, see \cite{BKM03}).  In dual string theory 
description~\cite{GKP,FT} 
they are identified as angular momenta of the string spinning on $\rm S^5$ and  
$\rm AdS_5$ part of the
background.

In general, anomalous dimensions in planar $\mathcal{N}=4$ theory in the $SL(2)$
sector  are nontrivial functions of 't Hooft coupling $g^2=g_{\rm YM}^2
N_c/(4\pi)^2$ and quantum numbers of Wilson operators -- twist $L$ and Lorentz
spin $N$. Significant simplification occurs in the limit~\cite{BGK06} when the Lorentz spin
grows exponentially with the twist,   $L\sim \ln N$ with
$N\to\infty$. In this limit, the anomalous dimensions scale logarithmically with
$N$ for arbitrary coupling and the minimal anomalous dimension has  the following
scaling behavior~\cite{BGK06,AM07,FRS,FTT06,CK07}
\be\label{anom-dim}
\gamma_{N,L}(g) =\left[ 2 \Gamma_{\rm cusp}(g) + \epsilon(g,j) \right] \ln N + \ldots\,,
\ee
where $j=L/\ln N$ is an appropriate scaling variable and ellipses denote terms
suppressed by powers of $1/L$. Here, the coefficient in front of $\ln N$ is split
into the sum of two functions in such a way that $\epsilon(g,j)$ carries the
dependence on the twist and it vanishes for $j=0$. The first term inside the
square brackets in \re{anom-dim} has a universal, twist independent
form~\cite{KR87,BGK03}.  It involves the function of the coupling constant  known
as the cusp anomalous dimension. This anomalous dimension was introduced in
\cite{KR87} to describe specific (cusp) ultraviolet divergences of 
Wilson loops~\cite{P80,Brandt:1981kf} with a light-like cusp on the integration 
contour~\cite{KK92}. The
cusp anomalous dimension plays a distinguished r\^ole in  $\mathcal{N}=4$ theory
and, in general, in four-dimensional Yang-Mills theories since, aside from
logarithmic scaling of the anomalous dimension \re{anom-dim}, it also controls
infrared divergences of scattering amplitudes~\cite{Korchemsky:1985xj}, Sudakov
asymptotics of elastic form factors~\cite{Korchemsky:1988hd}, gluon Regge
trajectories~\cite{Korchemskaya:1996je} etc.

According to \re{anom-dim}, asymptotic behavior of the minimal anomalous
dimension is determined  by two independent functions, $\Gamma_{\rm cusp}(g)$ and
$ \epsilon(g,j)$. At weak coupling, these functions are given by series in powers
of $g^2$ and the first few terms of the expansion can be computed in perturbation
theory.  At strong coupling, the AdS/CFT correspondence allows us to obtain
expansion of $\Gamma_{\rm cusp}(g)$ and $ \epsilon(g,j)$ in powers of $1/g$ from
the semiclassical expansion of the energy of the folded spinning string. Being
combined together, the weak and strong coupling expansions define asymptotic
behavior of these functions at the boundaries of (semi-infinite) interval $0\le g
< \infty$. The following questions arise: What are the corresponding
interpolating functions
for arbitrary $g$?  How does the transition from the weak to strong coupling
regimes occur? These are the questions that we address in this paper.

At weak coupling,  the functions $\Gamma_{\rm cusp}(g)$ and $\epsilon(g,j)$ can
be found in a generic (supersymmetric) Yang-Mills theory in the planar limit by
making use of the remarkable property of integrability. The Bethe Ansatz approach
to computing these functions at weak coupling was developed in
\cite{K95,BGK06,BGK03}. It was extended in  \cite{FRS,BES} to all loops in
$\mathcal{N}=4$ SYM theory leading to integral BES/FRS equations for $\Gamma_{\rm
cusp}(g)$ and $\epsilon(g,j)$ valid in the planar limit for arbitrary values of
the scaling parameter $j$ and the coupling constant $g$. For the cusp anomalous
dimension, the solution to the BES equation at weak coupling is in agreement with
the most advanced explicit four-loop perturbative calculation \cite{cusp-4loop}
and it yields a perturbative series for $\Gamma_{\rm cusp}(g)$  which has a finite
radius of convergence~\cite{BES}. The BES equation was also analyzed at strong
coupling~\cite{Benna06,Kotikov:2006ts,Alday07,KSV07} but constructing its solution for
$\Gamma_{\rm cusp}(g)$ turned out to be a nontrivial task. 

The problem was solved
in Refs.~\cite{BKK07,KSV08}, where the cusp anomalous dimension was found in the
form of an asymptotic series in $1/g$. It turned out that the coefficients of this
expansion have the same sign and grow factorially at higher orders. As a result, 
the asymptotic $1/g$
expansion of $\Gamma_{\rm cusp}(g)$ is given by a non-Borel summable series which
suffers from ambiguities that are exponentially small for $g\to\infty$. This
suggests that the cusp anomalous dimension receives  {\it nonperturbative
corrections} at strong coupling~\cite{BKK07}
\be\label{cusp=pt+npt}
\Gamma_{\rm cusp}(g) = \sum_{k=-1}^\infty c_k/g^{k} -\frac{\sigma}{4\sqrt{2}}  \,
m^2_{\rm cusp}  + o(m^2_{\rm cusp} )\,.
\ee
Here the dependence of the nonperturbative scale $m^2_{\rm cusp}$ on the coupling
constant $m_{\rm cusp} \sim  g^{1/4} \e^{-\pi g}$ follows, through a standard
analysis~\cite{Zinn-Justin,Bender}, from the large order behavior of the
expansion coefficients, $c_k\sim\Gamma(k+\ft12)$ for $k \to\infty$. The
value of the coefficient $\sigma$ in \re{cusp=pt+npt} depends on the regularization of Borel
singularities in the perturbative $1/g$ expansion and the numerical prefactor was introduced
for the later convenience. 

Notice that the expression for the nonperturbative scale $m^2_{\rm cusp}$ looks similar
to that for the mass gap in an asymptotically free field theory with 
the coupling constant $\sim 1/g$. An important difference is, however, that 
$m^2_{\rm cusp}$ is a dimensionless function of the 't
Hooft coupling. This is perfectly consistent with the fact that $\mathcal{N}=4$
model is a conformal field theory and, therefore, it does not involve any
dimensionfull scale.  Nevertheless, as we will show in this paper, the nonperturbative 
scale $m^2_{\rm cusp}$ is indeed related to the mass gap in the {\it two-dimensional}
bosonic O(6) sigma-model.

The relation \re{cusp=pt+npt} sheds light on the properties
of $\Gamma_{\rm cusp}(g)$ in the transition region $g\sim 1$. Going from $g\gg 1$
to $g=1$, we find that $m^2_{\rm cusp}$ increases and, as a consequence,
nonperturbative $O(m^2_{\rm cusp})$ corrections to $\Gamma_{\rm cusp}(g)$ become
comparable with perturbative $O(1/g)$ corrections. We will argue in this paper
that the nonperturbative corrections play a crucial role in the transition from the strong 
to weak coupling regime. To describe the transition, we present a simplified
model for the cusp anomalous dimension. This model correctly captures the properties 
of $\Gamma_{\rm cusp}(g)$ at strong coupling and, most importantly, it allows us to obtain a closed expression for 
the cusp anomalous dimension which turns out to be remarkably close to the exact value of 
$\Gamma_{\rm cusp}(g)$ throughout the entire range of the coupling constant.

In the AdS/CFT correspondence, the relation \re{cusp=pt+npt} should follow from
the semiclassical expansion of the energy of quantized folded spinning string~\cite{GKP,FT}. 
In the right-hand side of \re{cusp=pt+npt}, the coefficient
$c_{-1}$ corresponds to the classical energy and $c_{k}$ describes
$(k+1)$th loop correction. Indeed, the explicit two-loop stringy calculation~\cite{Roiban:2007dq}
yields the expressions for $c_{-1}$, $c_0$ and $c_1$ which are in a perfect agreement
with \re{cusp=pt+npt}.%
\footnote{The same result was obtained using different approach from the quantum string Bethe
Ansatz in Refs.~\cite{CK07,G08}.} 
 However, the semiclassical approach does not allow us
to calculate  nonperturbative corrections to  $\Gamma_{\rm cusp}(g)$ and verification of \re{cusp=pt+npt} 
remains a challenge for the string theory. 

Recently,  Alday and Maldacena~\cite{AM07} put forward an interesting proposal that
the scaling function $\epsilon(g,j)$ entering \re{anom-dim} can be found exactly at strong
coupling in terms of nonlinear O(6) bosonic sigma model embedded
into $\rm AdS_5\times S^5$ model.  More precisely, using  the dual description of
Wilson operators as folded strings spinning on ${\rm AdS_5\times S^5}$ and taking
into account the one-loop stringy corrections to these states~\ci{FTT06}, they
conjectured that the scaling function $\epsilon(g,j)$ should be related at strong
coupling to the energy density $\epsilon_{\rm O(6)}$ in the ground state of the
$\rm O(6)$ model corresponding to the particle density $ \rho_{\rm O(6)} ={j}/2$
\be\label{O6-epsilon}
\epsilon_{\rm O(6)}  = \frac{\epsilon(g,j) + j}2\,,\qqqquad m_{\rm O(6)} = k
g^{1/4} \e^{-\pi g} \left[ 1 + O(1/g)\right]\,.
\ee
This relation should hold at strong coupling and $j/m_{\rm O(6)}={\rm fixed}$.
Here the scale $m_{\rm O(6)} $  is identified as the dynamically generated mass gap
in the O(6) model with $k={2^{3/4} \pi^{1/4}}/{\Gamma(\ft54)}$ being the
normalization factor.

The $\rm O(6)$ sigma model is an exactly solvable
theory~\cite{ZZ78,PW83,FR85,HMN90} and the dependence of $\epsilon_{\rm O(6)}$ on the mass scale
$m_{\rm O(6)} $ and the density of particles $ \rho_{\rm O(6)}$ can be found
exactly with a help of thermodynamical Bethe ansatz equations. Together with
\re{O6-epsilon} this allows us to determine the scaling function $\epsilon(g,j)$
at strong coupling. In particular, for $j/m_{\rm O(6)} \ll 1$, the asymptotic behavior
of $\epsilon(g,j)$ follows from the known expression for the energy density of the O(6) model
in the (nonperturbative) regime of small density of particles~\cite{HMN90,AM07,BK08,FGR08a}
\be\label{11}
\epsilon(j,g) +j  =  m^2\left[\frac{j}{m} + \frac{\pi^2}{24} \lr{\frac{j}{m}}^3+
O\left(j^4/m^4\right)\right] ,
\ee
with $m\equiv m_{\rm O(6)} $.  For $j/m_{\rm O(6)} \gg 1$, the scaling function $\epsilon(g,j)$ admits a perturbative expansion in inverse powers of $g$ with the coefficients enhanced by powers of $\ln\ell$
(with  $\ell = j/(4g)\ll 1$)~\cite{FTT06,AM07}
\be\label{e-2-loop}
\epsilon(g,j)+j = 2\ell^2\left[ g +\frac{1}{\pi}\left(\frac{3}{4}-\ln{\ell}\right) + \frac{1}{4\pi^2
g}\left(\frac{q_{02}}2-3\ln{\ell}+4(\ln{\ell})^2\right) +
\mathcal{O}\left({1}/{g^2}\right)\right]+O(\ell^4)\,.
\ee
This expansion was derived both in string theory~\cite{RT07} and in gauge theory~\cite{G08,BBBKP08,V08} yielding however different results for the constant $q_{02}$. The reason
for the disagreement remains unclear.

Remarkably enough,  the relation \re{O6-epsilon} was established in planar
$\mathcal{N}=4$ SYM theory at strong coupling \cite{BK08} using the conjectured
integrability of the dilatation operator \cite{FRS}. The mass scale $m_{\rm
O(6)}$ was computed both numerically \cite{FGR08a} and analytically
\cite{BK08,BBBKP08} and it was found to be in a perfect agreement with
\re{O6-epsilon}. This result is an extremely nontrivial given the fact that
the scale $m_{\rm O(6)}$ has a different origin in gauge  and in string theory
sides of the AdS/CFT. In string theory, it is generated by the dimensional
transmutation mechanism in {\it two-dimensional} effective theory describing
dynamics of massless modes in the $\rm AdS_5\times S^5$ sigma model. In gauge
theory, the same scale parameterizes nonperturbative corrections to anomalous
dimensions in {\it four-dimensional} Yang-Mills theory at strong coupling.  It is
interesting to note that similar phenomenon, when two different quantities
computed in four-dimensional gauge theory and in dual two-dimensional sigma model
coincide, has already been observed in the BPS spectrum in $\mathcal{N}=2$
supersymmetric Yang-Mills theory~\cite{Dorey:1998yh,Shifman:2004dr}.
We would like to mention that the precise matching of  the leading  coefficients 
in perturbative expansion of spinning string energy  and 
anomalous dimensions  on gauge  side was previously found in Refs.~\cite{BMN02,Frolov:2003,BT05}.
The relation \re{O6-epsilon} implies that for the anomalous dimensions
\re{anom-dim} the gauge/string correspondence holds
 at the level of  nonperturbative corrections.

As we just explained, the functions $\Gamma_{\rm cusp}(g)$ and $\epsilon(g,j)$
entering \re{anom-dim} receive nonperturbative contributions at strong coupling
described by the scales  $m_{\rm cusp}$ and $m_{\rm O(6)} $, respectively. In
$\mathcal{N}=4$ SYM theory, these functions satisfy two different integral
equations \cite{FRS,BES} and there is no a priori reason  why the scales  $m_{\rm cusp}$ and
$m_{\rm O(6)} $ should be related to each other. Nevertheless, examining their
leading order expressions, Eqs.~\re{cusp=pt+npt} and \re{O6-epsilon}, we notice
that they have the same dependence on the coupling constant. One may wonder
whether subleading $O(1/g)$ corrections are also related to each other.  In this
paper, we show that  the two scales coincide at strong coupling to any
order of $1/g$ expansion
\be\label{m=m}
m_{\rm cusp} = m_{\rm O(6)} \,,
\ee
thus proving that nonperturbative corrections to the cusp anomalous dimension
\re{cusp=pt+npt} and to the scaling function \re{11} are parameterized by the same scale. 

The relations \re{cusp=pt+npt} and  \re{m=m} 
also have an interpretation in string theory. The cusp anomalous dimension  has the meaning of the
energy density of a  folded string spinning on $\rm AdS_3$~\cite{GKP,AM07}.  As such, it receives quantum corrections
from both massive and massless excitations of this string in the $\rm AdS_5\times S^5$ 
sigma model. The O(6) model emerges in this context as the effective theory describing 
the dynamics of massless modes.  In distinction with the scaling function $\epsilon(g,j)$, for which the massive modes decouple
in the  limit $j/m_{\rm O(6)}={\rm fixed}$ and $g\to\infty$, the cusp anomalous dimension is not
described entirely by the O(6) model. Nevertheless, it is expected that the leading nonperturbative
corrections to  $\Gamma_{\rm cusp}(g)$ should originate from nontrivial infrared dynamics of the massless excitations and, therefore, they should be related to nonperturbative corrections to the
vacuum energy density in the O(6) model. As a consequence, $\Gamma_{\rm cusp}(g)$  should
receive exponentially suppressed corrections proportional 
to square of the O(6) mass gap  $\sim m_{\rm O(6)}^2$. We show in this paper by explicit calculation that this is indeed the case.

The paper is organized as follows. In Section~2, we revisit the calculation of
the cusp anomalous dimension in planar $\mathcal{N}=4$ SYM theory and construct
the exact solution for $\Gamma_{\rm cusp}(g)$. In Section~3, we analyze the
obtained expressions at strong coupling and identify nonperturbative corrections
to $\Gamma_{\rm cusp}(g)$. In Section~4, we compute subleading corrections to the
nonperturbative scales   $m_{\rm cusp}$ and $m_{\rm O(6)} $ and show that they
are the same for the two scales. Then, we extend our analysis to higher orders  in $1/g$ and demonstrate that the two scales coincide.  Section~5 contains concluding remarks. Some
technical details of our calculations are presented in Appendices.

\section{Cusp anomalous dimension in $\mathcal{N}=4$ SYM}

The cusp anomalous dimension can be found in planar $\mathcal{N}=4$ SYM theory for
arbitrary coupling as solution to the BES equation \cite{BES}. At strong
coupling,  $\Gamma_{\rm cusp}(g)$ was constructed in \cite{BKK07,KSV08} in the form of
perturbative expansion in $1/g$.  The coefficients of this series grow
factorially at higher orders thus indicating that $\Gamma_{\rm cusp}(g)$ receives
nonperturbative corrections which are exponentially small at strong coupling,
Eq.~\re{cusp=pt+npt}. To identity such corrections, we revisit in this section
the calculation of the cusp anomalous dimension and construct the exact solution
to the BES equation for arbitrary coupling.

\subsection{Integral equation and mass scale}

In the Bethe ansatz approach,  the cusp anomalous dimension  is determined by the
behavior around the origin of the auxiliary function $\gamma(t)$ related to
density of Bethe roots 
\be\label{cusp0}
\Gamma_{\rm cusp}(g) = -8i g^2 \lim_{t\to 0} \gamma(t)/t \,.
\ee
The function $\gamma(t)$ depends on 't Hooft coupling and has the form
\be\label{g_pm}
\gamma(t) = \gamma^{\BES}_+(t) + i\gamma^{\BES}_-(t)\,,
\ee
where $\gamma^{\BES}_{\pm}(t)$ are real functions of $t$ with a definite parity
$\gamma^{\BES}_{\pm}(\pm t)=\pm \gamma^{\BES}_{\pm}(t)$. For arbitrary coupling,
the functions $\gamma^{\BES}_{\pm}(t)$ satisfy the (infinite-dimensional) system
of integral equations
\begin{align}\label{FRS2}
& \int_{0}^{\infty}\frac{dt}{t} \, J_{2n-1}(t) \left[
\frac{\gamma^{\BES}_{-}(t)}{1-\e^{-t/(2g)}}+\frac{\gamma^{\BES}_{+}(t)}{\e^{t/(2g)}-1}\right]
= \frac{1}{2} \ \delta_{n, 1}  \,,
\\ \notag
& \int_{0}^{\infty}\frac{dt}{t} \, J_{2n}(t) \left[
\frac{\gamma^{\BES}_{+}(t)}{1-\e^{-t/(2g)}}-\frac{\gamma^{\BES}_{-}(t)}{\e^{t/(2g)}-1}\right] = 0\,,
\end{align}
with $n\ge 1$  and $J_n(t)$ being the Bessel functions. These relations are
equivalent to BES equation~\cite{BES} provided that $\gamma^{\BES}_{\pm}(t)$
verify certain analyticity conditions specified below in Sect.~2.2.

As was shown in \cite{BKK07,BK08},  the equations \re{FRS2} can be significantly simplified
with a help of the transformation $\gamma(t) \to \Gamma(t)$:
\footnote{With a slight abuse of notations, we use here the same notation as for
Euler gamma-function.}
\be\label{rel}
\Gamma(t) =   \lr{1+ i \coth \frac{t}{4g}}\gamma(t) \equiv \Gamma^{\BES}_+(t) + i\Gamma^{\BES}_-(t)\,.
\ee
We find from \re{cusp0} and \re{rel} the following representation for the cusp
anomalous dimension
\be\label{cusp}
\Gamma_{\rm cusp}(g) = - 2g \Gamma(0)\,.
\ee
It follows from \re{g_pm} and   \re{FRS2} that $\Gamma^{\BES}_{\pm}(t)$ are real
functions  with a definite parity, $\Gamma^{\BES}_{\pm}(-t) =
\pm\Gamma^{\BES}_{\pm}(t)$,  satisfying the system of integral equations
\begin{align}\label{sys}
& \int_0^\infty dt\, \cos(ut)\bigg[\Gamma^{\BES}_-(t) -\Gamma^{\BES}_+(t)\bigg] = 2
\,,
\\\notag
& \int_0^\infty dt\, \sin(ut)\bigg[  \Gamma^{\BES}_-(t) +\Gamma^{\BES}_+(t)\bigg] = 0
\,,
\end{align}
with $u$ being arbitrary real parameter such that $-1\le u\le 1$. Since
$\Gamma^{\BES}_{\pm}(t)$ take real values,  we can rewrite these relations  in
a compact form
\be\label{G-int}
\int_0^\infty dt\, \bigg[ \e^{iut} \Gamma^{\BES}_-(t) - \e^{-iut}\Gamma^{\BES}_+(t)\bigg] = 2
\,. 
\ee
To recover \re{FRS2}, we apply \re{rel}, replace in \re{sys} trigonometric
functions by their Bessel series expansions
\begin{align}\notag
\cos(ut) &= 2\sum_{n\ge 1}  (2n-1) \frac{\cos((2n-1)\varphi)}{\cos \varphi }
\frac{J_{2n-1}(t)}{t}\,,
\\
\sin(ut) &= 2\sum_{n\ge 1}  (2n) \frac{\sin(2n\varphi)}{\cos \varphi }
\frac{J_{2n}(t)}{t}\,,
\end{align}
with $u=\sin \varphi$, and finally compare coefficients in front of
${\cos((2n-1)\varphi)}/{\cos \varphi}$ and ${\sin(2n\varphi)}/{\cos \varphi}$ in
both sides of \re{sys}. It is important to stress that, doing this calculation,
we interchanged the sum over $n$ with the integral over $t$. This is only
justified for $\varphi$ real and, therefore, the relation \re{sys} only holds for
$-1\le u\le 1$.

Comparing \re{G-int} and \re{FRS2} we observe that the transformation
$\gamma_\pm^{\BES} \to \Gamma_\pm^{\BES}$ eliminates the dependence of the
integral kernel in the left-hand side of \re{G-int} on the coupling constant. One
may then wonder where does the dependence of the functions $\Gamma_\pm^{\BES}(t)$
on the coupling constant come from? We will show in the next subsection that it
is dictated by additional conditions imposed on  analytical properties of
solutions to \re{G-int}.

The relations \re{cusp} and \re{sys} were used in \cite{BKK07} to derive
asymptotic (perturbative) expansion of $\Gamma_{\rm cusp}(g)$ in powers of $1/g$.
This series suffers however from Borel singularities and we expect that the cusp
anomalous dimension should receive nonperturbative corrections $\sim\e^{-2\pi g}$
exponentially small at strong coupling.  As was already mentioned in the
Introduction, similar corrections are also present in the scaling function
$\epsilon(g,j)$ which controls asymptotic behavior of the anomalous dimensions
\re{anom-dim} in the limit when Lorentz spin of Wilson operators grows
exponentially with their  twist. According to   \re{O6-epsilon}, for $j/m_{\rm
O(6)}={\rm fixed}$ and $g\to\infty$, the scaling function coincides with the
energy density of the O(6) model embedded into $\rm AdS_5\times S^5$.  The mass
gap of this model defines a new nonperturbative scale  $m_{\rm O(6)}$ in the
AdS/CFT. Its dependence on the coupling $g$ follows univocally from the FRS
equation and it has the following form~\cite{BK08,BBBKP08}
\be\label{m=int}
m_{\rm O(6)} =  \frac{8\sqrt{2}}{\pi^2}\e^{-\pi g}-\frac{8g}{\pi}\e^{-\pi
g}\Re\left[ \int_0^\infty \frac{dt\, \e^{i(t-\pi/4)}}{t+i\pi
g}\left(\Gamma^{\BES}_{+}(t)+ i \Gamma^{\BES}_{-}(t)\right)\right],
\ee
where $\Gamma^{\BES}_{\pm}(t)$ are solutions to \re{G-int}. To compute the mass
gap \re{m=int}, we have to solve the integral equation \re{G-int} and, then,
substitute the resulting expression for $\Gamma_\pm^{\BES}(t)$ into \re{m=int}.
Notice that the same functions also determine the cusp anomalous dimension
\re{cusp}.

Later in the paper, we will construct the solution to the integral equation
\re{G-int} and, then,  apply \re{cusp} to compute nonperturbative corrections to
$\Gamma_{\rm cusp}(g)$ at strong coupling.

\subsection{Analyticity conditions}

The integral equations \re{G-int} and \re{FRS2}  determine $\Gamma_\pm^{\BES}(t)$
and $\gamma_\pm^{\BES}(t)$, or equivalently the functions $\Gamma(t)$ and
$\gamma(t)$, up to a contribution of zero modes. The latter satisfy the same
integral equations  \re{G-int} and \re{FRS2} but without inhomogeneous term in
the right-hand side.

To fix the zero modes, we have to impose additional conditions on solutions to
\re{G-int} and \re{FRS2}. These conditions follow unambiguously from the BES 
equation~\cite{Alday07,BKK07} and they can be
formulated as a requirement that $\gamma^{\BES}_\pm(t)$ should be entire
functions of $t$ which admit a representation in the form of Neumann series over
Bessel functions
\begin{align}\label{Bessel}
\gamma^{\BES}_{-}(t) &=  2 \sum_{n\geqslant 1} \  (2n-1)  J_{2n-1}(t) \gamma_{2n-1} \,,
\\ \notag
\gamma^{\BES}_{+}(t) &=   2\sum_{n\geqslant 1} \  (2n) \ J_{2n}(t) \gamma_{2n} \,,
\end{align}
with the expansion coefficients $\gamma_{2n-1}$ and $\gamma_{2n}$ depending on
the coupling constant. This implies in particular that the series on the right-hand
side of \re{Bessel} are convergent on the real axis.
Using orthogonality
conditions for the Bessel functions, we obtain from \re{Bessel}
\be\label{inverse}
\gamma_{2n-1} = \int_0^\infty \frac{dt}{t} J_{2n-1}(t) \gamma^{\BES}_-(t) \,,\qquad \gamma_{2n} =
\int_0^\infty \frac{dt}{t} J_{2n}(t) \gamma^{\BES}_+(t) \,.
\ee
Here we assumed that the sum over $n$ in the right-hand side of \re{Bessel}
can be interchanged with the integral over $t$. 
We will show below that the relations \re{Bessel} and \re{inverse} determine a unique solution to the system \re{FRS2}.

The coefficient $\gamma_1$ plays a special role in our analysis since it
determines the cusp anomalous dimension \re{cusp0},
\be\label{cusp1}
\Gamma_{\rm cusp}(g) = 8 g^2 \gamma_1(g)\,.
\ee
Here we applied \re{g_pm} and \re{Bessel} and took into account small$-t$
behavior of the Bessel functions, $J_n(t)\sim t^n$ as $t\to 0$.

Let us now translate \re{Bessel} and \re{inverse} into properties of the
functions $\Gamma_\pm^{\BES}(t)$, or equivalently $\Gamma(t)$. It is convenient
to rewrite the relation \re{rel} as
\be\label{sin-sin}
\Gamma(it) =  \gamma(it) \frac{\sin(\frac{t}{4g}+
\frac{\pi}4)}{\sin(\frac{t}{4g})\sin(\frac{\pi}4)} =\gamma(it)
\sqrt{2}\prod_{k=-\infty}^\infty\frac{t-4\pi g \lr{k-\ft14}}{t-4\pi g k} \,.
\ee
Since $\gamma(it)$ is an entire function in the complex $t-$plane, we conclude
from \re{sin-sin} that $\Gamma(it)$ has an infinite number of zeros,
$\Gamma(it_{\rm zeros})=0$, and poles, $\Gamma(it)\sim 1/(t-t_{\rm poles})$, on
real $t-$axis located at
\be\label{poles}
  t_{\rm zeros}= 4\pi g \lr{\ell-\ft14} \,,\qqqquad t_{\rm poles}=4\pi g \ell'\,,
\ee
where $\ell,\ell' \in \mathbb{Z} $ {and} $\ell'\neq 0$ so that $\Gamma(it)$ is
regular at the origin (see Eq.~\re{cusp0}). Notice that $\Gamma(it)$ has an
additional (infinite) set of  zeros coming from the function $\gamma(it)$ but, in
distinction with \re{poles}, their position is not  fixed. Later in the paper we will
construct the solution to  the integral equation \re{sys}  which satisfies 
the relations \re{poles}.

\subsection{Toy model}\label{Sect:toy}

To understand the relationship between analytical properties of $\Gamma(it)$ and
properties of  the cusp anomalous dimension, it is instructive to slightly
simplify the problem and consider a `toy' model in which the function
$\Gamma(it)$ is replaced with $\Gamma^{\toy}(it)$.

We require that  $\Gamma^{\toy}(it)$ satisfies the same integral equation
\re{sys} and define, following \re{cusp}, the cusp anomalous dimension in the
toy model as
\be
\Gamma^{\toy}_{\rm cusp}(g) = - 2g \Gamma^{\toy}(0)\,.
\ee
The only difference compared to $\Gamma(it)$ is that $\Gamma^{\toy}(it)$  has
different analytical properties dictated by the relation
\be\label{toy}
\Gamma^{\toy}(it) =  \gamma^{\toy}(it)   \frac{t+\pi g}{t } \,,
\ee
while $\gamma^{\toy}(it)$ has the same analytical properties as the function
$\gamma(it)$.%
\footnote{Notice that the function $\gamma^{\toy}(t)$ does not satisfy the
integral equation \re{FRS2} anymore. Substitution of \re{toy} into \re{G-int}
yields integral equation for $\gamma^{\toy}(t)$ which can be obtained from
\re{FRS2} by replacing $1/\lr{1-\e^{- t/(2g)}} \to \frac{\pi g}{2t}+\frac12$ and
$ 1/\lr{\e^{t/(2g)}-1} \to  \frac{\pi g}{2t}-\frac12$
in the kernel in the left-hand side of \re{FRS2}. } This relation can be
considered as a simplified version of \re{sin-sin}. Indeed, it can be obtained
from \re{sin-sin} if we retained in the product  only one term with $k=0$. As
compared with \re{poles}, the function $\Gamma^{\toy}(it)$ does not have poles
and it vanishes for $t=-\pi g$.

The main advantage of the toy model is that, as we will show in
Sect.~\ref{cusp-toy},  the expression  for  $\Gamma^{\toy}_{\rm cusp}(g)$ can be
found  in a closed form for arbitrary value of the coupling constant (see
Eq.~\re{cusp-toy-fin} below). We will then compare it with the exact expression
for $\Gamma_{\rm cusp}(g)$ and identify the difference between the two functions.

\subsection{Exact bounds and unicity of the solution}

Before we turn to finding the solution to \re{sys}, let us demonstrate
that this integral equation supplemented with the additional conditions
\re{Bessel} and \re{inverse} on its solutions, leads to nontrivial constraints
for the cusp anomalous dimension valid for arbitrary coupling $g$.

Let us multiply both sides of the two relations in \re{FRS2} by
$2(2n-1)\gamma_{2n-1}$ and $2(2n)\gamma_{2n}$, respectively, and perform
summation over $n \ge 1$. Then, we convert the sums into the functions
$\gamma^{\BES}_\pm(t)$ using \re{Bessel} and add the second relation to the first
one to obtain \footnote{Our analysis here goes along the same lines as in
Appendix A of \cite{BK08}. }
\be\label{uni}
\gamma_1 = \int_{0}^{\infty}\frac{dt}{t}
\frac{(\gamma^{\BES}_{+}(t))^2+(\gamma^{\BES}_{-}(t))^2}{1-\e^{-t/(2g)}}  \,.
\ee
Since $\gamma^{\BES}_{\pm}(t)$ are real functions of $t$ and the denominator is
positively definite for $0\le t<\infty$,  this relation leads to the following
inequality
\be\label{ineq}
\gamma_1 \ge \int_{0}^{\infty}\frac{dt}{t}
 (\gamma^{\BES}_{-}(t))^2 \ge 2 \gamma^2_1  \ge 0\,.
\ee
Here we replaced the function $\gamma^{\BES}_{-}(t)$ by its Bessel series
\re{Bessel} and made use of the orthogonality condition for the Bessel functions
with odd indices. We deduce from \re{ineq} that
\be
0\le \gamma_1  \le \frac12
\ee
and, then, apply \re{cusp1} to translate this inequality into the following
relation for the cusp anomalous dimension
\begin{align}\label{bounds}
0 \le \Gamma_{\rm cusp}(g) \le 4 g^2\,.
\end{align}
We would like to stress that this relation should hold in planar $\mathcal{N}=4$
SYM theory for arbitrary coupling $g$.

Notice that the lower bound on the cusp anomalous dimension, $\Gamma_{\rm
cusp}(g)\ge 0$, holds in any gauge theory~\cite{BGK03}.  It is the upper bound $
\Gamma_{\rm cusp}(g) \le 4 g^2$ that is a distinguished feature of ${\cal N}=4$
theory. Let us verify the validity of \re{bounds}. At weak coupling  $\Gamma_{\rm
cusp}(g)$ admits perturbative expansion in powers of $g^2$~\cite{cusp-4loop}
\be\label{weak}
\Gamma_{\rm cusp}(g) = 4 g^2\bigg[1-\frac{1}{3} \pi^2 g^2 +\frac{11}{45} \pi^4 g^4 - 2
\lr{\frac{73}{630}\pi^6 + 4\zeta_3^2} g^6
+\ldots\bigg]\,,
\ee
while at strong coupling it has the form~\cite{BKK07,Roiban:2007dq,KSV08}
\be\label{strong}
\Gamma_{\rm cusp}(g) = 
2g \left[1 - \frac{3\ln 2}{4\pi} g^{-1} - \frac{\rm K}{16\pi^2}g^{-2}- \lr{\frac{3 {\rm K}\ln 2 }{64\pi^3}+\frac{27\zeta_3}{2048 \pi^3}}g^{-3}+ O(g^{-4})\right]
\,,
\ee
with ${\rm K}$ being the Catalan constant. It is easy to see that the relations
\re{weak} and \re{strong} are in an agreement with \re{bounds}.

For arbitrary $g$ we can verify the relation \re{bounds} by using the results for
the cusp anomalous dimension obtained from numerical solution of the BES
equation~\cite{BKK07,K08}. The comparison is shown in Figure~1. We observe that
the upper bound condition $\Gamma_{\rm cusp}(g)/(2g) \le 2g$ is indeed satisfied
for arbitrary $g>0$.

\newcommand{\insertfig}[2]{\mbox{\epsfysize=#1cm \epsfbox{#2.eps}}}%
\begin{figure}[h]%
\begin{center}%
\mbox{\begin{picture}(0,240)(150,0)%
\put(0,0){\insertfig{8}{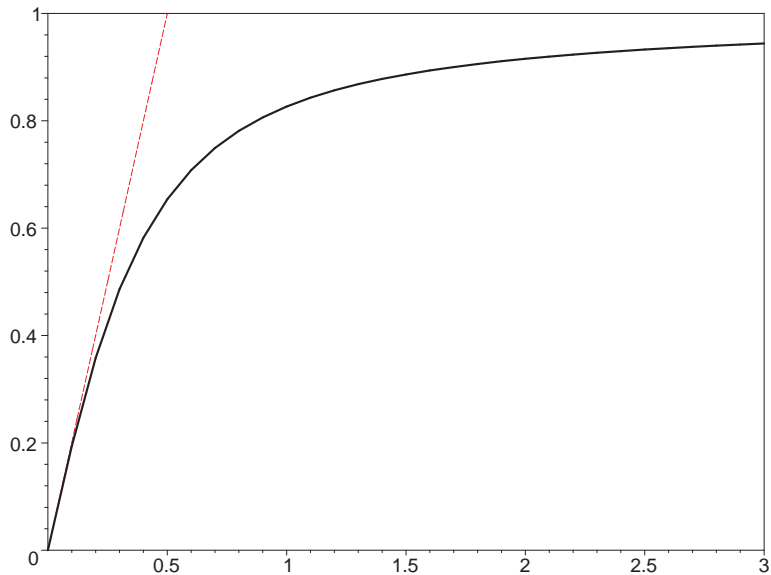}}
\end{picture}}%
\end{center}%
\caption{\small Dependence of the cusp anomalous dimension $\Gamma_{\rm cusp}(g)/(2g)$
on the coupling constant. Dashed line denotes the upper bound $2g$.}%
\end{figure}%

We are ready to show that the analyticity conditions formulated in Sect.~2.2 specify a unique
solution to \re{FRS2}. As was already mentioned, solutions to \re{FRS2} are defined modulo
contribution of zero modes, $\gamma(t) \to \gamma(t)+\gamma^{(0)}(t)$, with $\gamma^{(0)}(t)$ being
solution  to homogenous equations. Going through the same steps that led us to \re{uni} we obtain 
\be\label{uni1}
0 = \int_{0}^{\infty}\frac{dt}{t}
\frac{(\gamma^{(0)}_{+}(t))^2+(\gamma^{(0)}_{-}(t))^2}{1-\e^{-t/(2g)}}  \,,
\ee
where zero on the left-hand side is due to absence of the inhomogeneous term. Since the integrand
is a positively definite function, we immediately deduce that $\gamma^{(0)}(t)=0$ and, therefore,
the solution for $\gamma(t)$ is unique.

\subsection{Riemann-Hilbert problem}

Let us now construct the exact solution to the integral equations \re{G-int} and
\re{FRS2}. To this end, it is convenient to Fourier transform the functions
\re{g_pm} and \re{rel}
\be\label{Fourier}
\widetilde \Gamma(k) = \int_{-\infty}^\infty\frac{dt}{2\pi}\e^{ikt}\Gamma(t)\,,\qquad \widetilde
\gamma(k) = \int_{-\infty}^\infty\frac{dt}{2\pi}\e^{ikt}\gamma(t)\,.
\ee
According to \re{g_pm} and \re{Bessel}, the function $\gamma(t)$ is given by the
Neumann series over Bessel functions. Then, we perform the Fourier transform on
both sides of \re{Bessel} and use the well-known fact that the Fourier transform
of the Bessel function $J_n(t)$ vanishes for $k^2 > 1$ to deduce that the same is
true for $\gamma(t)$ leading to
\be\label{g-zero}
\widetilde\gamma(k) = 0\,,\qquad \text{for $k^2 >1$.}
\ee
This implies that  the Fourier integral for $\gamma(t)$ only involves modes with
$-1 \le k \le 1$ and, therefore, the function  $\gamma(t)$ behaves at large
(complex) $t$ as
\be\label{infinity}
\gamma(t) \sim \e^{|t|}\,,\qqquad \mbox{for $|t|\to\infty$.}
\ee
Let us now examine the function $\widetilde \Gamma(k)$. We find from \re{Fourier} and
\re{sin-sin} that  $\widetilde \Gamma(k)$ admits the following representation
\be\label{int1}
\widetilde \Gamma(k) = \int_{-\infty}^\infty \frac{dt}{2\pi}\e^{ikt}
\frac{\sinh(\frac{t}{4g}+i\frac{\pi}4)}{\sinh(\frac{t}{4g})\sin(\frac{\pi}4)}\gamma(t)\,.
\ee
Here the integrand has poles along the imaginary axis at $t=4\pi i g n$ (with
$n=\pm 1, \pm 2,\ldots$). \footnote{We recall that $\gamma(t) = O(t)$ and,
therefore, the integrand is regular at $t=0$.}

It is suggestive to evaluate the integral \re{int1} by deforming the integration
contour to infinity and by picking up residues at the poles. However, taking into
account the relation \re{infinity}, we find that the contribution to \re{int1} at
infinity can be neglected for $k^2 > 1$ only. In this case, closing the
integration contour into the upper (or lower) half-plane for $k>1$ (or $k< -1$)
we find
\be\label{G-exponential}
\widetilde \Gamma(k) \stackrel{k^2>1}{=} \theta(k-1)\sum_{n\ge 1} c_+(n,g) \e^{-4\pi n g (k-1)} +\,
\theta(-k-1)\sum_{n\ge 1} c_-(n,g) \e^{-4\pi n g (-k-1)}\,.
\ee
Here the notation was introduced for $k-$independent expansion coefficients
\be\label{c-gamma}
c_\pm (n,g) = \mp 4g \gamma(\pm 4\pi i gn) \e^{-4\pi ng}\,,
\ee
where the factor $\e^{-4\pi ng}$ is inserted to compensate exponential growth of
$\gamma(\pm 4\pi i gn)\sim \e^{4\pi ng}$ at large $n$ (see Eq.~\re{infinity}).
For $k^2\le 1$, we are not allowed to neglect the contribution to \re{int1} at
infinity and the relation \re{G-exponential} does not hold anymore. As we will
see in a moment, for $k^2\le 1$ the function $\widetilde \Gamma(k)$ can be found
from \re{G-int}.

Comparing the relations \re{g-zero} and  \re{G-exponential}, we conclude that, in
distinction with $\widetilde\gamma(k)$, the function $\widetilde \Gamma(k)$ does
not vanish for $k^2>1$. Moreover, each term in the right-hand side of
\re{G-exponential} is exponentially small at strong coupling and the function
scales at large $k$ as $\widetilde \Gamma(k)\sim \e^{-4\pi g (|k|-1)}$. This
implies that nonzero values of $\widetilde \Gamma(k)$ for $k^2 > 1$ are of
nonperturbative origin. Indeed, in perturbative approach of \cite{BKK07},  the
function $\Gamma(t)$ is given by the Bessel function series analogous to
\re{Bessel} and, similar to \re{g-zero}, the function $\widetilde\Gamma(k)$
vanishes for $k^2 > 1$ to any order in $1/g$ expansion.

We note that the sum in the right-hand side of \re{G-exponential} runs over poles
of the function $\Gamma(it)$ specified in \re{poles}. We recall that in the toy
model \re{toy}, $\Gamma^{\toy}(it)$ and $\gamma^{\toy}(it)$  are entire functions
of $t$. At large $t$ they have the same asymptotic behavior as the Bessel
functions, $\Gamma^{\toy}(it)\sim\gamma^{\toy}(it)\sim \e^{\pm it} $. Performing
their Fourier transformation \re{Fourier}, we find
\be
\widetilde\gamma^{\toy}(k) =\widetilde\Gamma^{\toy}(k) = 0\,,\qquad \text{for $k^2 >1$}\,,
\ee
in a close analogy with \re{g-zero}. Comparison with \re{G-exponential} shows
that the coefficients \re{c-gamma} vanish in the toy model for arbitrary $n$ and
$g$
\be\label{c-toy}
c^{\toy}_+(n,g) = c^{\toy}_-(n,g) = 0\,.
\ee

The relation \re{G-exponential} defines the function $\widetilde\Gamma(k)$ for
$k^2> 1$ but it involves the coefficients $c_\pm(n,g)$ that need to be
determined. In addition, we have to construct the same function for $k^2 \le 1$.
To achieve both goals, let us return to the integral equations \re{sys} and
replace $\Gamma^{\BES}_\pm(t)$ by Fourier integrals (see Eqs.~\re{Fourier} and
\re{rel})
\begin{align}
\Gamma^{\BES}_+(t) &= \int_{-\infty}^\infty dk\,\cos(kt)\, \widetilde \Gamma(k)\,, 
\\ \notag
\Gamma^{\BES}_-(t) &= -\int_{-\infty}^\infty dk\,\sin(kt)\, \widetilde \Gamma(k)\,.
\end{align}
In this way, we obtain from \re{sys} the following remarkably
simple integral equation for $\widetilde\Gamma(k)$
\be\label{master}
\dashint_{-\infty}^\infty \frac{dk\,\widetilde \Gamma(k)}{k-u}+\pi  \widetilde
\Gamma(u)=-2\,,\qqqquad (-1\le u \le 1)\,,
\ee
where the integral is defined using the principal value prescription. 
This relation is equivalent to the functional equation obtained in \cite{KSV08} (see Eq.~(55) there).

Let us
split the integral in \re{master} into $k^2\le 1$ and $k^2 > 1$ and rewrite
\re{master} in the form of singular integral equation 
for the function $\widetilde \Gamma(k)$ on the interval $-1 \le k \le 1$
\be\label{RH}
\widetilde \Gamma(u)+ \frac1{\pi}\dashint_{-1}^1 \frac{dk\,\widetilde \Gamma(k)}{k-u} =\phi(u)
\,,\qqqquad (-1\le u \le 1)\,,
\ee
where the inhomogeneous term is given by
\be\label{f}
\phi(u) = -\frac1{\pi}\lr{2+\int_{-\infty}^{-1} \frac{dk\,\widetilde
\Gamma(k)}{k-u}+\int^{\infty}_1 \frac{dk\,\widetilde \Gamma(k)}{k-u}}\,.
\ee
Since integration in \re{f} goes over $k^2>1$, the function $\widetilde
\Gamma(k)$ can be replaced in the right-hand side of \re{f} by its expression
\re{G-exponential} in terms of the coefficients $c_\pm(n,g)$.

The integral equation \re{RH} can be solved by standard methods \cite{Mikhlin}. A
general solution for $\widetilde \Gamma(k)$ reads (for $-1\le k \le 1$)
\be
\widetilde \Gamma(k) = \frac1{2}\phi(k)-\frac1{2\pi}\lr{\frac{1+k}{1-k}}^{1/4}
\dashint_{-1}^1\frac{du\, \phi(u)}{u-k}\lr{\frac{1-u}{1+u}}^{1/4}
-\frac{\sqrt{2}}{\pi}\lr{\frac{1+k}{1-k}}^{1/4}{\frac{c}{1+k}}\,,
\ee
where the last term describes the zero mode contribution with $c$ being an
arbitrary function of the coupling. We replace $\phi(u)$ by its expression
\re{f}, interchange the order of integration and find after some algebra
\be\label{G-small}
\widetilde \Gamma(k) \stackrel{k^2\leqslant 1}{=}
-\frac{\sqrt{2}}{\pi}\lr{\frac{1+k}{1-k}}^{1/4}\left[
 {1+\frac{c}{1+k}}
 +\frac12\int_{-\infty}^\infty
 \frac{dp\,\widetilde\Gamma(p)}{p-k}\lr{\frac{p-1}{p+1}}^{1/4}\theta(p^2-1)\right]\,.
\ee
Notice that the integral in the right-hand side of \re{G-small} goes along the
real axis except the interval $[-1,1]$ and, therefore, $\widetilde\Gamma(p)$ can
be replaced by its expression \re{G-exponential}.

Being combined together, the relations \re{G-exponential} and \re{G-small} define
the function $\widetilde \Gamma(k)$ for $-\infty < k < \infty$ in terms of (an
infinite) set of yet unknown coefficients $c_\pm(n,g)$ and $c(g)$. To fix these
coefficients we will first perform Fourier transform of $\widetilde \Gamma(k)$ to
obtain the function $\Gamma(t)$ and, then, require that $\Gamma(t)$ should have
correct analytical properties \re{poles}.

\subsection{General solution}

We are now ready to write down a general expression for the function $\Gamma(t)$.
According to \re{Fourier}, it is related to $\widetilde \Gamma(k)$ through the
inverse Fourier transformation
\begin{align}\label{three-terms}
\Gamma(t)  = \int_{-1}^1 dk\e^{-ikt} \widetilde \Gamma(k) +\int^{-1}_{-\infty} dk\e^{-ikt}
\widetilde \Gamma(k) +\int_{1}^\infty dk\e^{-ikt} \widetilde \Gamma(k)\,,
\end{align}
where we split the integral into three terms since $\widetilde \Gamma(k)$ has a
different form for $k<-1$, $-1\le k\le 1$ and $k>1$. Then, we use the obtained
expressions for $\widetilde \Gamma(k)$, Eqs.~\re{G-exponential} and \re{G-small},
to find after some algebra the following remarkable relation (see
Appendix~\ref{App:B} for details)
\begin{align} \label{G-gen}
\Gamma(it) &= f_0(t)\I_0(t)+f_1(t) \I_1(t)\,.
\end{align}
Here the notation was introduced for
\begin{align}\label{ff}
f_0(t) & = -1+  \sum_{n\ge 1} t \left[{c_+(n,g)}\frac{  U_1^+(4\pi ng) }{4\pi ng-t}+
  {c_-(n,g)}  \frac{   U_1^-(4\pi ng)}{4\pi ng+t}\right]\,,
\\ \notag
f_1(t) &= -c(g) +\sum_{n\ge 1}   4\pi ng\left[{c_+(n,g)}\frac{  U_0^+(4\pi ng) }{4\pi ng-t} +
  {c_-(n,g)}  \frac{ U_0^-(4\pi ng) }{4\pi ng+t}\right] \,.
\end{align}
Also, $\I_n$ and $U_n^\pm$ (with $n=0,1$) stand for integrals
\begin{align}\label{I-U}
\I_n(x)&= \frac{\sqrt{2}}{\pi}\int_{-1}^1du\,
(1+u)^{1/4-n}(1-u)^{-1/4}\e^{u x}\,,
\\ \notag
U_n^\pm(x) &=   \frac1{2} \int_{1}^\infty
du\, (u\pm 1)^{-1/4}(u\mp1)^{1/4-n} \e^{-(u-1)x}\,,
\end{align}
which can be expressed in terms of Whittaker functions of 1st and 2nd kind
\cite{WW} (see Appendix~\ref{App:D}). We would like to emphasize that the
solution \re{G-gen} is exact for arbitrary coupling $g>0$ and that
the only undetermined ingredients in \re{G-gen} are  the expansion coefficients
$c_\pm(n,g)$ and $c(g)$. 

In the special case of the toy model, Eq.~\re{c-toy}, the expansion coefficients
vanish, $c_\pm^{\toy}(n,g)=0$,  and the relation \re{ff} takes a simple form
\be\label{f-toy}
f^{\toy}_0(t) = -1\,,\qqquad f_1^{\toy}(t) = -c^{\toy}(g)\,.
\ee
Substituting these expressions into \re{G-gen} we obtain a general solution to
the integral equation \re{G-int} in the toy model
\begin{align} \label{G-toy}
\Gamma^{\toy}(it) &= -\I_0(t)-c^{\toy}(g) \I_1(t)\,.
\end{align}
It involves an arbitrary $g-$dependent constant $c^{\toy}$ which will be
determined in Sect.~\ref{cusp-toy}.

\subsection{Quantization conditions}

The relation \re{G-gen} defines a general solution to the integral equation
\re{G-int}. It still depends on the coefficients  $c_\pm(n,g)$ and $c(g)$ that need
to be determined. We recall that $\Gamma(it)$ should have poles and
zeros specified in \re{poles}.

Let us first examine poles in the right-hand side of \re{G-gen}. It follows from
\re{I-U} that $\I_0(t)$ and $\I_1(t)$ are entire functions of $t$ and, therefore,
poles can only come from the functions $f_0(t)$ and $f_1(t)$. Indeed, the sums
entering  \re{ff} produce an infinite sequence of poles located at $t=\pm 4\pi n$
(with $n\ge 1$) and, as a result, the solution \re{G-gen} has a correct pole
structure \re{poles}. Let us now require that  $\Gamma(it)$ should vanish  for
$t=t_{\rm zero}$ specified in \re{poles}. This leads to an infinite set of
relations
\be\label{QC0}
\Gamma\left(4\pi ig \lr{\ell-\ft14}\right) 
= 0 \,,\qqqquad  {\ell\in \mathbb{Z}}\,.
\ee
Replacing $\Gamma(it)$ by its expression \re{G-gen}, we rewrite these relations
in equivalent form
\begin{align} \label{QC01}
 f_0\left(t_{\ell}\right)\I_0\left(t_{\ell}\right)
 +f_1(t_{\ell}) \I_1(t_{\ell}) = 0 \,,\qqqquad  t_{\ell}  =4\pi g \lr{\ell-\ft14}\,.
\end{align}
The relations \re{QC0} and \re{QC01} provide the quantization conditions for the
coefficients  $c(g)$ and $c_\pm(n,g)$ that we will analyze  in
Sect.~\ref{QUANT}.

Let us substitute \re{G-gen} into the expression \re{cusp} for the cusp anomalous
dimension. The result involves the functions  $\I_n(t)$ and $f_n(t)$ (with
$n=1,2$) evaluated at $t=0$. It is easy to see from \re{I-U} that $\I_0(0) = 1$
and $\I_1(0)=2$. In addition, we obtain from \re{ff} that $f_0(0)=-1$ for
arbitrary coupling leading to
\begin{align}\label{cusp=f}
\Gamma_{\rm cusp}(g) =  2g\big[ 1 - 2 f_1(0)\big]\,.
\end{align}
Replacing $f_1(0)$ by its expression \re{ff} we find
the following relation for the cusp anomalous dimension in terms of the coefficients $c$ and $c_\pm$
\be\label{cusp-c}
\Gamma_{\rm cusp}(g) =  2g\left\{1+2c(g) - 2\sum_{n\ge 1}\big[{ c_-(n,g) U_0^-(4\pi
ng) +c_+(n,g) U_0^+(4\pi ng)}\big] \right\}\,.
\ee
We would like to stress that the relations \re{cusp=f} and \re{cusp-c} are exact and hold for arbitrary
coupling $g$. This implies that, at weak coupling, it should reproduce the known
expansion of $\Gamma_{\rm cusp}(g)$ in \textit{positive integer} powers of
$g^2$~\cite{cusp-4loop}. Similarly, at strong coupling, it should reproduce the
known $1/g$ expansion \cite{BKK07,KSV08} and, most importantly, describe
nonperturbative, exponentially suppressed corrections to $\Gamma_{\rm cusp}(g)$.

\subsection{Cusp anomalous dimension in the toy model}\label{cusp-toy}

As before, the situation simplifies for the toy model \re{G-toy}. In this case,
we have only one quantization condition $\Gamma^{\toy}(-\pi ig) = 0$ which
follows from \re{toy}. Together with \re{G-toy} it allows us to fix the
coefficient $c^{\toy}(g)$ as
\be\label{c1-toy}
c^{\toy}(g) = - \frac{\I_0(-\pi g)}{\I_1(-\pi g)}\,.
\ee
Then, we substitute the relations
\re{c1-toy} and \re{c-toy} into \re{cusp-c} and obtain
\begin{align}\label{cusp-toy-I}
\Gamma^{\toy}_{\rm cusp}(g) & = 2g\left[{1+2c^{\toy}(g)}\right] = 2g\left[1 - 2\frac{\I_0(-\pi
g)}{\I_1(-\pi g)}\right]\,.
\end{align}
Replacing $\I_0(-\pi g)$ and $\I_1(-\pi g)$ by their expressions in terms of
Whittaker function of the first kind (see Eq.~\re{I=M}), we find the following
remarkable relation
\begin{align}\label{cusp-toy-fin}
\Gamma^{\toy}_{\rm cusp}(g)   =  2g\left[1 -  (2\pi g)^{-1/2}\frac{M_{1/4,1/2}(2\pi
g)}{M_{-1/4,\,0}(2\pi g)}\right]\,,
\end{align}
which defines the cusp anomalous dimension in the toy model for arbitrary coupling $g>0$.

Using \re{cusp-toy-fin} it is straightforward to compute  $\Gamma^{\toy}_{\rm
cusp}(g)$ for arbitrary positive $g$. By construction,
$\Gamma^{\toy}_{\rm cusp}(g)$ should be different from $\Gamma_{\rm cusp}(g)$.
Nevertheless, evaluating \re{cusp-toy-fin} for $0\le g \le 3$, we found that
the numerical values of $\Gamma^{\toy}_{\rm cusp}(g)$ are very close to the exact
values of the cusp anomalous dimension shown by the solid line in Figure~1. Also, as
we will show in a moment, the two functions have similar properties at strong 
coupling. To compare these functions, it is instructive to
examine the asymptotic behavior of $\Gamma^{\toy}_{\rm cusp}(g) $ at weak and at strong coupling.

\subsubsection{Weak coupling}

At weak coupling, we find from  \re{cusp-toy-fin}
\begin{align}\label{toy-weak}
\Gamma^{\toy}_{\rm cusp}(g)   =\frac32\,\pi \,{g}^{2}-\frac12\,{\pi
}^{2}{g}^{3}-{\frac {1}{64}}\,{\pi }^{3}{ g}^{4}+{\frac {5}{64}}\,{\pi
}^{4}{g}^{5}-{\frac {11}{512}}\,{\pi }^{5 }{g}^{6}-{\frac {3}{512}}\,{\pi
}^{6}{g}^{7}
+O({g}^{8})\,.
\end{align}
Comparison with \re{weak} shows that this expansion is quite different from the
weak coupling expansion of the cusp anomalous dimension. 
In distinction with
$\Gamma_{\rm cusp}(g)$, the expansion in \re{toy-weak} runs both in even and odd
powers of the coupling. In addition, the coefficient in front of $g^n$ in the
right-hand side of \re{toy-weak} has transcendentality $(n-1)$ while for
$\Gamma_{\rm cusp}(g)$ it equals $(n-2)$  (with $n$ taking even values only).

Despite of this and similarly to the weak coupling expansion of the cusp anomalous
dimension~\cite{BES}, the series \re{toy-weak} has a finite radius of convergence $|g_0|=0.796$. 
It is determined by the position of the zero of the Whittaker function closest to the origin,
$M_{-1/4,0}(2\pi g_0) =0$ for $g_0 = -0.297 \pm i\, 0.739$.  Moreover, numerical analysis indicates that  $\Gamma^{\toy}_{\rm cusp}(g)$ has  an infinite number of poles in the complex $g-$plane. The poles are located in the left-half side of the complex plane, $\textrm{Re}\, g < 0$, symmetrically with 
respect to the real axis, and they approach progressively the imaginary axis as one goes away from
the origin.

\subsubsection{Strong coupling}

At strong coupling, we can replace the Whittaker functions in \re{cusp-toy-fin}
by their  asymptotic expansion for $g\gg 1$. It is convenient however to apply
\re{cusp-toy-I} and replace the functions $\I_0(-\pi
g)$ and $\I_1(-\pi g)$  by their expressions given  in \re{I0-sep} and \re{I1-sep}, respectively. In
particular, we have (see Eq.~\re{I0-sep})
\begin{align}\label{I0-pos}
\I_0(-\pi g)  & = \e^{1/(2\alpha)} \frac{\alpha^{5/4}}{\Gamma(\ft34)}
\bigg[F\lr{\ft14,\ft54|\alpha+i0} +   \Lambda^2
F\lr{-\ft14,\ft34|-\alpha}\bigg]\,,\qquad {\alpha=1/(2\pi g)}\,,
\end{align}
where the parameter $\Lambda^2$ is defined as
\be\label{Lambda1}
\Lambda^2 = \sigma\, \alpha^{-1/2}
\e^{-1/\alpha}\frac{\Gamma(\ft34)}{\Gamma(\ft54)}\,,\qqqquad
\sigma=\e^{-3i\pi/4}\,.
\ee
Here, $F\lr{a,b|-\alpha}$ is expressed in terms of the confluent 
hypergeometric function of the second kind (see Eqs.~\re{U-large} and \re{U's} in Appendix~D and Eq.~\re{F-Borel} below)~\cite{WW} 
\begin{align}\label{F-series0}
F\lr{\ft14,\ft54|-\alpha} \phantom{-}&=  {\alpha^{-5/4}}  U_{0}^{+}\left(1/(2\alpha)\right)/\Gamma(\ft54)\,, \qquad
\\[3mm] \notag
F\lr{-\ft14,\ft34|-\alpha} & = \alpha^{-3/4}U_{0}^{-}\left(1/(2\alpha)\right)/\Gamma(\ft34)
\,.
\end{align}
The function $F\lr{a,b|-\alpha}$ defined in this way is an analytical 
function of $\alpha$  with a cut along the negative semi-axis. 

 For positive $\alpha=1/(2\pi g)$, the function
$F\lr{-\ft14,\ft34|-\alpha}$ entering \re{I0-pos} is defined away from the cut and its large $g$
expansion is given by Borel summable asymptotic series (for $a=-\ft14$ and $b=\ft34$)
\be\label{F-series}
F\lr{a,b|-\alpha} = \sum_{k\ge 0} \frac{(-\alpha)^k}{k!}
 \frac{\Gamma(a+k)\Gamma(b+k)}{\Gamma(a)\Gamma(b)}  
= 1- \alpha ab +
O(\alpha^2)\,,
\ee
with the expansion coefficients growing factorially to higher orders in $\alpha$. This series can be immediately resummed 
 by means of the Borel resummation method. Namely, replacing $\Gamma(a+k)$ by its integral representation and performing the sum over $k$ we find for $\textrm{Re}\, \alpha > 0$
\be\label{F-Borel}
F\lr{a,b|-\alpha} = \frac{\alpha^{-a}}{\Gamma(a)}\int_{0}^{\infty}ds\, s^{a-1}(1+s)^{-b} \e^{-s/\alpha}\,,
\ee
in agreement with \re{F-series0} and \re{I-U}. 

The relation \re{F-series} holds in fact for arbitrary complex $\alpha$ and the functions $F\lr{a,b|\alpha \pm i0}$, defined  for $\alpha > 0$ above and below the cut, respectively, are given  by the same asymptotic expansion \re{F-series} with $\alpha$ replaced by $-\alpha$.  The important difference is that now the series \re{F-series} is not Borel summable anymore. Indeed, if one attempted to resum this series using the Borel summation method, one would immediately find a branch point singularity 
along the integration contour at $s=1$
\be\label{F-deform}
F\lr{a, b| \alpha\pm i0}  = \frac{\alpha^{-a}}{\Gamma(a)}\int_{0}^{\infty}ds\, s^{a-1}(1-s\mp i0)^{-b}\e^{-s/\alpha}\,. 
\ee
The ambiguity related to the choice of the prescription to integrate over the singularity is known as Borel ambiguity.  In particular, deforming the $s-$integration contour above or below the cut, one obtains
two different functions $F\lr{a, b| \alpha\pm i0}$. They define analytical continuation of the same
function $F\lr{a,b|-\alpha}$ from $\Re \alpha>0$ to the upper and lower edge of the cut running along
the negative semi-axis. Its discontinuity across the cut, $F\lr{a, b| \alpha+ i0} -F\lr{a, b| \alpha-i0}$ is 
exponentially suppressed at small $\alpha>0$ and is proportional to the nonperturbative
scale $\Lambda^2$ (see Eq.~\re{disc-F}). This property is perfectly consistent with the fact
that the function \re{I0-pos} is an entire function of $\alpha$. Indeed, it takes the same form
if one used $\alpha-i0$ prescription in the first term in the right-hand side of \re{I0-pos} and
replaced $\sigma$ in \re{Lambda1} by its complex conjugated value. 
  
We can now elucidate the reason for decomposing the  entire $V_0-$function in \re{I0-pos}  into the
sum of two $F-$functions. In spite of the fact that analytical properties of the former function are
simpler compared to the latter functions, its asymptotic behavior at large $g$ is more complicated.
Indeed, the $F-$functions admit asymptotic expansions in the whole complex $g-$plane and they can be unambiguously defined through the Borel resummation once their analytical properties are specified
(we recall that the  function  $F\lr{a, b| \alpha}$  has a cut along positive semi-axis). In distinction with this, the entire function $V_0(-\pi g)$ admits different asymptotic behaviors for positive and negative values of $g$ in virtue of the Stokes phenomenon.
Not only does it restrict the domain of validity of each asymptotic expansion, but it also forces us to keep track of both perturbative and nonperturbative contributions in the transition region from positive
to negative $g$,  including the transition from the strong to weak coupling.

We are now in position to discuss the strong coupling expansion of the cusp anomalous dimension in the toy model, including into our consideration both perturbative and nonperturbative contributions. Substituting \re{I0-pos} and similar relation for $\I_1(-\pi g)$  (see
Eq.~\re{I1-sep})  into \re{cusp-toy-I} we find (for $\alpha^+\equiv\alpha+i0$ and
$\alpha=1/(2\pi g)$)
\begin{align}\label{toy-ratio}
\Gamma^{\toy}_{\rm cusp}(g) /( 2g) = 1-  \alpha
\frac{F\lr{\frac14,\frac54|\alpha^+}+ \Lambda^2
F\lr{-\frac14,\frac34|-\alpha}}{F\lr{\frac14,\frac14|\alpha^+} + \frac14
\Lambda^2 {\alpha}  F\lr{\frac34,\frac34|-\alpha}}\,.
\end{align}
Since the parameter $\Lambda^2$  is exponentially suppressed at strong coupling, Eq.~\re{Lambda1}, and, at the same time, the $F-$functions are all of the same order, it makes sense to expand the right-hand side of \re{toy-ratio} in powers of $\Lambda^2$ and, then, study separately each coefficient function.  In this way, we  identify the leading, $\Lambda^2$ independent term as perturbative contribution to $\Gamma^{\toy}_{\rm cusp}(g)$ and the $O(\Lambda^2)$ term as the leading nonperturbative correction. More precisely, expanding the right-hand side of \re{toy-ratio} in powers of $\Lambda^2$ we obtain
\be\label{toy-f}
\Gamma^{\toy}_{\rm cusp}(g) /( 2g) = C_0(\alpha) - \alpha\Lambda^2 C_2(\alpha) + \frac14\alpha^2\Lambda^4 C_4(\alpha) + O(\Lambda^6)\,.
\ee
Here the expansion runs in even powers of $\Lambda$ and the coefficient functions
$C_k(\alpha)$ are given by algebraic combinations of $F-$functions
\begin{align} \label{toy-cc}
C_0 & =1-  \alpha \frac{ F\lr{\frac14,\frac54|\alpha^+}}{F\lr{\frac14,\frac14|\alpha^+}}\,,
\qqquad
C_2 =
\frac{1}{\left[F\lr{\frac14,\frac14|\alpha^+}\right]^2}\,, \qqquad 
C_4= \frac{F\lr{\frac34,\frac34|-\alpha}}{\left[F\lr{\frac14,\frac14|\alpha^+}\right]^3}\,,
\end{align}
where we applied   \re{U-Wr} and \re{U-large} to simplify the last two relations.
Since the coefficient functions are expressed in terms of the functions $F(a,b|\alpha^+)$ and
$F(a,b|-\alpha)$ having the cut along the positive and negative semi-axis, respectively, $C_k(\alpha)$ are analytical functions of $\alpha$ in the upper-half place.
 
Let us now examine the strong coupling expansion of the coefficient functions \re{toy-cc}.
Replacing $F-$functions in \re{toy-cc} by their asymptotic series representation \re{F-series} we get
\begin{align} \label{c0-series}
C_0 &=  1-\alpha-{\frac {1}{4}}{\alpha}^{2}-{\frac {3}{8}}{\alpha}^{3}-{
\frac {61}{64}}{\alpha}^{4}-{\frac {433}{128}}{\alpha}^{5} +O
 \left( {\alpha}^{6} \right) \,,
 \\[2mm]\notag
 C_2&=1-{\frac {1}{8}}{\alpha} -{\frac {11}{128}}{\alpha}^{2}-{\frac
{151}{1024}}{\alpha}^{3}-{\frac {13085}{32768}}{\alpha}^{4}+  O \left(
{\alpha}^{5} \right)\,,
\\[2mm] \notag
 C_4&=1-{\frac {3}{4}}{\alpha} -{\frac {27}{32}}{\alpha}^{2}-{\frac
{317}{128}}{\alpha}^{3} 
+  O \left({\alpha}^{4} \right)\,.
\end{align}
Not surprisingly, these expressions inherit the properties of the  $F-$functions -- the series
\re{c0-series} are asymptotic and non-Borel summable. If one simply substituted the relations \re{c0-series} into the right-hand side of \re{toy-f}, one would then worry about the meaning
of nonperturbative $O(\Lambda^2)$ corrections to \re{toy-f} given the fact that the strong coupling expansion of
perturbative contribution $C_0(\alpha)$ suffers from Borel ambiguity. We recall that appearance
of exponentially suppressed corrections to $\Gamma^{\toy}_{\rm cusp}(g)$ is ultimately
related to the Stokes phenomenon for the function $V_0(-\pi g)$, Eq.~\re{I0-pos}. As was already mentioned,  this does not happen for the $F-$function and, as a consequence, its asymptotic
expansion, supplemented with the additional analyticity conditions, allows us to reconstruct
the $F-$function through the Borel transformation, Eqs.~\re{F-Borel} and \re{F-deform}. Since the 
coefficient functions \re{toy-cc} are expressed in terms of the $F-$functions, we may expect that
the same should be true for the $C-$functions. Indeed, it follows from the unicity condition of
asymptotic expansion \cite{Zinn-Justin},  that the functions $C_{0}(\alpha)$, $C_{2}(\alpha)$,  $C_{4}(\alpha)$, $\ldots$ are uniquely determined by their series representations \re{c0-series} as soon as the latter are understood as asymptotic expansions for the functions  analytical in the upper half plane $\Im \alpha \ge 0$. This implies that the exact expressions for the functions  \re{toy-cc} can be unambiguously constructed by means of the Borel resummation but the explicit construction remains beyond the scope of the present study.

Since the expression \re{toy-ratio} is exact for arbitrary coupling $g$ we may now address the question
formulated in the Introduction: how does the transition from the strong to the weak coupling regime
occur? We recall that,  in the toy model, $\Gamma^{\toy}_{\rm cusp}(g) /(2g)$ is given for $g\ll 1$ and $g\gg 1$ by the relations \re{toy-weak} and \re{toy-f}, respectively.  Let us choose some sufficiently small value of the coupling constant, say $g=1/4$, and compute  $\Gamma^{\toy}_{\rm cusp}(g) /(2g)$ using three different representations. Firstly, we substitute $g=0.25$ into \re{toy-ratio} and find the exact value as $0.4424(3)$. Then, we use the weak coupling expansion \re{toy-weak}
and obtain a close value $0.4420(2)$. Finally, we use the strong coupling expansion \re{toy-f} and
evaluate the first few terms in the right-hand side of \re{toy-f} for $g=0.25$ to get
\begin{align} \notag
\rm{Eq.}~\re{toy-f} & = (0.2902-0.1434 \,i) + (0.1517+0.1345\, i) 
\\ \label{sum-comp}
&  + (0.0008+0.0086\, i) - (0.0002-0.0003\, i) + \ldots   = 0.4425 + \ldots
\end{align}
Here the four expressions inside the round brackets correspond  to contributions proportional to $\Lambda^0$,  $\Lambda^2$,  $\Lambda^4$ and  $\Lambda^6$, respectively, with $\Lambda^2(g=0.25)=0.3522\times \e^{-3i\pi/4}$ being the nonperturbative scale \re{Lambda1}. 

We observe that each
term in \re{sum-comp} takes complex values and their sum is remarkably
close to the exact value. In addition, the leading $O(\Lambda^2)$ nonperturbative correction (the second term) is comparable with the perturbative correction (the first term). Moreover, the former term starts to dominate over the latter one as we go to smaller values of the coupling constant. Thus, the transition from the strong to weak coupling regime is driven by nonperturbative corrections parameterized by the scale $\Lambda^2$. Moreover, the numerical analysis indicates that the expansion of $\Gamma^{\toy}_{\rm cusp}(g)$ in powers of $\Lambda^2$ is convergent for $\textrm{Re}\, g>0$.
 
\subsubsection{From toy model to the exact solution} 

The relation \re{toy-f} is remarkably similar to the expected strong coupling
expansion of the cusp anomalous dimension \re{cusp=pt+npt} with the function
$C_0(\alpha)$ providing  perturbative contribution and $\Lambda^2$ defining the
leading nonperturbative contribution. Let us compare $C_0(\alpha)$ with the known
perturbative expansion \re{strong} of $\Gamma_{\rm cusp}(g)$. In terms of the
coupling $\alpha=1/(2\pi g)$, the first few terms of this expansion look as
\be\label{cusp-alpha}
\Gamma_{\rm cusp}(g) /( 2g) = 1 - \frac{3\ln 2}{2} \alpha -\frac{\rm K}{4}\alpha^2
- \lr{\frac{3 {\rm K}\ln 2 }{8}+\frac{27\zeta_3}{256}}\alpha^3
+ \ldots \,,
\ee
where ellipses denote both higher order corrections in $\alpha$ and
nonperturbative corrections in $\Lambda^2$. Comparing \re{cusp-alpha} and the
first term, $C_0(\alpha)$, in the right-hand side of \re{toy-f}, we observe that
both expressions  approach the same value $1$ as $\alpha\to 0$.

As was already mentioned, the expansion coefficients of the two series have
different transcendentality -- they are rational for the toy model, Eq.~\re{c0-series}, and have
maximal transcendentality for the cusp anomalous dimension, Eq.~\re{cusp-alpha}.  
Notice that the two series would coincide if one formally replaced the transcendental numbers in
\re{cusp-alpha} by appropriate rational constants. In particular, replacing
\be\label{recipe}
\frac{3\ln 2}{2}  \to 1 \,,\qqqquad \frac{\rm K}2 \to \frac12\,, \qqqquad  \frac{9\zeta_3}{32} \to \frac13\,, \quad \ldots\,,
\ee
one obtains from \re{cusp-alpha} the first few terms of perturbative expansion
\re{c0-series} of the function $C_0$ in the toy model. This rule can be generalized to all
loops as follows. Introducing an auxiliary parameter $\tau$, we define the generating function
for the transcendental numbers in \re{recipe} and rewrite \re{recipe} as
\be\label{Sub1}
\exp{\left[\frac{3\ln{2}}{2}\,\tau - \frac{\textrm{K}}{2}\,\tau^2 + \frac{9\zeta_3}{32}\,\tau^{3} + \ldots\right]} \ \rightarrow\  \exp{\left[\tau-\frac{\tau^2}{2} + \frac{\tau^3}{3}+\ldots\right]}\,.
\ee
Going to higher loops, we have to add higher order terms in $\tau$ to both exponents. In the right-hand
side, these terms are resummed into $\exp(\ln(1+\tau))=1+\tau$, while in the left-hand side they produce the ratio of Euler gamma-functions leading to 
\be\label{Sub}
\frac{\Gamma(\frac{1}{4})\Gamma(1+\frac{\tau}{4})\Gamma(\frac{3}{4}-\frac{\tau}{4})}{\Gamma(\frac{3}{4})\Gamma(1-\frac{\tau}{4})\Gamma(\frac{1}{4}+\frac{\tau}{4})} \ \rightarrow\ \left(1+{\tau}\right).
\ee
Taking logarithms in both sides of this relation and subsequently expanding them in powers of $\tau$,
we obtain the subtitution rules which generalize \re{recipe} to the complete family of transcendental numbers entering into the strong coupling expansion \re{cusp-alpha}. 
At this point, the relation \re{Sub} can be thought of as an empirical rule, which allows us to map the strong coupling expansion of the cusp anomalous dimension \re{cusp-alpha} into
that in the toy model, Eq.~\re{c0-series}. We will clarify its  origin in Sect.~4.2. 

In spite of the fact that the numbers entering both sides of \re{recipe} have different transcendentality,
we may compare their numerical values. Taking into account that $3\ln 2/2 = 1.0397(2)$, ${\rm K}/2=0.4579(8)$ and $ {9\zeta_3}/{32} =0.3380(7)$ we observe that the relation \re{recipe} defines
a meaningful approximation to the transcendental numbers. Moreover, examining the coefficients in front of $\tau^n$ in both sides of \re{Sub1}  at large $n$, we find that the accuracy of approximation increases as $n\to\infty$. {This is in agreement with the observation made in the beginning
of Sect.~2.8, that  the cusp anomalous dimension in the toy model $\Gamma_{\rm cusp}^{\toy}(g)$ is close numerically to the exact expression $\Gamma_{\rm cusp}(g)$}. In addition, the same property suggests that the coefficients in the strong
coupling expansion of $\Gamma_{\rm cusp}^{\toy}(g)$ and $\Gamma_{\rm cusp}(g)$ should have the same
large order behavior.
It was found in \cite{BKK07} that the expansion coefficients in the right-hand
side of \re{cusp-alpha} grow at higher orders as $\Gamma_{\rm cusp}(g) \sim
\sum_{k}\Gamma(k+\ft12)\alpha^k$. It is straightforward to verify using
\re{toy-cc} and \re{F-series} that the expansion coefficients of $C_0(\alpha)$ in
the toy model have the same behavior. This suggests that nonperturbative
corrections to $\Gamma_{\rm cusp}(g)$ and $\Gamma_{\rm cusp}^{\toy}(g)$ are
parameterized by the same scale $\Lambda^2$ defined in \re{Lambda1}. Indeed we
will show this in the next section by explicit calculation.

We demonstrated in this section that nonperturbative corrections in the toy model
follow unambiguously from the exact solution \re{cusp-toy-fin}. In the next
section, we will extend analysis to the cusp anomalous dimension and  work out
the strong coupling expansion of $\Gamma_{\rm cusp}(g) /( 2g)$ analogous to
\re{toy-f}.

\section{Solving the quantization conditions}\label{QUANT}

Let us now solve the quantization conditions \re{QC01} for the cusp anomalous
dimension. The relation \re{QC01} involves two sets of functions. The functions
$\I_0(t)$ and $\I_1(t)$ are given by the Whittaker function of 1st kind (see
Eq.~\re{I=M}). At the same time, the functions $f_0(t)$ and $f_1(t)$ are defined
in \re{ff} and they depend on the (infinite) set of expansion coefficients $c(g)$
and $c_\pm(n,g)$. Having determined these coefficients from the quantization
conditions \re{QC01}, we can then compute the cusp anomalous dimension for
arbitrary coupling  with a help of \re{cusp-c}.

We expect that at strong coupling the resulting expression for $\Gamma_{\rm
cusp}(g)$ will have the form \re{cusp=pt+npt}.  Examining \re{cusp-c} we observe
that the dependence on the coupling resides both in the expansion coefficients
and in the functions $U_0^\pm(4\pi g)$. The latter are given by the Whittaker
functions of 2nd kind (see Eq.~\re{U's}) and, as such, they are given by Borel
summable sign-alternating asymptotic series in $1/g$. Therefore, nonperturbative
corrections to the cusp anomalous dimension \re{cusp-c} could only come from the
coefficients $c_\pm(n,g)$ and $c(g)$.

\subsection{Quantization conditions}

Let us replace $f_0(t)$ and $f_1(t)$ in \re{QC01} by their explicit expressions
\re{ff} and rewrite the quantization conditions \re{QC01} as
\begin{align}\label{QC}
 \I_0(4\pi g x_\ell)+c(g) \I_1(4\pi g x_\ell) & =
 \sum_{n\ge 1} \left[ {c_+(n,g)} A_+(n,x_\ell) +  {c_-(n,g)}A_-(n,x_\ell) \right] \,,
\end{align}
where $x_\ell=\ell-\ft14$ (with $\ell=0,\pm 1,\pm 2,\ldots$) and  the notation
was introduced for
\be
A_\pm(n,x_\ell) = \frac{n \I_1(4\pi g x_\ell) U_0^\pm (4\pi ng)+x_\ell \I_0(4\pi
g x_\ell) U_1^\pm (4\pi ng) } {n \mp x_\ell}  \,.
\ee
The relation \re{QC} provides an infinite system of linear equations for
$c_\pm(g,n)$ and $c(g)$. The coefficients in this system depend on $\I_{0,1}(4\pi
gx_\ell)$ and $U^\pm_{0,1}(4\pi ng)$ which are known functions defined in
Appendix~\ref{App:D}. We would like to stress that the relation \re{QC} holds for
arbitrary $g>0$.

Let us show that the quantization conditions \re{QC} lead to $c(g)=0$ for
arbitrary coupling. To this end, we examine \re{QC} for $|x_\ell|\gg 1$. In this
limit, for $g={\rm fixed}$ we are allowed to replace the functions $\I_0(4\pi g
x_\ell)$ and $\I_1(4\pi g x_\ell)$ in both sides of \re{QC} by their asymptotic
behavior at infinity. Making use of \re{I-U-relation} and \re{U-large}, we find
for $|x_\ell|\gg 1$
\begin{align}\label{r}
 r(x_\ell)\equiv \frac{\I_1(4\pi gx_\ell)}{\I_0(4\pi gx_\ell)}=\left\{
\begin{array}{ll}
  -16\pi gx_\ell
+\ldots \,, & (x_\ell<0)\\[3mm]
 \phantom{+}\frac12 +\ldots \,, & (x_\ell>0) \\
\end{array}
\right.
\end{align}
where ellipses denote terms suppressed by powers of $1/(gx_\ell)$ and $\e^{-8\pi
g|x_\ell|}$.  We divide both sides of \re{QC} by $\I_1(4\pi g x_\ell)$ and
observe that for $x_\ell \to -\infty$ the first term in the left-hand side of \re{QC} is
subleading and can be safely neglected. In the similar manner, one has
$A_\pm(n,x_\ell)/\I_1(4\pi g x_\ell) =O(1/x_\ell)$ for fixed $n$ in the
right-hand side of \re{QC}. Therefore, going to the limit $x_\ell\to -\infty$ in
both sides of \re{QC} we get
\be\label{c-zero}
c(g)=0
\ee
for arbitrary $g$. We verify in Appendix~\ref{App:A} by explicit calculation that
this relation indeed holds at weak coupling.

Arriving at \re{c-zero}, we tacitly assumed that the sum over $n$ in \re{QC}
remains finite in the limit $x_\ell\to -\infty$. Taking into account large $n$
behavior of the functions $U_0^\pm (4\pi ng)$ and $U_1^\pm (4\pi ng)$ (see
Eq.~\re{U-large}), we obtain that this condition translates into the following
condition for asymptotic behavior of the coefficients at large $n$
\be\label{c=o}
c_+(n,g) = o(n^{1/4}) \,,\qqqquad c_-(n,g) = o(n^{-1/4})\,.
\ee
These relations also ensure that the sum in the expression \re{cusp-c} for the
cusp anomalous dimension is convergent.

\subsection{Numerical solution}

To begin with, let us solve the infinite system of linear equations \re{QC}
numerically. In order to verify \re{c-zero}, we decided to do it in two steps: we
first solve \re{QC} for $c_\pm(n,g)$ assuming $c(g)=0$ and, then, repeat the same
analysis by relaxing the condition \re{c-zero} and treating $c(g)$ as unknown.

For $c(g) = 0$, we truncate the infinite sums on the right-hand side of \re{QC} at
some large $n_{\rm max}$ and, then, use \re{QC} for $\ell=1-n_{\rm
max},\ldots,n_{\rm max}$ to find numerical values of $c_\pm(n,g)$  with $1\le n
\le n_{\rm max}$  for given coupling $g$. Substituting the resulting expressions
for $c_\pm(n,g)$ into \re{cusp-c} we compute the cusp anomalous dimension. Taking
the limit $n_{\rm max}\to\infty$ we expect to recover the exact result. Results
of our analysis are summarized in two tables. Table~1 shows the dependence of the
cusp anomalous dimension on the coupling constant. Table~2 shows the dependence
of the cusp anomalous dimension on the truncation parameter $n_{\rm max}$ for
fixed coupling.

For $c(g)$ arbitrary, we use \re{QC} for $\ell=-n_{\rm max},\ldots,n_{\rm max}$
to find numerical values of $c(g)$ and $c_\pm(n,g)$  with $1\le n \le n_{\rm
max}$  for given coupling $g$. In this manner, we compute $\Gamma_{\rm
cusp}(g)/(2g)$ and $c(g)$ and, then, compare them with the exact expressions
corresponding to $n_{\rm max}\to\infty$. For the cusp anomalous dimension, our
results for $\Gamma_{\rm cusp}(g)/(2g)$ are in remarkable agreement with the
exact expression. Namely, for $n_{\rm max}=40$ their difference equals $5.480
\times 10^{-6}$ for $g=1$ and it decreases down to $8.028\times 10^{-7}$ for
$g=1.8$. The reason why agreement is better compared to the $c(g)=0$ case (see
Table~1) is that $c(g)$ takes effectively into account a reminder of the sum in
the right-hand side of \re{QC} corresponding to $n>n_{\rm max}$. The dependence
of the obtained expression for $c(g)$ on the truncation parameter $n_{\rm max}$
is shown in Table~3. We observe that, in agreement with \re{c-zero}, $c(g)$
vanishes as $n_{\rm max}\to\infty$.

\begin{table}[h]
\begin{center}
\begin{tabular}{|c||c|c|c|c|c|c|c|c|c|c|}
\hline $g$ & $0.1$ & $0.2$ &  $0.4$ &
     $0.6$ &  $0.8$ &
     $1.0$ & $1.2$ & $1.4$ & $1.6$ & $1.8$
\\
\hline numer & $0.1976$  & $0.3616$ & $0.5843$ & $0.7096$ & $0.7825$ & $0.8276$ & $0.8576$ &
$0.8787$ & $0.8944$ & $0.9065$
\\
\hline exact & $0.1939$ & $0.3584$ & $0.5821$ & $0.7080$ & $0.7813$ & $0.8267$ & $0.8568$ & $0.8781$
& $0.8938$ & $0.9059$
\\
\hline
\end{tabular}
\end{center}
\caption{Comparison of the numerical value of $\Gamma_{\rm cusp}(g)/(2g)$ found
from \re{QC} and \re{cusp-c} for $n_{\rm max}=40$ with the exact one
\cite{BKK07,K08} for different values of the coupling constant $g$.}
\end{table}%

\begin{table}[h]
\begin{center}
\begin{tabular}{|c||c|c|c|c|c|c|c||c|}
\hline $n_{\rm max}$ & $10$ & $20$ &  $30$ &
     $40$ &  $50$ &
     $60$ & $70$ & $\infty$
\\
\hline numer & $0.8305$  & $0.8286$ & $0.8279$ & $0.8276$ & $0.8274$ & $0.8273$ & $0.8272$ & 0.8267
\\
\hline
\end{tabular}
\end{center}
\caption{Dependence of $\Gamma_{\rm cusp}(g)/(2g)$ on the truncation parameter $n_{\rm max}$ for
$g=1$ and $c(g)=0$. The last column describes the exact result.}
\end{table}

\begin{table}[h]
\begin{center}
\begin{tabular}{|c||c|c|c|c|c|c|c||c|}
\hline $n_{\rm max}$ & $10$ & $20$ &  $30$ &
     $40$ &  $50$ &
     $60$ & $70$ & $\infty$
\\
\hline $-c(g)$ & $0.0421$ & $0.0357$ &  $0.0323$ & $0.0301$ & $0.0285$ & $0.0272$ & $0.0262$ & 0
\\
\hline
\end{tabular}
\end{center}
\caption{Dependence of $c(g)$ on the truncation parameter  $n_{\rm max}$ for $g=1$ derived from the
quantization condition \re{QC}.}
\end{table}

Our numerical analysis shows that  the cusp anomalous dimension  \re{cusp-c} can
be determined from the quantization conditions \re{QC} and \re{c-zero}  for
arbitrary coupling $g$. In distinction with the toy model \re{cusp-toy-fin}, the
resulting expression for $\Gamma_{\rm cusp}(g)$ does not admit a closed form
representation. Still, as we will show in the next subsection, the quantization
conditions \re{QC} can be solved analytically for $g\gg 1$ leading to asymptotic
expansion for the cusp anomalous dimension at strong coupling.

\subsection{Strong coupling solution}

Let us divide both sides of \re{QC} by $\I_0(4\pi gx_\ell)$ and use \re{c-zero}
to get (for $x_\ell=\ell-\ft14$ and $\ell \in \mathbb{Z}$)
\begin{align}\label{QC1}
  1 & =
   \sum_{n\ge 1}  {c_+(n,g)} \left[\frac{n  U_0^+(4\pi ng)r(x_\ell)+  U_1^+(4\pi ng) x_\ell }
 {n-x_\ell} \right]
\\& \notag
+  \sum_{n\ge 1}
  {c_-(n,g)}\left[ \frac{n U_0^-(4\pi ng)r(x_\ell) +  U_1^-(4\pi ng)x_\ell}{n+x_\ell}\right]\,,
\end{align}
where the function $r(x_\ell)$ was defined in \re{r}.

Let us now examine the large $g$ asymptotics of the coefficient functions accompanying
$c_\pm(n,g)$ in the right-hand side of \re{QC1}.  The functions $U_0^\pm(4\pi
ng)$ and $U_1^\pm(4\pi ng)$ admit asymptotic expansion in $1/g$ given by
\re{U-large}. For the function $r(x_\ell)$  the situation is different. As
follows from its definition, Eqs.~\re{r} and \re{I-U-relation}, large $g$
expansion of $r(x_\ell)$ runs in two parameters: perturbative $1/g$ and
nonperturbative exponentially small parameter $\Lambda^2\sim g^{1/2}\e^{-2\pi g}$
which we already encountered in the toy model, Eq.~\re{Lambda1}. Moreover, we
deduce from \re{r} and \re{I-U-relation} that the leading nonperturbative
correction to $r(x_\ell)$ scales  as
\be
\delta r(x_\ell)= O\big(\Lambda^{|8\ell-2|}\big)\,,\qqqquad (x_\ell=\ell-\ft14\,,\ \ell \in
\mathbb{Z})\,,
\ee
so that the power of $\Lambda$ grows with $\ell$. We observe that $O(\Lambda^2)$
corrections are only present in $r(x_\ell)$ for $\ell=0$. Therefore, as far as
the leading $O(\Lambda^2)$ correction to the solutions to \re{QC1} are concerned,
we are allowed to neglect nonperturbative ($\Lambda^2-$dependent) corrections to
$r(x_\ell)$ in the right-hand side of \re{QC1} for  $\ell \neq 0$ and retain them
for $\ell=0$ only.

Since the coefficient functions in the linear equations \re{QC1} admit a double
series expansion in powers of $1/g$ and $\Lambda^2$, we expect that the same
should be true for their solutions $c_\pm(n,g)$. Let us determine the first few
terms of this expansion using the following ansatz:
\begin{align}\label{c=a+b}
c_\pm(n,g) = (8\pi gn)^{\pm 1/4} \bigg\{ \left[ {a_\pm(n) +  \frac{b_\pm (n)}{4\pi g} +
\ldots}\right] + \Lambda^2 \left[{\alpha_\pm(n) + \frac{\beta_\pm (n)}{4\pi g} + \ldots}
\right]+O(\Lambda^4)\bigg\}\,,
\end{align}
where  $\Lambda^2$ is a nonperturbative parameter defined in \re{Lambda1}
\be\label{Lambda2}
\Lambda^2  = \sigma (2\pi g)^{1/2} \e^{-2\pi g}\frac{\Gamma(\ft34)}{\Gamma(\ft54)}\,,
\ee
and ellipses denote terms suppressed by powers of $1/g$. Here the functions
$a_\pm(n), b_\pm(n),\ldots$ are assumed to be $g-$independent. We recall that the
functions $c_\pm(n,g)$ have to verify the relation \re{c=o}. This implies that
the functions $a_\pm(n),  b_\pm(n), \ldots$ should vanish as $n\to\infty$. To
determine them we substitute \re{c=a+b} into \re{QC1} and compare the
coefficients in front of powers of $1/g$ and $\Lambda^2$ in both sides of
\re{QC1}.

\subsubsection{Perturbative corrections}

Let us start with `perturbative', $\Lambda^2-$independent part of \re{c=a+b} and
compute the functions $a_\pm(n)$ and $b_\pm(n)$.

To determine $a_\pm(n)$, we substitute \re{c=a+b} into \re{QC1}, replace the
functions $U_{0,1}^\pm(4\pi gn)$ and $r(x_\ell)$ by their large $g$ asymptotic
expansion, Eqs.~\re{U-large} and \re{r}, respectively, neglect corrections in
$\Lambda^2$ and compare the leading $O(g^0)$ terms in both sides of \re{QC1}. In
this way, we obtain from \re{QC1} the following relations for $a_\pm(n)$ (with
$x_\ell=\ell-\ft14$)
\begin{align}
2x_\ell\, \Gamma(\ft54)\sum_{n\ge 1}  \frac{a_+(n)}{n-x_\ell} &= 1 \,, \qqqquad (\ell\ge 1)
\\ \notag
-2x_\ell\, \Gamma(\ft34) \sum_{n\ge 1} \frac{a_-(n)}{n+x_\ell} &= 1 \,, \qqqquad (\ell\le 0)
\end{align}
One can verify that the solutions to this system satisfying  $a_\pm(n)\to 0$ for $n\to\infty$ have
the form
\begin{align}\label{a}
a_+(n) &=\frac{2\Gamma(n+\ft14)}{\Gamma(n+1)\Gamma^2(\ft14)} \,,
 \\ \notag
a_-(n) &= \frac{\Gamma(n+\ft34)}{2\Gamma(n+1)\Gamma^2(\ft34)}\,.
\end{align}
In the similar manner, we compare the subleading $O(1/g)$ terms in both sides of
\re{QC1} and find that the functions $b_\pm(n)$ satisfy the following relations
(with $x_\ell=\ell-\ft14$)
\begin{align}
 2x_\ell\,\Gamma(\ft54)\sum_{n\ge 1}    \frac{{b_+(n)}}
 {n-x_\ell} &  =    - \frac3{32x_\ell} - \frac{3\pi}{64}-\frac{15}{32}\ln 2\,, \qquad (\ell\ge 1)
\\ \notag
  -2x_\ell\,\Gamma(\ft34)\sum_{n\ge 1}
   \frac{{b_-(n)}}{n+x_\ell} & =  - \frac{5}{32x_\ell} -\frac{5\pi}{64}+\frac{9}{32}\ln 2\,, \qquad (\ell\le 0)
\end{align}
where in the right-hand side we made use of \re{a}. Solutions to these relations are
\begin{align}\label{NLO-PT}
b_+(n) & = -  a_+(n)\bigg( \frac{3\ln 2}4 + \frac{3}{32n}\bigg)\,,
\\\notag
b_-(n) & = \phantom{-}  a_-(n)\bigg( \frac{3\ln 2}4 + \frac{5}{32n}\bigg)\,.
\end{align}
It is straightforward to extend analysis to subleading perturbative corrections to $c_\pm(n,g)$.

Let us substitute \re{c=a+b} into expression \re{cusp-c} for the cusp anomalous
dimension. Taking into account the identities \re{U-large} we find the `perturbative'
contribution to $\Gamma_{\rm cusp}(g)$ as
\begin{align}
\Gamma_{\rm cusp}(g)  =  2g -  \sum_{n\ge 1} (2\pi n)^{-1}\bigg[& \Gamma(\ft54)\lr{a_+(n)+
\frac{b_+(n)}{4\pi g} +\ldots}\lr{1- \frac{5}{128\pi g n} +\ldots}
\\\notag
+&  \Gamma(\ft34)\lr{a_-(n)+ \frac{b_-(n)}{4\pi g} +\ldots}\lr{1+ \frac{3}{128\pi g n} +\ldots}
\bigg]+O(\Lambda^2)  \,.
\end{align}
Replacing $a_\pm(n)$ and $b_\pm(n)$ by their expressions \re{a} and \re{NLO-PT}, we find after some
algebra
\begin{align}\label{cusp-NLO}
\Gamma_{\rm cusp}(g) = 2g\bigg[1 -\frac{3\ln 2}{4\pi g} - \frac{\textrm{K}}{16\pi^2 g^2} +
O(1/g^3)\bigg]  +O(\Lambda^2)\,,
\end{align}
where $\textrm{K}$ is the Catalan number. This relation is in agreement with the known result
obtained both in $\mathcal{N}=4$ SYM theory~\cite{BKK07,KSV08} and in string theory~\cite{Roiban:2007dq}.

\subsubsection{Nonperturbative corrections}

Let us now compute the leading $O(\Lambda^2)$ nonperturbative correction to the
coefficients $c_\pm(n,g)$. According to
\re{c=a+b}, it is described by the functions $\alpha_\pm(n)$ and $\beta_\pm(n)$. To
determine them from \re{QC1}, we have to retain in $r(x_\ell)$ corrections proportional to
$\Lambda^2$. As was already explained, they only appear for $\ell=0$.
Combining together the relations \re{r}, \re{I-U-relation} and \re{U-large} we find after some algebra
\be
\delta r(x_\ell) =  - \delta_{\ell, 0}\Lambda^2 \left[{4\pi g-\frac5{4}  + O(g^{-1})}\right] +O(\Lambda^4)\,.
\ee
Let us substitute this relation into \re{QC1} and equate to zero the coefficient in front of $\Lambda^2$
in the right-hand side of  \re{QC1}. This coefficient is given by series in $1/g$ and, examining the
first two terms, we obtain the relations for the functions $\alpha_\pm(n)$ and $\beta_\pm(n)$.

In this way, we find that the leading  functions $\alpha_\pm(n)$ satisfy the relations
 (with $x_\ell=\ell-\ft14$)
\begin{align} \notag
 2x_\ell\,\Gamma(\ft54)&\sum_{n\ge 1} \frac{\alpha_+(n)}{n-x_\ell}   = 0 \,,    &&     (\ell\ge 1) 
 \\ \label{b-rel}
- 2x_\ell\,\Gamma(\ft34)& \sum_{n\ge 1} \frac{\alpha_-(n)}{n+x_\ell}  =  \frac{\pi}{2\sqrt{2}}\,
 \delta_{\ell,0}\,,  &&     (\ell\le 0) 
\end{align}
where in the right-hand side we applied \re{a}. Solution to \re{b-rel} satisfying $\alpha_\pm(n)\to 0$
as $n\to\infty$ reads
\begin{align}\notag
\alpha_+(n) &= 0 \,,\qqqquad
\\[2mm] \label{al}
\alpha_-(n) & =   a_-(n-1)\,.
\end{align}
with $a_-(n)$ defined in \re{a}.
For subleading functions $\beta_\pm(n)$ we have similar relations
\begin{align} \notag
   2x_\ell\,\Gamma(\ft54)\sum_{n\ge 1}    \frac{{\beta_+(n)}}
 {n-x_\ell} & =
- \frac12   \,,\hspace*{49mm} (\ell\ge 1)
\\ \label{Eq-NLO-NPT}
  -2x_\ell\, \Gamma(\ft34)\sum_{n\ge 1}
   \frac{{\beta_-(n)}}{n+x_\ell}
& = -\frac18+\frac{3\pi}{16\sqrt{2}}\lr{1-2\ln 2}\delta_{\ell,0} \,,\qquad (\ell\le 0)
\end{align}
In a close analogy with \re{NLO-PT}, the solutions to these relations can be written in
terms of leading-order functions $a_\pm(n)$  defined in \re{a}
\begin{align}\notag
\beta_+(n)
  & = -\frac12\,a_+(n)\,,
\\ \label{be}
\beta_-(n)
  & =a_-(n-1)\left( \frac14-\frac{3\ln 2}{4}+ \frac1{32n}\right)
  \,.
\end{align}
It is straightforward to extend analysis and compute subleading $O(\Lambda^2)$ corrections to
\re{c=a+b}.

The relation  \re{c=a+b} supplemented with  \re{a}, \re{NLO-PT}, \re{al} and
\re{be} defines the solution to the quantization condition \re{QC1} to leading
order in both perturbative, $1/g$, and nonperturbative, $\Lambda^2$, expansion
parameters. We are now ready to compute nonperturbative correction to the cusp
anomalous dimension \re{cusp-c}.  Substituting \re{c=a+b} into \re{cusp-c} we
obtain
\begin{align}
\delta\Gamma_{\rm cusp}(g) =    - \Lambda^2 \sum_{n\ge 1} (2\pi n)^{-1}\bigg[& \Gamma(\ft54)\lr{\alpha_+(n)+
\frac{\beta_+(n)}{4\pi g} +\ldots}\lr{1- \frac{5}{128\pi g n} +\ldots}
\\\notag
+&  \Gamma(\ft34)\lr{\alpha_-(n)+ \frac{\beta_-(n)}{4\pi g} +\ldots}\lr{1+ \frac{3}{128\pi g n} +\ldots}
\bigg]+O(\Lambda^4)  \,.
\end{align}
We replace $\alpha_\pm(n)$ and $\beta_\pm(n)$ by their explicit expressions \re{al} and \re{be}, evaluate
the sums and find
\begin{align}\label{cusp-nonPT}
\delta\Gamma_{\rm cusp}(g) = - \frac{\Lambda^2}{\pi}\left[ 1+ \frac{3-6\ln 2}{16\pi g}  + O(1/g^2)\right]
+O(\Lambda^4)  \,,
\end{align}
with $\Lambda^2$ defined in \re{Lambda2}.

The relations \re{cusp-NLO} and \re{cusp-nonPT} describe, correspondingly,
perturbative and nonperturbative corrections to the cusp anomalous dimension. Let
us define a new nonperturbative parameter $m_{\rm cusp}^2$ whose meaning will be
clear in a moment
\be\label{m^2}
m_{\rm cusp}^2 = \frac{4\sqrt{2}} {\pi \sigma}\Lambda^2\left[ 1+\frac{3-6\ln 2}{16\pi g}  + O(1/g^2)\right]+ O(\Lambda^4)\,.
\ee
Then, the obtained expressions \re{cusp-NLO} and \re{cusp-nonPT} for
the cusp anomalous dimension takes the form
\begin{align}\label{cusp-NLO1}
\Gamma_{\rm cusp}(g) = \left[2g -\frac{3\ln 2}{2\pi} - \frac{\textrm{K}}{8\pi^2 g} + O(1/g^2)\right]
-  \frac{\sigma}{4\sqrt{2}} m_{\rm cusp}^2 + O(m_{\rm cusp}^4)\,.
\end{align}
We recall that another nonperturbative parameter was already
introduced in Sect.~2.1 as defining the mass gap $m_{\rm O(6)}$ in the O(6) model. We will show in the next section, that the two scales, $m_{\rm cusp}$
and $m_{\rm O(6)}$,  coincide to any order in $1/g$.

\section{Mass scale}

The cusp anomalous dimension  controls the leading logarithmic scaling behavior
of the anomalous dimensions \re{anom-dim} in the double scaling limit $L\,, N\to
\infty$ and $j=L/\ln N = {\rm fixed}$. The subleading corrections to this
behavior are described by the scaling function $\epsilon(j,g)$. At strong
coupling, this function coincides with the energy density of the ground state of the
bosonic O(6) model \re{O6-epsilon}. The mass gap in this model $m_{\rm O(6)}$ is
given by expression \re{m=int} which involves the functions $\Gamma_\pm(t)$
constructed in Section~2.

\subsection{General expression}

Let us apply \re{m=int} and compute the mass gap $m_{\rm O(6)}$ at strong
coupling. At large $g$ the integral in \re{m=int} receives a dominant
contribution from $t\sim g$. In order to evaluate \re{m=int} it is convenient to
change the integration variable as $t\to 4\pi g i t$  
\be\label{m=int1}
m_{\rm O(6)} =  \frac{8\sqrt{2}}{\pi^2}\e^{-\pi g}-\frac{8g}{\pi}\e^{-\pi g}\Re\left[  \int_0^{-i\infty}  {dt\,
\e^{-4\pi g t-i\pi/4}} \frac{\Gamma(4\pi g it)}{t+\ft14} \right]\,,
\ee
where integration goes along the imaginary axis. We find from \re{G-gen} that $\Gamma(4\pi g it)$ takes
the form
\begin{align} \label{G-gen1}
\Gamma(4\pi g it) &= f_0(4\pi g t)\I_0(4\pi g t)+f_1(4\pi g t) \I_1(4\pi g t)\,,
\end{align}
where $\I_{0,1}(4\pi g t)$  are given by the Whittaker functions of first kind,
Eq.~\re{I=M}, and $f_{0,1}(4\pi g t)$  admit the following representation (see
Eqs.~\re{ff} and \re{c-zero})
\begin{align}\label{ff1}
f_0(4\pi g t) & = \sum_{n\ge 1}  t\left[{c_+(n,g)}\frac{  U_1^+(4\pi ng) }{n-t}+
  {c_-(n,g)}  \frac{   U_1^-(4\pi ng)}{n+t}\right] -1 \,,
\\ \notag
f_1(4\pi g t) &=  \sum_{n\ge 1}  n \left[{c_+(n,g)}\frac{  U_0^+(4\pi ng) }{n-t} +
  {c_-(n,g)}  \frac{ U_0^-(4\pi ng) }{n+t}\right].
\end{align}
Here the functions $U_{0,1}^\pm(4\pi ng)$ are expressed in terms of  Whittaker functions of first
kind, Eq.~\re{U's}, and the expansion coefficients $c_\pm(n,g)$ are solutions to the quantization
conditions \re{QC01}.

Replacing $\Gamma(4\pi g it)$ in \re{m=int1} by its expression \re{G-gen1}, we
evaluate the $t-$integral and find after some algebra (see Appendix~\ref{App-m} for details)~\cite{BBBKP08}
\begin{align}\label{m=U}
m_{\rm O(6)}  & = - \frac{16\sqrt{2}}{\pi}g \e^{-\pi g} \left[f_0(-\pi g) U_0^-(\pi g)+   f_1(-\pi g )U_1^-(\pi
g)
 \right]\,.
\end{align}
This relation can be further simplified with a help of the quantization conditions \re{QC01}. For
$\ell = 0$, we obtain from \re{QC01} that $f_0(-\pi g)\I_0(-\pi g)+f_1(-\pi g) \I_1(-\pi g)=0$.
Together with the Wronskian relation for the Whittaker functions \re{Wh-Wr} this leads to the
following remarkable relation for the mass gap
\begin{align}\label{m=f1}
m_{\rm O(6)} & = \frac{16\sqrt{2}}{\pi^2} \frac{f_1(-\pi g )}{\I_0(-\pi g)}\,.
\end{align}
It is instructive to compare this relation with similar relation \re{cusp=f} for the cusp anomalous
dimension. We observe that both quantities involve the same function $f_1(4\pi g t)$ but evaluated
for different values of its argument, that is $t=-1/4$ for the mass gap  and $t=0$ for the cusp anomalous
dimension. As a consequence, there are no reasons to expect that the two functions, $m(g)$ and
$\Gamma_{\rm cusp}(g)$, could be related to each other in a simple way. Nevertheless, we will
demonstrate in this subsection, that $m_{\rm O(6)}^2$ determines the leading nonperturbative correction to
$\Gamma_{\rm cusp}(g)$ at strong coupling.

\subsection{Strong coupling expansion}

Let us now determine the strong coupling expansion of the functions \re{ff1}. We
replace coefficients $c_\pm(n,g)$ in \re{ff1} by their expression
\re{c=a+b} and take into account the obtained results for the functions
$a_\pm, b_\pm,\ldots$, Eqs.~\re{a}, \re{NLO-PT}, \re{al} and \re{be}. In
addition, we replace in \re{ff1} the functions $U_{0,1}^\pm(4\pi ng)$  by their
strong coupling expansion \re{U-large}. We recall that the coefficients
$c_\pm(n,g)$ admit the double series expansion \re{c=a+b} in powers of $1/g$ and
$\Lambda^2\sim \e^{-2\pi g}$, Eq.~\re{Lambda2}. As a consequence, the
functions $f_0(4\pi gt)$ and $f_1(4\pi gt)$ have the form
\be\label{dec}
f_n(4\pi gt) = f_n^{\scriptscriptstyle \rm (PT)} (4\pi gt) + \delta f_n(4\pi gt)\,,\qqqquad (n=0,1)\,,
\ee
where $f_n^{\scriptscriptstyle \rm (PT)}$ is given by asymptotic (non-Borel
summable) series in $1/g$ and $\delta f_n$ takes into account nonperturbative
corrections in  $\Lambda^2$.

Evaluating sums in the right-hand side of \re{ff1} we find that $f_0(4\pi gt)$
and $f_1(4\pi gt)$ can be expressed in terms of two sums involving functions
$a_\pm(n)$ defined in \re{a}
\begin{align} \notag
2\Gamma(\ft54) \sum_{n \ge 1} \frac{a_+(n)}{t-n} =
\frac1{t}\left[\frac{\Gamma(\ft34)\Gamma(1-t)}{\Gamma(\ft34-t)}-1\right]\,,
\\
2\Gamma(\ft34) \sum_{n \ge 1} \frac{a_-(n)}{t+n} =
\frac1{t}\left[\frac{\Gamma(\ft14)\Gamma(1+t)}{\Gamma(\ft14+t)}-1\right]\,.
\end{align}
Going through calculation of \re{ff1}, we find after some algebra that
perturbative corrections to $f_0(4\pi gt)$ and $f_1(4\pi gt)$ are given by linear
combinations of the ratios of Euler gamma-functions
\begin{align}\notag
f_0^{\scriptscriptstyle \rm (PT)}(4\pi gt)&= - \frac{{\Gamma}(\ft34)\Gamma(1-t)}{\Gamma(\ft34-t)} \\ & \notag +
\frac{1}{4\pi g}\left[\lr{\frac{3\ln 2}{4}+\frac1{8t}}
\frac{\Gamma(\ft34)\Gamma(1-t)}{\Gamma(\ft34-t)}- \frac{\Gamma(\ft14)\Gamma(1+t)}{8t\,
\Gamma(\ft14+t)} \right]+O(g^{-2})\,,
\\[3mm]\label{ff2}
f_1^{\scriptscriptstyle \rm (PT)} (4\pi gt)&= \frac{1}{4\pi
g}\left[\frac{\Gamma(\ft14)\Gamma(1+t)}{4t\,\Gamma(\ft14+t)}
-\frac{\Gamma(\ft34)\Gamma(1-t)}{4t\,\Gamma(\ft34-t)} \right]
\\ \notag
& -\frac1{(4\pi g)^2}\left[\frac{\Gamma(\ft14)\Gamma(1+t)}{4t\,\Gamma(\ft14+t)}\lr{\frac1{4t}-\frac{3\ln
2}{4}}-\frac{\Gamma(\ft34)\Gamma(1-t)}{4t\,\Gamma(\ft34-t)}\lr{\frac1{4t}+\frac{3\ln
2}{4}}\right]+O(g^{-3})\,.
\end{align}
Notice that $f_1(t)$ is suppressed by factor $1/(4\pi g)$ compared to $f_0(t)$.
In the similar manner, we compute
nonperturbative corrections to  \re{dec}
\begin{align}\notag
\delta f_0 (4\pi gt) =\Lambda^2 & \left\{ \frac{1}{4\pi
g}\left[\frac{\Gamma(\ft34)\Gamma(1-t)}{2\,\Gamma(\ft34-t)}-\frac{\Gamma(\ft54)\Gamma(1+t)}{2\,\Gamma(\ft54+t)}
\right] + O(g^{-2})\right\} +\ldots \,,
\\[3mm] \label{ff3}
\delta f_1(4\pi gt)  =\Lambda^2 & \left\{\frac{1}{4\pi
g}\frac{\Gamma(\ft54)\Gamma(1+t)}{\Gamma(\ft54+t)}\right.
\\\notag
& \hspace*{-3mm}+ \frac1{(4\pi g)^2} \left[ \frac{\Gamma(\ft34)\Gamma(1-t)}{8t\Gamma(\ft34-t)} -
\frac{\Gamma(\ft54)\Gamma(1+t)}{\Gamma(\ft54+t)}\lr{\frac1{8t}+\frac34\ln2 -\frac14} \right]+
O(g^{-3}) \bigg\} +\ldots \,,
\end{align}
where ellipses denote $O(\Lambda^4)$ terms.

Substituting \re{ff2} and \re{ff3} into \re{G-gen1} we obtain the strong coupling
expansion of the function $\Gamma(4\pi i gt)$. To
verify the obtained expressions, we apply \re{cusp=f} to calculate the cusp
anomalous dimension
\be\label{cusp=pt+npt1}
\Gamma_{\rm cusp}(g) = 2g - 4 g f_1^{\scriptscriptstyle \rm (PT)}(0) - 4 g \,\delta f_1(0)\,.
\ee
Replacing $f_1^{\scriptscriptstyle \rm (PT)}(0)$ and $\delta f_1(0)$ by their expressions,
Eqs.~\re{ff2} and \re{ff3}, we obtain
\begin{align}\label{cusp-full}
\Gamma_{\rm cusp}(g) & = 2g \left[1 -\frac{3\ln 2}{4\pi g} - \frac{\textrm{K}}{(4\pi g)^2} + \ldots
\right] - \frac{\Lambda^2}{\pi}\left[ 1+ \frac{3-6\ln 2}{16\pi g}  + \ldots \right]
+O(\Lambda^4) \,, 
\end{align}
in a perfect agreement with \re{cusp-NLO} and \re{cusp-nonPT}, respectively.

Let us obtain the strong coupling expansion of the mass gap \re{m=f1}. We replace
$\I_0(-\pi g)$ by its asymptotic series, Eqs.~\re{I0-sep}  and \re{U-large}, and take into account
\re{ff2} and \re{ff3} to get
\begin{align}\notag
m_{\rm O(6)} = \frac{\sqrt{2}}{\Gamma(\ft54)} (2\pi g)^{1/4} \e^{-\pi g} \bigg\{
\left[ 1 +   \frac{3-6\ln 2}{32\pi g} +\frac{-63 + 108\, \ln 2 -108 (\ln 2)^2 +16 {\rm
K}}{2048 (\pi g)^2}+\ldots\right]
  \\\label{mm}
 -   \frac{\Lambda^2}{8\pi g}\left[1 -  \frac{15-6\ln 2}{32 \pi g} +\ldots \right]+
 O(\Lambda^4)\bigg\}\,.
\end{align}
Here, in order to determine $O(1/g^{2})$ and $O(\Lambda^2/g^{2})$ terms inside
the curly brackets, we computed in addition the subleading $O(g^{-3})$
corrections to $f_1^{\scriptscriptstyle \rm (PT)} $ and $\delta f_1$ in
Eqs.~\re{ff2} and \re{ff3}, respectively. The leading $O(1/g)$ correction 
to $m_{\rm O(6)} $ (the second term inside the first 
square bracket in the r.h.s.\ of \re{mm}) is in agreement with
both analytical \cite{BK08,BBBKP08} and numerical calculations \cite{FGR08a}.

We are now ready to clarify the origin of the `substitution rule' \re{Sub} that establishes 
the relation between the cusp anomalous dimension in the toy model and the exact solution. 
To this end, we compare the expressions for the functions $f_n(4\pi gt)$ given by \re{dec}, \re{ff2}
and \re{ff3} with those in the toy model, Eqs.~\re{f-toy} and \re{c1-toy}.%
\footnote{It worth mentioning that the functions $f_0^{\rm (toy)}$ and $f_1^{\rm (toy)}$ in the toy model are, in fact, $t-$independent.}
 It is straightforward
to verify that upon the substitution  \re{Sub} and \re{recipe} the two set of functions
coincide up to an overall $t-$dependent factor%
\footnote{Roughly speaking, this substitution simplifies the complicated structure of poles
and zeros of the exact solution, Eqs.~\re{ff2} and \re{ff3}, encoded in the ratio of the 
gamma-functions to match simple analytical properties of the same functions  in the toy model (compare \re{sin-sin} and \re{toy}).}   
\begin{align}
f_n(4\pi gt) \frac{\Gamma(\ft34-t)}{{\Gamma}(\ft34)\Gamma(1-t)}\quad \to \quad
f_n^{\rm (toy)}(4\pi gt)\,, \qqqquad (n=0,1)\,.
\end{align} 
Since the cusp anomalous dimension \re{cusp=f} is determined by the $f_1-$function evaluated at 
$t=0$,  the additional factor does not affect its value. 

\subsection{Nonperturbative corrections to the cusp anomalous dimension}

The relation \re{mm} defines strong coupling corrections to the mass gap. In a
close analogy with the cusp anomalous dimension \re{cusp-full}, it runs in two
parameters: perturbative $1/g$ and nonperturbative $\Lambda^2$. We would like to
stress  that the separation of the corrections to $m_{\rm O(6)} $ into
perturbative and nonperturbative ones is ambiguous since the `perturbative'
series inside the square brackets in the right-hand side of \re{mm} is non-Borel
summable and, therefore, it suffers from Borel ambiguity. It is only the sum of
perturbative and nonperturbative corrections that is a unambiguously defined function of
the coupling constant. In distinction with the mass scale $m_{\rm O(6)} $, the
definition \re{Lambda1} of the nonperturbative scale $\Lambda^2$ involves a
complex  parameter $\sigma$ whose value depends on the prescription employed to
regularize singularities of the `perturbative' series.

To illustrate the underlying mechanism of cancellation of Borel ambiguity
inside $m_{\rm O(6)}$, let us examine the expression for the mass gap \re{m=f1}
in the toy model. As was already explained in Sect.~2.8, the toy model
captures the main features of the exact solution at strong coupling and, at the same time, it allows
us to obtain expressions for various quantities in a closed analytical form. The
mass gap in the toy model is given by the relation \re{m=f1} with $f_1(-\pi g)$ 
replaced with $f_1^{\toy}(-\pi g)$ 
defined in  \re{f-toy} and \re{c1-toy}. In this way, we obtain
\be\label{m-toy}
m_{\rm toy} = \frac{16\sqrt{2}}{\pi^2} \frac{f_1^{\toy}(-\pi g )}{\I_0(-\pi g)} =
\frac{16\sqrt{2}}{\pi^2} \frac{1}{\I_1(-\pi g)}\,.
\ee
Here $\I_1(-\pi g)$ is an entire function of the coupling constant (see
Eq.~\re{I-U}). {Its large $g$ asymptotic expansion can be easily deduced from 
\re{I1-sep} and it involves the nonperturbative parameter $\Lambda^2$. }

Making use of \re{I0-pos}
we obtain from \re{m-toy}
\begin{align} \notag
m_{\rm toy} = \frac{4}{\pi \Gamma(\ft54)} (2\pi g)^{1/4} \e^{-\pi g} & \bigg\{\bigg[1-\frac{1}{32\pi
g}-\frac{23}{2048 (\pi g)^2}+\ldots\bigg]
\\ \label{m-toy1}
& \qquad
- \frac{\Lambda^2}{8\pi g}\left[1 -  \frac{11}{32 \pi g} +\ldots \right]+
 O(\Lambda^4)\bigg\}\,,
\end{align}
where ellipses denote terms with higher power of $1/g$. By construction, $m_{\rm
toy}$ is a unambiguous function of the coupling constant whereas the asymptotic
series inside the square brackets are non-Borel summable. It is easy to verify
that `perturbative' corrections to $m_{\rm toy}^2$ are described by the
asymptotic series $C_2(\alpha)$ given by \re{c0-series}. Together with \re{toy-f}
this allows us to identify the leading nonperturbative correction to \re{toy-f}
in the toy model as
\begin{align}\label{delta-toy}
\delta \Gamma_{\rm cusp}^{\toy}  = - \frac{\Lambda^2}{\pi} C_2(\alpha) +
O(\Lambda^4) = -\frac{\pi^2}{32\sqrt{2}}\sigma m_{\rm toy}^2 + O(m_{\rm toy}^4)\,,
\end{align}
with $\Lambda^2$ given by \re{Lambda2}.

Comparing the relations \re{mm} and \re{m-toy1} we observe that $m_{\rm O(6)}$
and $m_{\rm toy}$ have the same leading asymptotics while subleading $1/g$
corrections to the two scales have different transcendentality. Namely, the
perturbative coefficients in $m_{\rm toy}$ are rational numbers while for $m_{\rm
O(6)}$  their transcendentality increases with order in $1/g$. We recall that we
already encountered the same property for the cusp anomalous dimension,
Eqs.~\re{toy-f} and \re{cusp-alpha}. There, we have observed that the two
expressions \re{toy-f} and \re{cusp-alpha} coincide upon the substitution
\re{recipe}. Performing the same substitution in \re{mm} we find that, remarkably
enough, the two expressions for the mass gap indeed coincide up to an overall
normalization factor
\be\label{O(6)-toy}
m_{\rm O(6)}\stackrel{\rm Eq.\re{recipe}}{=} \frac{\pi}{2\sqrt{2}}\, m_{\rm toy}\,.
\ee

The expressions for the cusp anomalous dimension \re{cusp-full} and for the mass
scale \re{mm} can be further simplified if one redefines the coupling constant as
\be
 g' = g - c_1\,,\qqqquad c_1 = \frac{3\ln 2}{4\pi}\,,
\ee
and re-expands both quantities in $1/g'$. As was observed in \cite{BKK07}, such
redefinition allows one to eliminate `$\ln 2$' terms in perturbative expansion of
the cusp anomalous dimension. Repeating the same analysis for \re{cusp-full} we
find that the same is also true for nonperturbative corrections
\begin{align}\label{cusp-full1}
\Gamma_{\rm cusp}\left(g+c_1\right) & = 2g \left[1  - \frac{\textrm{K}}{(4\pi g)^2} + \ldots
\right] - \frac{\Lambda^2}{2\sqrt{2}\pi}\left[ 1+ \frac{3}{16\pi g}  + \ldots \right]
+O(\Lambda^4)\,,
\end{align}
with $\Lambda^2$ defined in \re{Lambda2}. In the similar manner, the expression
for the mass scale \re{mm} takes the form
\begin{align}\notag
\left[ m_{\rm O(6)}(g+c_1)\right]^2 =  \frac{2\Lambda^2}{\pi\sigma} &
\bigg[ \lr{1+\frac{3}{16\pi g}+\frac{16{\rm K} - 54}{512(\pi g)^2}+\ldots}
\\\label{mm2}
&\qquad
- \frac{\Lambda^2}{8\sqrt{2}\pi g}\lr{1-\frac{3}{8\pi g} + \ldots}+ O(\Lambda^4)\bigg]\,.
\end{align}
Comparing the relations \re{cusp-full1} and \re{mm2} we immediately recognize
that, within an accuracy of the obtained expressions, nonperturbative
$O(\Lambda^2)$  correction to the cusp anomalous dimension is given by $m_{\rm
O(6)}^2$
\begin{align}\label{all-loop}
 \delta \Gamma_{\rm cusp} = -\frac{\sigma}{4\sqrt{2}} m_{\rm O(6)}^2 + O(m_{\rm O(6)}^4)\,.
\end{align}
It worth mentioning that, upon identification of the scales \re{O(6)-toy}, this
relation coincides with \re{delta-toy}.

We will show in the next subsection that the relation \re{all-loop} holds at
strong coupling to all orders in $1/g$.

\subsection{Relation between cusp anomalous dimension and mass gap}

We demonstrated that the strong coupling expansion of the cusp anomalous
dimension has the form \re{cusp-NLO1} with the leading nonperturbative correction
given to the first few orders in $1/g$ expansion by the mass scale of the O(6)
model, $m_{\rm cusp}^2=m_{\rm O(6)}^2$. Let us show that this relation is in fact
exact at strong coupling.

According to \re{cusp=pt+npt1}, the leading nonperturbative correction to the
cusp anomalous dimension is given by
\be\label{delta-cusp}
\delta \Gamma_{\rm cusp} = -4g \, \delta f_1(0)\,,
\ee
with $\delta f_1(0)$ denoting $O(\Lambda^2)$ correction to the function
$f_1(t=0)$, Eq.~\re{dec}. We recall that this function verifies the quantization
conditions \re{QC01}. As was explained in Section~3.3, the leading $O(\Lambda^2)$
corrections to solutions of  \re{QC01} originate from subleading, exponentially
suppressed terms in the strong coupling expansion of the functions $\I_0(-\pi g)$
and $\I_1(-\pi g)$ that we shall denote as $\delta \I_0(-\pi g)$ and $\delta
\I_1(-\pi g)$, respectively. Using the identities \re{I0-sep} and \re{I1-sep}, we
find
\begin{align}\label{delta_I}
\delta \I_0(-\pi g) & = \sigma\frac{2\sqrt{2}}{\pi}\e^{-\pi g} U_0^-(\pi g)\,, 
\\ \notag
 \delta \I_1(-\pi
g) & = \sigma\frac{2\sqrt{2}}{\pi} \e^{-\pi g} U_1^-(\pi g)\,,
\end{align}
where the functions $U_0^-(\pi g)$ and $U_1^-(\pi g)$ are defined in \re{U's}.
Then, we split the functions $f_0(t)$ and $f_1(t)$ entering the quantization
conditions \re{QC01} into perturbative and nonperturbative parts according to
\re{dec} and compare exponentially small terms in both sides of \re{QC01} to get
\begin{align}\label{delta-QC}
 \delta f_0(t_{\ell})\I_0(t_{\ell})
 + \delta f_1(t_{\ell}) \I_1(t_{\ell}) = -m' \delta_{\ell,0} \,,
\end{align}
where $t_{\ell}  =4\pi g
 \lr{\ell-\ft14}$ and the notation was introduced for
\begin{align}
m' =   f_0(-\pi g)\delta \I_0(-\pi g)
 + f_1(-\pi g) \delta \I_1(-\pi g)\,.
\end{align}
Taking into account the relations \re{delta_I} and comparing the resulting
expression for $m'$ with \re{m=U} we find that
\be\label{alpha}
m' = -\frac{\sigma}{8g} m_{\rm O(6)}\,,
\ee
with $m_{\rm O(6)}$ being the mass scale \re{m=U}.

To compute nonperturbative $O(\Lambda^2)$ correction to the cusp anomalous
dimension, we have to solve the system of relations \re{delta-QC}, determine the
function $\delta f_1(t)$ and, then, apply \re{delta-cusp}. We will show in this
subsection that the result reads
\be\label{f1-des}
\delta f_1(0) = -\frac{\sqrt{2}}{4}m' m_{\rm O(6)} =\sigma\frac{\sqrt{2}}{32 g}
\, m_{\rm O(6)}^2\,,
\ee
to all orders in strong coupling expansion. Together with \re{delta-cusp} this
leads to the desired expression \re{all-loop} for leading nonperturbative
correction to the cusp anomalous dimension.

To begin with, let us introduce a new function analogous to \re{G-gen}
\begin{align} \label{G-hat-gen}
\delta \Gamma(it) &=\delta  f_0(t)\I_0(t)+\delta f_1(t) \I_1(t)\,.
\end{align}
Here $\delta  f_0(t)$ and $\delta  f_1(t)$ are given by the same expressions as
before, Eq.~\re{ff}, with the only difference that the coefficients $c_\pm(n,g)$
are replaced in \re{ff} by their leading nonperturbative correction $\delta
c_\pm(n,g) = O(\Lambda^2)$ and the relation \re{c-zero} is taken into account.
This  implies that various relations for $\Gamma(it)$ can be immediately
translated into those for the function $\delta \Gamma(it)$.  In particular, for
$t=0$ we find from \re{ff} that $\delta  f_0(0)=0$ for arbitrary coupling,
leading to
\be\label{G-delta-f}
\delta \Gamma(0) = 2 \delta f_1(0)
\ee
In addition, we recall that, for arbitrary $c_\pm(n,g)$, the function \re{G-gen}
satisfies the inhomogeneous integral equation \re{G-int}. In other words, the
$c_\pm(n,g)-$dependent terms in the expression for the function  $\Gamma(it)$ are
zero modes for the integral equation \re{G-int}. Since the function
\re{G-hat-gen} is just given by the sum of such terms, it automatically satisfies
the homogenous equation
\begin{align}\label{hat-eq}
\int_0^\infty dt\, \bigg[ \e^{itu} \delta\Gamma_-(t) - \e^{-itu}\delta\Gamma_+(t)\bigg] = 0
\,,\qqqquad (-1\le u\le 1)\,,
\end{align}
where $\delta\Gamma(t) = \delta\Gamma_+(t)+i\delta\Gamma_-(t)$ and
$\delta\Gamma_\pm(-t)=\pm \delta\Gamma_\pm(t)$.

As before, in order to construct the solution to \re{hat-eq}, we have to specify
additional conditions for  $\delta\Gamma(t)$. Since the substitution
$c_\pm(n,g) \to \delta c_\pm(n,g)$ does not affect analytical properties of the
functions \re{ff}, the function \re{G-hat-gen} shares with $\Gamma(it)$  an
infinite set of simple poles located at the same position \re{poles}
\begin{align}
\delta \Gamma(it) \sim \frac1{t-4\pi g\ell}\,,\qqquad (\ell\in \mathbb{Z}/0)\,.
\end{align}
In addition, we deduce from \re{delta-QC} that it also satisfies the relation (with $x_\ell=\ell-\ft14$)
\begin{align}
\delta \Gamma(4\pi i g x_\ell) =-m'\delta_{\ell,0} \,,\qqquad (\ell\in \mathbb{Z})\,,
\end{align}
and, therefore, has an infinite number of zeros. An important difference with
$\Gamma(it)$ is that $\delta\Gamma(it)$ does not vanish at $t=-\pi g$ and its
value is fixed by the parameter $m'$ defined in \re{alpha}.

Having in mind similarity between the functions $\Gamma(it)$ and $\delta
\Gamma(it)$ we follow \re{sin-sin} and define a new function
\be\label{hat-gamma}
\delta\gamma(it) = \frac{\sin(t/4g)}{\sqrt{2}\sin{(t/4g+\pi/4)}}\delta\Gamma(i t)\,.
\ee
As before, the poles and zeros of  $\widehat\Gamma(i t)$ are compensated by the
ratio of sinus functions. However, in distinction with $\gamma(it)$ and in virtue
of $\delta \Gamma(-\pi i g ) = -m'$, the function $\delta\gamma(it)$ has a
single pole at $t=-\pi g$ with the residue equal to $2gm'$. For $t\to 0$ we find
from \re{hat-gamma} that $\delta\gamma(it)$ vanishes as
\be\label{hat-gamma-small}
\delta\gamma(it) = \frac{t}{4g}\delta\Gamma(0) + O(t^2) = \frac{t}{2g}\delta f_1(0) + O(t^2)\,,
\ee
where in the second relation we applied \re{G-delta-f}. It is convenient to split
the function $\delta\gamma(t) $ into the sum of two terms of a definite parity,
$\delta\gamma(t)=\delta\gamma_+(t)+ i \delta\gamma_-(t)$ with
$\delta\gamma_\pm(-t)=\pm \delta\gamma_\pm(t)$.
Then, combining together \re{hat-eq} and \re{hat-gamma} we obtain that the functions $\delta\gamma_\pm(t)$
satisfy the infinite system of homogenous equations (for $n\geqslant 1$)
\begin{align} \notag
& \int_{0}^{\infty}\frac{dt}{t} \, J_{2n-1}(t)
\bigg[\frac{\delta\gamma_{-}(t)}{1-\e^{-t/2g}}+\frac{\delta\gamma_{+}(t)}{\e^{t/2g}-1}\bigg]   = 0\,,
\\ \label{hat-gamma-eq}
& \int_{0}^{\infty}\frac{dt}{t} \, J_{2n}(t)
\bigg[\frac{\delta\gamma_{+}(t)}{1-\e^{-t/2g}}-\frac{\delta\gamma_{-}(t)}{\e^{t/2g}-1}\bigg]  = 0\,.
\end{align}
By construction, the solution to this system $\delta\gamma(t)$ should vanish at $t=0$ and have a
simple pole at $t=-i\pi g$.

As was already explained, the functions $\delta\Gamma_\pm(t)$ satisfy the same
integral equation \re{hat-eq} as  the function $\Gamma_\pm(t)$ up to
inhomogeneous term in the right-hand side of \re{G-int}. Therefore, it  should
not be surprising that the system \re{hat-gamma-eq} coincides with the relations
\re{FRS2} after one neglects the inhomogeneous term in the right-hand side of
\re{FRS2}. As we show in Appendix~\ref{App:C}, this fact allows us to derive
Wronskian like relations between the functions $\delta\gamma(t)$ and
$\gamma(t)$. These relations turn out to be powerful enough to determine the
small $t$ asymptotics of the function $\delta\gamma(t)$ at small $t$ in terms
of $\gamma(t)$, or equivalently $\Gamma(t)$. In this way we obtain (see
Appendix~\ref{App:C} for more detail)
\be\label{hat-gamma-m}
\delta\gamma(it) = - m' t\bigg[\frac{2}{\pi^2g}\e^{-\pi
g}-\frac{\sqrt{2}}{\pi}\e^{-\pi g}\Re \int_{0}^{\infty}\frac{dt'}{t'+i \pi g}\, \e^{i
(t'-\pi/4)} \Gamma (t') \bigg] + O(t^2)\,.
\ee
Comparing this relation with \re{m=int} we realize that the expression inside the
square brackets is proportional to the mass scale $m_{\rm O(6)}$ leading to
\be\label{hat-gamma-small1}
\delta\gamma(it) =-  m' m_{\rm O(6)}\frac{t \sqrt{2}}{8g}+ O(t^2)\,.
\ee
Matching this relation into  \re{hat-gamma-small}, we obtain the desired
expression for $\delta f_1(0)$, Eq.~\re{f1-des}. Then, we substitute it into
\re{delta-cusp} and compute the leading nonperturbative correction to the cusp
anomalous dimension, Eq.~\re{all-loop}, leading to
\be
m_{\rm cusp}(g) = m_{\rm O(6)}(g)\,.
\ee
Thus, we demonstrated in this section that nonperturbative, exponentially small
corrections to the cusp anomalous dimensions at strong coupling are determined to
all orders in $1/g$ by the mass gap of the two-dimensional bosonic O(6) model
embedded into $\rm AdS_5\times S^5$ sigma-model.

\section{Conclusions}

In this paper, we have studied anomalous dimensions of Wilson operators in  the
$SL(2)$ sector of planar $\mathcal{N}=4$ SYM theory in the double scaling limit
when Lorentz spin of the operators grows exponentially with their twist. In this
limit, the asymptotic behavior of the anomalous dimensions is determined by the
cusp anomalous dimension $\Gamma_{\rm cusp}(g)$ and the scaling function
$\epsilon(g,j)$. We found that  at strong coupling both functions receive
exponentially small corrections which are parameterized by the same
nonperturbative scale. It is remarkable that this scale appears on both sides of
the AdS/CFT correspondence. In string theory it emerges as the mass gap of the
two-dimensional bosonic O(6) sigma model which describes the effective dynamics
of massless excitations for folded spinning string in the $\rm AdS_5\times S^5$
sigma model~\cite{AM07}.

The dependence on $\Gamma_{\rm cusp}(g)$ and $\epsilon(g,j)$ on the coupling
constant is governed by integral BES/FRS equations which follow from the
conjectured all-loop integrability of the dilatation operator of $\mathcal{N}=4$
model. At weak coupling, their solutions agree with results of explicit
perturbative calculations. At strong coupling, systematic expansion of  the cusp
anomalous dimension  in  powers of $1/g$ was derived in \cite{BKK07}. In
agreement with the AdS/CFT correspondence, the first few terms of this expansion
coincide with the energy of the semiclassically quantized folded spinning
strings. However, the expansion coefficients grow factorially at higher orders
and, as a consequence, `perturbative' $1/g$  expansion of the cusp anomalous
dimension suffers from Borel singularities which induce exponentially small
corrections to $\Gamma_{\rm cusp}(g)$. To identify such nonperturbative
corrections, we revisited the BES equation and constructed the exact solution for
the cusp anomalous dimension valid for arbitrary coupling constant.

At strong coupling, we found that the obtained expression for $\Gamma_{\rm
cusp}(g)$  depends on a new scale $m_{\rm cusp}(g)$ which is exponentially small
as $g\to\infty$. Nonperturbative corrections to $\Gamma_{\rm cusp}(g)$ at strong
coupling run in even powers of this scale and the coefficients of this expansion
depend on the prescription employed to regularize Borel singularities in
perturbative $1/g$ series. It is only the sum of perturbative and nonperturbative
contributions which is independent on the choice of the prescription.
For the
scaling function $\epsilon(g,j)$, the defining integral FRS equation can be
brought to the form of the thermodynamical Bethe ansatz equations for the energy
density of the ground state of the O(6) model. As a consequence, nonperturbative
contribution to $\epsilon(g,j)$ at strong coupling is described by the mass scale
of this model $m_{\rm O(6)}(g)$. We have shown that the two scales 
coincide, $m_{\rm cusp}(g)=m_{\rm O(6)}(g)$, 
and, therefore, nonperturbative contributions to $\Gamma_{\rm cusp}(g)$ and $\epsilon(g,j)$
are governed by the same scale $m_{\rm O(6)}(g)$.

This result agrees with the proposal by Alday-Maldacena that, in
string theory, the leading nonperturbative corrections to the cusp anomalous dimension 
coincide  with those to the vacuum energy
density of two-dimensional bosonic O(6) model embedded into the $\rm AdS_5\times
S^5$ sigma-model. These models have different properties: the former model has
asymptotic freedom at short distances and develops mass gap in the infrared while
the latter model is conformal. The O(6) model only describes an effective
dynamics of massless modes of $\rm AdS_5\times S^5$ and the mass of massive
excitations $\mu\sim 1$ defines a ultraviolet (UV) cut-off for this model. The
coupling constants  in the two models are related to each other as
$\bar g^2(\mu)=1/(2g)$. The vacuum energy density in the O(6) model and, more
generally in the O($n$) model is a ultraviolet divergent quantity. It also
depends on the mass scale of model and has the following form
\begin{align}\label{vac}
\epsilon_{\rm vac}  
=  \mu^2 \epsilon(\bar g^2) + \kappa\, m_{{\rm O}(n)}^2 + O(m_{{\rm O}(n)}^4/\mu^2)\,.
\end{align}
Here $\mu^2$ is a UV cut-off, $\epsilon(\bar g^2)$ stands for perturbative series
in $\bar g^2$ and the mass gap $m_{{\rm O}(n)}^2$ is
\begin{align}\label{mass-On}
m^2_{{\rm O}(n)} = c\, \mu^2 \e^{-\frac1{\beta_0 \bar g^2}} \bar
g^{-2\beta_1/\beta_0^2} \left[1+ O(\bar g^2)\right]\,,
\end{align}
where $\beta_0$ and $\beta_1$ are the beta-function coefficients  for the O($n$)
model and the normalization factor $c$ ensures independence of $m_{{\rm O}(n)}$
on the renormalization scheme. For $n=6$ the relation \re{mass-On} coincides with
\re{O6-epsilon} and the expression for the vacuum energy density \re{vac} should
be compared with \re{cusp=pt+npt}.

The two terms in the right-hand side of \re{vac} describe perturbative and
nonperturbative corrections to $\epsilon_{\rm vac}$. For $n\to\infty$ each of
them is well-defined separately and can be computed exactly~\cite{David,NSVZ84}.
For $n$ finite, including $n=6$, the function $\epsilon(\bar g^2)$ is given in a
generic renormalization scheme by a non-Borel summable series and, therefore, is
not well-defined. In a close analogy with \re{cusp=pt+npt}, the coefficient
$\kappa$ in front of $m_{{\rm O}(n)}^2$ in the right-hand side of \re{vac}
depends on the regularization of Borel singularities in perturbative series for
$\epsilon(\bar g^2)$. Notice that $\epsilon_{\rm vac} $  is
related to the vacuum expectation value of the trace of the tensor energy-momentum in the
two-dimensional O($n$) sigma model~\cite{NSVZ84}. The AdS/CFT correspondence
implies that for $n=6$ the same quantity defines nonperturbative correction to
the cusp anomalous dimension  \re{cusp=pt+npt}. It would be interesting to obtain
its dual representation (if any)  in terms of  certain operators in four-dimensional
$\mathcal{N}=4$ SYM theory. Finally, one may wonder whether it is possible to
identify a restricted class of Feynman diagrams in $\mathcal{N}=4$ theory whose
resummation could produce  contribution to the cusp anomalous dimension
exponentially small as $g\to\infty$. As a relevant example, we would like to
mention that exponentially suppressed corrections were obtained in
Ref.~\cite{Broadhurst:1993ib} from exact resummation of ladder diagrams in
four-dimensional massless  $g\phi^3$ theory.

\section*{Acknowledgments}

We would like to thank Janek Kota\' nski for collaboration at the early stages of this work
and Zoltan Bajnok, Janos Balog, David Broadhurst,
Fran{\c{c}}ois David, Alexander Gorsky, Juan Maldacena,  Laszlo Palla, Mikhail
Shifman, Arkady Tseytlin and Dima Volin for interesting discussions. This work was supported
in part by the French Agence Nationale de la Recherche under grant
ANR-06-BLAN-0142 and by the CNRS/RFFI grant 09-02-00308.

\appendix

\section{Weak coupling expansion}\label{App:A}

In this Appendix, we work out the first few terms of the weak coupling expansion of the coefficient $c(g)$ entering \re{cusp-c} and show that they vanish in agreement with \re{c-zero}. To this end, we will not attempt
at solving the quantization conditions \re{QC01} at weak coupling but will use instead the fact
that the BES equation can be solved by iteration of the inhomogeneous term.

The system of integral equation \re{FRS2} can be easily solved at weak coupling
by looking for its solutions $\gamma_\pm(t)$ in the form of the Bessel series
\re{Bessel} and expanding the coefficients $\gamma_{2k}$ and $\gamma_{2k-1}$ in
powers of the coupling constant. For $g\to 0$ it follows from \re{FRS2} and from
orthogonality conditions for the Bessel functions, that $\gamma_-(t) = J_1(t) +
\ldots$ and $\gamma_+(t) = 0+\ldots$ with ellipses denoting subleading terms. To
determine such terms it is convenient to change the integration variable in
\re{FRS2} as $t\to tg$. Then, taking into account the relations $J_{k}(-gt) =
(-1)^k J_k(gt)$ we observe that the resulting equations are invariant under
substitution $g\to -g$ provided that  the functions $\gamma_\pm(gt)$ change
sign under this transformation. Since $\gamma_\pm(-t)=\pm\gamma_\pm(t)$, this
implies that the coefficients $\gamma_{2n-1}(g)$ and $\gamma_{2n}(g)$ entering
\re{Bessel} have a definite parity as functions of the coupling constant
\be
\gamma_{2n-1}(-g) = \gamma_{2n-1}(g)\,,\qqqquad \gamma_{2n}(-g) = - \gamma_{2n}(g)\,,
\ee
and, therefore, their weak coupling expansion runs in even and odd powers of $g$,
respectively. Expanding both sides of \re{FRS2} at weak coupling and comparing
the coefficients in front of powers of $g$ we find
\begin{align}\notag
\gamma_1 &= \frac12 - \frac{\pi^2}{6} g^2 +\frac{11\pi^4}{90} g^4 -
\lr{\frac{73\pi^6}{630}+4\zeta_3^2} g^6 +O(g^8)\,,
\\\notag
\gamma_2 &= \zeta_3 g^3 -\lr{\frac{\pi^2}{3} \zeta_3 + 10\zeta_5} g^5 +\lr{\frac{8\pi^4}{45}\zeta_3
+\frac{10\pi^2}{3} \zeta_5+105 \zeta_7} g^7 + O(g^9)\,, 
\\[2mm] \notag
\gamma_3  &= -\frac{\pi^4}{90} g^4 +\frac{37\pi^6}{1890} g^6 + O(g^8)\,,
&& \hspace*{-75mm}
\gamma_4  =\zeta_5 g^5 -\lr{\frac{\pi^2}{3}\zeta_5 + 21 \zeta_7}g^7 + O(g^9)\,,
\\[2mm]
\gamma_5  &=-\frac{\pi^6}{945} g^6 + O(g^8)\,,
&& \hspace*{-75mm} \label{weak-c}
\gamma_6 =\zeta_7 g^7 + O(g^9)\,.
\end{align}
We verify with a help of \re{cusp1} that the expression for the cusp anomalous
dimension
\be
\Gamma_{\rm cusp}(g) = 8g^2 \gamma_1(g) = 4g^2 - \frac{4\pi^2}{3}g^4
+\frac{44\pi^4}{45}g^6 - \lr{\frac{292\pi^6}{315}+32\zeta_3^2}g^8 + O(g^{10})
\ee
agrees with the known four loop result  in planar
$\mathcal{N}=4$ SYM theory~\cite{cusp-4loop}.

In our approach, the cusp anomalous dimension is given for arbitrary value of the
coupling constant by the expression \re{cusp-c} which involves the functions
$c(g)$ and $c_\pm(n,g)$. According to \re{c-gamma}, the latter functions are
related to the functions $\gamma(t)=\gamma_+(t)+i\gamma_-(t)$ evaluated at
$t=4\pi ign$
\begin{align} \notag
c_+(n,g) &= -4g \e^{-4\pi gn} \left[\gamma_+(4\pi i gn) + i\gamma_-(4\pi ign)\right]\,,
\\[2mm]\label{app-c}
c_-(n,g) &=\phantom{-}  4g \e^{-4\pi gn} \left[\gamma_+(4\pi i gn) - i\gamma_-(4\pi ign)\right]\,.
\end{align}
At strong coupling, we determined $c_\pm(n,g)$ by solving the quantization
conditions \re{QC1}. At weak coupling, we can compute $c_\pm(n,g)$ from
\re{app-c} by replacing $\gamma_\pm(t)$ with their Bessel series \re{Bessel}  and
making use of the obtained expressions for the expansion coefficients
\re{weak-c}.

The remaining function $c(g)$ can be found from comparison of two different
representations for the cusp anomalous dimension, Eqs.~\re{cusp-c} and
\re{cusp1},
\begin{align}
 c(g) =  -\frac12+2g \gamma_1(g)+\sum_{n\ge 1}\big[{ c_-(n,g)\, U_0^-(4\pi
ng) +c_+(n,g)\, U_0^+(4\pi ng)}\big] \,.
\end{align}
Taking into account the relations \re{app-c} and \re{Bessel} we find
\begin{align}\label{app-c-sum}
c(g) = -\frac12 + 2g\gamma_1(g) -\sum_{k\ge 1}(-1)^k\big[ (2k-1)\gamma_{2k-1}(g)
f_{2k-1}(g) +(2k) \gamma_{2k} (g) f_{2k}(g)\big]\,,
\end{align}
where the coefficients $\gamma_k$ are given by \re{weak-c} and
the notation was introduced for the functions
\be
f_k(g) = 8g \sum_{n\ge 1} \big[ U_0^+ (4\pi gn) -(-1)^k U_0^-(4\pi gn)\big]
I_k(4\pi gn) \e^{-4\pi gn}\,.
\ee
Here $I_k(x)$ is the modified Bessel function~\cite{WW} and the functions
$U_0^\pm(x)$ are defined in \re{U's}. At weak coupling, the sum over $n$ can be
evaluated with a help of  the Euler-Maclaurin summation formula.  Going through
lengthy calculation we find
\begin{align}\notag
f_1 &= 1-2g +\frac{\pi^2}{3}g^2+ 2 \zeta_3 g^3 -\frac{\pi^4}6 g^4 -23 \zeta_5 g^5 +\frac{17\pi^6}{108} g^6 +\frac{1107}{4}\zeta_7 g^7+ O(g^8)\,,
\\[2mm] \notag
f_2 &=-\frac12+2 \zeta_3 g^3 - \frac{\pi^4}{30}g^4 + O(g^5)\,,\hspace*{20mm}
  f_3 =\frac12 + O(g^4)\,,
\\[2mm]
f_4 &= -\frac38+O(g^5)\,,\hspace*{20mm}
f_5  =\frac38 + O(g^6)\,, \hspace*{20mm} f_6 =-\frac5{16}+ O(g^7)\,.
\end{align}
In this way, we obtain from \re{app-c-sum}
\begin{align}
c(g) = -\frac12+\lr{f_1+2g}\gamma_1 + 2 f_2 \gamma_2 - 3 f_3\gamma_3 -4 f_4
\gamma_4 +5 f_5 \gamma_5 + 6 f_6 \gamma_6 +\ldots = O(g^8)\,.
\end{align}
Thus, in agreement with \re{c-zero}, the function  $c(g)$ vanishes at weak
coupling. As was shown in Sect.~3.1, the relation $c(g)=0$ holds for arbitrary
coupling.

\section{Constructing general solution}\label{App:B}

By construction, the function $\Gamma(t)=\Gamma_+(t)+i\Gamma_-(t)$ defined as the
exact solution to the integral equation \re{G-int} is given by the Fourier
integral
\begin{align}\label{App-Fourier}
\Gamma(t)  = \int_{-\infty}^\infty dk\e^{-ikt} \widetilde \Gamma(k) \,,
\end{align}
with the function $\widetilde \Gamma(k)$ having different form for $k^2\le 1$ and
$k^2>1$:
\begin{itemize}
\item For $-\infty < k <-1$:
\be\label{Ap-exp1}
\widetilde \Gamma(k)  {=}  \sum_{n\ge 1} c_-(n,g) \e^{-4\pi n g (-k-1)}\,,
\ee
\item For $1< k < \infty$:
\be\label{Ap-exp2}
\widetilde \Gamma(k)  {=}\sum_{n\ge 1} c_+(n,g) \e^{-4\pi n g (k-1)}
 \,,
\ee
\item For $-1 \le k \le 1$:
\be\label{Ap-exp3}
\widetilde \Gamma(k) = -\frac{\sqrt{2}}{\pi}\lr{\frac{1+k}{1-k}}^{1/4}\left[
 {1+\frac{c(g)}{1+k}}
 +\frac12\lr{\int_{-\infty}^{-1}+\int^{\infty}_{1}}
 \frac{dp\,\widetilde\Gamma(p)}{p-k}\lr{\frac{p-1}{p+1}}^{1/4}\right]\,,
\ee
where $\widetilde\Gamma(p)$ inside the integral is replaced by \re{Ap-exp1} and
\re{Ap-exp2}.
\end{itemize}
Let us split the integral in \re{App-Fourier} into three terms as in
\re{three-terms} and evaluate them one after another. Integration over $k^2>1$
can be done immediately while the integral over $-1 \le k \le 1$ can be expressed
in terms of special functions
\begin{align} \notag
\Gamma(t)  &=  \sum_{n\ge 1}  c_+(n,g)\left[{ \frac{\e^{-it}}{4\pi ng+it}-\I_+(-it,4\pi ng)}\right]
\\& \label{App-Gamma}
+ \sum_{n\ge 1}
 c_-(n,g)\left[{\frac{\e^{it}}{4\pi ng-it} + \I_-(it,4\pi ng)}\right]  -\I_0(-it)-c(g) \I_1(-it)\,,
\end{align}
where the notation was introduced for the functions  
(with $n=0,1$)
\begin{align}\notag
\I_\pm(x,y) &= \frac{1}{\sqrt{2}\pi}\int_{-1}^1 dk\e^{\pm x k}
 \int_{1}^\infty
 \frac{dp\,\e^{-y (p-1)}}{p-k}\lr{\frac{1+k}{1-k}\frac{p-1}{p+1}}^{\pm 1/4}\,,
\\ \notag
\I_n(x)&=\frac{\sqrt{2}}{\pi}\int_{-1}^1 \frac{dk\,\e^{x k}}{(k+1)^n}
\lr{\frac{1+k}{1-k}}^{1/4}\,,
 \\  \label{fff}
 U_n^\pm(y) &= \frac12 \int_{1}^\infty \frac{dp\,\e^{-y(p-1)}}{(p\mp 1)^n} \left(
\frac{p+1}{p-1} \right)^{\mp 1/4}\,.
\end{align}
The reason why we also introduced $U_n^\pm(y) $ is that
the functions $\I_\pm(x,y)$ can be further simplified with a help of master identities  (we shall return to them in a moment)
\begin{align}\notag
(x+y)\I_-(x,y) &=   {x \I_0(x) U_1^-(y) + y \I_1(x) U_0^-(y)}- {\e^{-x}}\,,
\\[3mm] \label{I-iden}
(x-y) \I_+(x,y) & =   {x \I_0(x) U_1^+(y) + y \I_1(x) U_0^+(y)}-{\e^x}\,.
\end{align}
Combining together \re{I-iden} and \re{App-Gamma} we arrive at the following
expression for the function $\Gamma(it)$
\begin{align} \notag
\Gamma(it) &=  -\I_0(t)-c(g) \I_1(t)
\\[3mm] & \notag
 + \sum_{n\ge 1}  {c_+(n,g)} \left[\frac{4\pi ng \I_1(t) U_0^+(4\pi ng)+t \I_0(t) U_1^+(4\pi ng) }
 {4\pi ng-t} \right]
\\&
+ \sum_{n\ge 1}
  {c_-(n,g)}\left[ \frac{4\pi ng \I_1(t) U_0^-(4\pi ng)+t \I_0(t) U_1^-(4\pi ng)}{4\pi
  ng+t}\right],
\end{align}
which leads to \re{G-gen}.

We show in Appendix~\ref{App:D} that the functions $\I_{0,1}(t)$ and
$U_{0,1}^\pm(4\pi ng)$ can be expressed in terms of Whittaker functions of the
first and second kind, respectively. As follows from their integral
representation, $\I_0(t)$ and $\I_1(t)$ are holomorphic functions of $t$. As a
result,  $\Gamma(it)$  is a meromorphic function of $t$ with (an infinite) set of
poles located at $t=\pm 4\pi ng$ with $n$ positive integer.

Let us now prove the master identities \re{I-iden}. 
We start with the second relation in \re{I-iden}
and make use of \re{fff} to rewrite the expression in the left-hand side of  \re{I-iden}  as
\begin{align}\label{aux1}
(x-y) V_+(x,y)\e^{-y} = (x-y) \int_0^\infty ds\, V_0(x+s) U_0^+(y+s)\e^{-y-s}\,.
\end{align}
Let us introduce two auxiliary functions
\begin{align} \notag
 z_1(x) & = V_1(x)\,, &&   z_1(x) + z'_1(x) = V_0(x)\,,
 \\[2mm]
 z_2(x) &= \e^{-x}U_1^+(x)\,, && z_2(x) + z'_2(x) = - \e^{-x}U_0^+(x)\,,
\end{align}
 with $V_n(x)$ and $U_n^+(x)$ given by \re{fff}. They satisfy the second-order differential equation
\be
\frac{d}{dx} \lr{x z_i'(x)} = \left(x-\ft12\right) z_i(x) \,.
\ee
Applying this relation it  is straightforward to verify the following identity  
 \begin{align}\notag
  -(x-y) \big[z_1(x+s) & + z'_1(x+s)\big]  \big[z_2(y+s) + z'_2(y+s)\big]
\\[2mm] \notag
& = \frac{d}{ds} \big\{  (y+s)[z_2(y+s)+z'_2(y+s)]z_1(x+s) \big\}
\\ \label{aux3}
& - \frac{d}{ds} \big\{ (x+s)[z_1(x+s)+z'_1(x+s)]z_2(y+s) \big\}.
\end{align}
It is easy to see that the expression in the left-hand side coincides with the integrand in \re{aux1}.
Therefore, integrating both sides of \re{aux3} over $0\le s <\infty$, we obtain
\begin{align}\notag
 (x-y) V_+(x,y) 
  & =- \e^{-s} \left[ (x+s) V_0(x+s)U_1^+(y+s) 
 + (y+s) V_1(x+s)  U_0^+(y+s) \right]\big|_{s=0}^{s=\infty}
  \\[3mm] 
  &= - {\e^x} + xV_0(x)U_1^+(y) +y V_1(x)  U_0^+(y)\,,
\end{align}
where in the second relation we took into account the asymptotic behavior of the functions \re{fff} (see Eqs.~\re{I-U-relation} and \re{U-large}),
$V_n(s) \sim \e^s s^{-3/4}$ and $U_n^+(s) \sim  s^{n-5/4}$ as $s\to\infty$. 

The derivation of the first relation in \re{I-iden} goes along the same lines.
 
\section{Wronskian like relations}\label{App:C}

In this Appendix we present a detailed derivation of the relation
\re{hat-gamma-m} which determines the small $t$ expansion of the function
$\delta\gamma(t)$. This function satisfies the infinite system of integral
equations \re{hat-gamma-eq}. In addition, it should vanish at the origin, $t=0$
and have a simple pole at $t=-i\pi g$ with the residue $2i g m'$ (see
Eq.~\re{hat-gamma}). To fulfill these requirements, we  split
$\delta\gamma(it)$ into the sum of two functions
\be\label{gamma-pol}
 \delta\gamma(it) = \widehat\gamma(it)-\frac{2m'}{\pi}\frac{t}{t+\pi g}\,,
\ee
where, by the construction, $\widehat\gamma(it)$ is an entire function vanishing  at $t=0$ 
and its Fourier transform has a support on the interval $[-1,1]$. Similarly
to \re{g_pm}, we decompose $\delta\gamma(t)$ and $\widehat\gamma(t)$ into the sum
of two functions with a definite parity
\begin{align} \notag
\delta\gamma_+(t) & = \widehat\gamma_+(t)
-\frac{2m'}{\pi}\frac{t^2}{t^2+(\pi g)^2}\,,
\\
\delta\gamma_{-}(t) &= \widehat\gamma_{-}(t)+\frac{2gm' t}{t^2+\pi^2
g^2}\,.
\end{align}
Then, we substitute these relations into \re{hat-gamma-eq} and obtain the system
of inhomogeneous integral equations for the functions $\widehat\gamma_\pm(t)$
\begin{align}\notag
& \int_{0}^{\infty}\frac{dt}{t} \, J_{2n-1}(t) \bigg[\frac{
\widehat\gamma_{-}(t)}{1-\e^{-t/2g}}+\frac{
\widehat\gamma_{+}(t)}{\e^{t/2g}-1}\bigg] =  h_{2n-1} (g)\,,
\\ \label{g-h}
&\int_{0}^{\infty}\frac{dt}{t} \, J_{2n}(t)
\bigg[\frac{\widehat\gamma_{+}(t)}{1-e^{-t/2g}}-\frac{\widehat\gamma_{-}(t)}{e^{t/2g}-1}\bigg]
=h_{2n}(g)\,,
\end{align}
with inhomogeneous terms given by
\begin{align}\notag
h_{2n-1} &= \frac{2m'}{\pi}\int_0^\infty \frac{dt\, J_{2n-1}(t)}{t^2 + (\pi
g)^2}\bigg[\frac{t}{\e^{t/(2g)}-1}-\frac{\pi g}{1-\e^{-t/(2g)}} \bigg]\,,
\\ \label{hh}
h_{2n} &= \frac{2m'}{\pi}\int_0^\infty \frac{dt\, J_{2n}(t)}{t^2 + (\pi
g)^2}\bigg[\frac{\pi g}{\e^{t/(2g)}-1}+\frac{t}{1-\e^{-t/(2g)}} \bigg]\,.
\end{align}
Comparing these relations with \re{g-h} we observe that they  only differ by the
form of inhomogeneous terms and can be obtained one from another  through the
substitution
\be\label{exchange}
\widehat \gamma_\pm(t) \to \gamma_\pm(t)\,,\qquad h_{2n-1} \to \ft12
\delta_{n,1}\,,\qquad h_{2n}\to 0
\ee
In a close analogy with \re{Bessel}, we look for solution to \re{g-h} in the form
of Bessel series
\begin{align}\label{delta-Bessel}
\widehat \gamma_{-}(t) &=  2 \sum_{n\geqslant 1} \  (2n-1)  J_{2n-1}(t)
\widehat \gamma_{2n-1}(g) \,,
\\ \notag
\widehat \gamma^{\BES}_{+}(t) &=   2\sum_{n\geqslant 1} \  (2n) \ J_{2n}(t)
\widehat \gamma_{2n}(g) \,.
\end{align}
For small $t$ we have $\widehat\gamma_{-}(t)= t \widehat \gamma_1+O(t^2)$ and  $\widehat
\gamma_{+}(t) = O(t^2)$. Then it follows from \re{gamma-pol}
\be\label{small-t}
\delta \gamma(t) = i \widehat\gamma_{-}(t) + \frac{2im'}{\pi^2 g} t +
O(t^2) = i t \lr{\widehat \gamma_1 +\frac{2m'}{\pi^2 g}} +  O(t^2)\,,
\ee
so that the leading asymptotics is controlled by  the coefficient
$\widehat \gamma_1$.

Let us multiply both sides of the first relation in \re{g-h} by
$(2n-1)\gamma_{2n-1}$ and sum both sides over $n\ge 1$ with a help of
\re{Bessel}. In the similar manner, we multiply the second relation in \re{g-h}
by $(2n)\gamma_{2n}$ and follow the same steps. Then, we subtract the second
relation from the first one and obtain
\begin{align}\notag
  \int_{0}^{\infty} \frac{dt}{t}
\bigg[\frac{
\gamma_-(t)\widehat\gamma_{-}(t)-\gamma_+(t)\widehat\gamma_{+}(t)}{1-\e^{-t/2g}}+\frac{
\gamma_-(t)\widehat\gamma_{+}(t)+\gamma_+(t)\widehat\gamma_{-}(t)}{\e^{t/2g}-1}\bigg]
\\
= 2 \sum_{n\ge 1}  \left[ (2n-1) \gamma_{2n-1} h_{2n-1}   -(2n) \gamma_{2n}
h_{2n}  \right].
 \end{align}
We notice that  the expression in the left-hand side of this relation is
invariant under exchange $\widehat \gamma_\pm(t) \leftrightarrow \gamma_\pm(t)$.
Therefore, the  right-hand side should be also invariant under \re{exchange}
leading to
\be
\widehat \gamma_1 = 2 \sum_{n\ge 1}  \left[ (2n-1) \gamma_{2n-1} h_{2n-1}   -(2n)
\gamma_{2n} h_{2n}  \right]\,.
\ee
Replacing $h_{2n-1}$ and $h_{2n}$ by their expressions \re{hh} and taking into
account \re{Bessel} we obtain that $\widehat \gamma_1$ is given by  the integral
involving the functions $\gamma_\pm(t)$. It takes much simpler form when
expressed in terms of the functions $\Gamma_\pm(t)$ defined in \re{rel}
\be
 \widehat \gamma_{1}  = - \frac{m'}{\pi}\int_{0}^{\infty}dt\, \bigg[\frac{\pi g}{t^2+\pi^2
g^2}\left(\Gamma_{-}(t)-\Gamma_{+}(t)\right) + \frac{t}{t^2+\pi^2
g^2}\left(\Gamma_{-}(t)+\Gamma_{+}(t)\right)\bigg].
\ee
Making use of identities
\begin{align}
\frac{\pi g}{t^2+\pi^2g^2} &= \int_{0}^{\infty}du \, \e^{-\pi g u} \cos{(ut)}\,,
\\ \notag
\frac{t}{t^2+\pi^2g^2} &= \int_{0}^{\infty}du \, \e^{-\pi g u} \sin{(ut)}\,,
\end{align}
we rewrite $\widehat \gamma_{1}(g)$ as
\begin{align}\notag
\widehat \gamma_{1}  = - \frac{m'}{\pi}\int_{0}^{\infty}du\,\e^{-\pi gu}
\bigg[& \int_{0}^{\infty}dt\,\cos(ut) \left(\Gamma_{-}(t)-\Gamma_{+}(t)\right)
\\
+& \int_{0}^{\infty}dt\,\sin(ut)
\left(\Gamma_{-}(t)+\Gamma_{+}(t)\right)\bigg]\,.
\end{align}
Let us spit the $u-$integral into $0\le u \le 1$ and $u>1$. We observe that for
$u^2 \le 1$ the $t-$integrals in this relation are given by \re{sys}. Then, we
perform  integration over $u\ge 1$ and find after some algebra (with
$\Gamma(t)=\Gamma_+(t) +i\Gamma_-(t)$)
\be
\widehat \gamma_{1}  = - \frac{2m'}{\pi^2g}\lr{1-\e^{-\pi
g}}-\frac{\sqrt{2}m'}{\pi}\e^{-\pi g}\Re\bigg[\int_{0}^{\infty}\frac{dt}{t+i
\pi g}\, \e^{i (t-\pi/4)} \Gamma(t) \bigg].
\ee
Substituting this relation into
\re{small-t} we arrive at \re{hat-gamma-m}.

\section{Relation to Whittaker functions}\label{App:D}

In this appendix we summarize properties of special functions that we encountered
in our analysis.

\subsection*{Integral representations}

Let us first consider the functions $\I_n(x)$ (with $n=0, 1$) introduced in
\re{I-U}. As follows from their integral representation, $\I_0(x)$ and $\I_1(x)$
are entire function on a complex $x-$plane. Changing the integration variable in
\re{I-U} as $u=2t-1$ and $u=1-2t$ we obtain two equivalent representations
\begin{align} \notag
\I_n(x)  &= \frac{1}{\pi}2^{3/2-n} \e^x \int_0^1 dt\, t^{-1/4}
(1-t)^{1/4-n}\e^{-2tx}\,,
\\
& = \frac{1}{\pi}2^{3/2-n} \e^{-x} \int_0^1 dt\, t^{1/4-n} (1-t)^{-1/4}\e^{2tx}\,,
\end{align}
which give rise to the following expressions for $\I_n(x)$ (with $n=0,1$) in terms
of Whittaker functions of the first kind
\begin{align}\notag
\I_n(x) &= 2^{-n}\frac{\Gamma(\ft54-n)}{\Gamma(\ft54)\Gamma(2-n)} (2x)^{n/2-1}
M_{n/2-1/4,1/2-n/2}(2x)\,,
\\ \label{I=M}
&= 2^{-n}\frac{\Gamma(\ft54-n)}{\Gamma(\ft54)\Gamma(2-n)} (-2x)^{n/2-1}
M_{1/4-n/2,1/2-n/2}(-2x)\,.
\end{align}
In distinction with $\I_n(x)$, the Whittaker function $M_{n/2-1/4,1/2-n/2}(2x)$
is an analytical function of $x$ on the complex plane with the cut  along
negative semi-axis. The same is true for the factor $(2x)^{n/2-1}$ so that the
product of two functions in the right-hand side of \re{I=M} is a single-valued
analytical function in the whole complex plane. The two representations \re{I=M}
are equivalent in virtue of the relation
\begin{align}
M_{n/2-1/4,1/2-n/2}(2x) = \e^{\pm i\pi\lr{1-n/2}} M_{1/4-n/2,1/2-n/2}(-2x) \qquad \text{(for $\Im x \gtrless 0$)}\,,
\end{align}
where the upper and lower signs in the exponent correspond to $\Im x>0$ and $\Im
x<0$, respectively.

Let us know consider the functions $U_0^\pm(x)$ and $U_1^\pm(x)$. For real
positive $x$ they have an integral representation \re{I-U}. It is easy to see
that four different integrals in \re{I-U} can be found as special cases of the
following generic integral
\begin{align}\label{Uab-gen}
U_{ab}(x) = \frac12 \int_1^\infty du\, \e^{-x(u-1)} (u+1)^{a+b-1/2} (u-1)^{b-a-1/2}\,,
\end{align}
defined for $x>0$. Changing the integration variable as $u = t/x+1$ we obtain
\begin{align}
U_{ab}(x) = 2^{a+b-3/2} x^{a-b-1/2} \int_0^\infty dt\, \e^{-t} t^{b-a-1/2}
\lr{1+\frac{t}{2x}}^{a+b-1/2}\,.
\end{align}
The integral entering this relation can be expressed in terms of Whittaker
functions of second kind or equivalently confluent hypergeometric function of the
second kind
\begin{align}\label{U-ab}
U_{ab}(x) &= 2^{b-3/2} \Gamma(\ft12-a+b) x^{-b-1/2} \e^xW_{ab}(2x)\,,
\\[3mm] \notag
&= \ft12 \Gamma(\ft12-a+b) {\rm U}\lr{\ft12-a+b,1+2b; 2x}\,.
\end{align}
This relation can be used to analytically  continue $U_{ab}(x)$  from $x>0$  to
the whole complex $x-$plane with the cut along negative semi-axis. Matching
\re{Uab-gen} into \re{I-U} we obtain the following relations for the functions
$U_0^\pm(x)$ and $U_1^\pm(x)$
\begin{align}\notag
U_0^+(x) &= \ft12 \Gamma(\ft54) x^{-1} \e^x W_{-1/4,1/2}(2x)\,, && U_1^+(x)  =
\ft12 \Gamma(\ft14) (2x)^{-1/2} \e^x W_{1/4,0}(2x)\,,
\\[3mm] \label{U's}
U_0^-(x) &= \ft12 \Gamma(\ft34) x^{-1} \e^x W_{1/4,1/2}(2x)\,, && U_1^-(x)  =
\ft12 \Gamma(\ft34) (2x)^{-1/2} \e^x W_{-1/4,0}(2x) \,.
\end{align}
The functions $\I_1(\pm x), U_1^\pm(x)$ and  $\I_0(\pm x), U_0^\pm(x)$ satisfy the
same Whittaker differential equation and, as a consequence, they satisfy Wronskian relations
\begin{align}\label{Wh-Wr}
\I_1(-x) U_0^-(x) - \I_0(-x) U_1^-(x) =
\I_1(x) U_0^+(x) + \I_0(x) U_1^+(x) = \frac{\e^{x}}{x}\,.
\end{align}
The same relations also follow from \re{I-iden} for $x=\pm y$. In addition, 
\begin{align}\label{U-Wr}
U_0^+(x) U_1^-(-x) + U_1^+(x) U_0^-(-x) =  \frac{\pi}{2  \sqrt{2}x}\e^{\pm \frac{3 i\pi}{4}} \,,
 \qquad \text{(for $\Im x \gtrless 0$)}\,.
\end{align}
Combining together \re{Wh-Wr} and \re{U-Wr} we obtain the following relations between the functions
\begin{align} \notag
\I_0(x) & =\frac{2 \sqrt{2}}{\pi}\e^{\mp \frac{3 i\pi}{4}} \left[ \e^x U_0^-(-x) + \e^{-x} U_0^+(x)\right],
\\ \label{I-U-relation}
\I_1(x) & =\frac{2 \sqrt{2}}{\pi}\e^{\mp \frac{3 i\pi}{4}} \left[ \e^x U_1^-(-x) - \e^{-x} U_1^+(x)\right],
\end{align}
where the upper and lower signs correspond to $\Im x > 0$ and $\Im x < 0$, respectively.

At first sight, the relations \re{I-U-relation} look surprising since $\I_0(x)$
and $\I_1(x)$ are entire functions in the complex $x-$plane, while $U_0^\pm(x)$
and $U_1^\pm(x)$ are single-valued functions  in the same plane but with the cut
along the negative semi-axis. Indeed, one can use the relations \re{Wh-Wr} and
\re{U-Wr} to compute the discontinuity of the these functions across the cut as
\begin{align}\notag
\Delta U_0^\pm (-x) &= {\pm} \frac{\pi}{4}\e^{-x} \I_0(\mp x)\, \theta(x)\,,
\\ \label{disc}
\Delta U_1^\pm(-x) & =- \frac{\pi}{4}\e^{-x} \I_1(\mp x)\, \theta(x)\,,
\end{align}
where $\Delta U(-x) \equiv\lim_{\epsilon \to 0}
[U(-x+i\epsilon)-U(-x-i\epsilon)]/(2i)$ and $\theta(x)$ is a step function. Then,
one verifies with a help of these identities that the linear combinations of
$U-$functions in the right-hand side of \re{I-U-relation} have zero discontinuity
across the cut and, therefore, they are well-defined in the whole complex plane.

\subsection*{Asymptotic expansions}

For our purposes, we need asymptotic expansion of functions $\I_n(x)$ and
$U_n^\pm(x)$ at large real $x$. Let us start with the latter functions and
consider a generic integral \re{U-ab}.

To find asymptotic expansion of the function $U_{ab}(x)$ at large $x$, it
suffices to replace the last factor in the integrand \re{U-ab} in powers of
$t/(2x)$ and integrate term by term. In this way, we find from \re{U-ab} and
\re{U's}
\begin{align}\notag
U_0^+(x) &= (2x)^{-5/4}\Gamma(\ft54) F\lr{\ft14,\ft54|-\ft1{2x}}= (2x)^{-5/4} \Gamma(\ft54)
\left[ 1 -\frac5{32 x}+\ldots\right]\,,
\\ \notag
U_0^-(x) &= (2x)^{-3/4} \Gamma(\ft34) F\lr{-\ft14,\ft34|-\ft1{2x}}=
(2x)^{-3/4} \Gamma(\ft34) \left[ 1 +\frac3{32 x}+\ldots\right]\,,
\\ \notag
U_1^+(x) &= (2x)^{-1/4} \ft12\Gamma(\ft14) F\lr{\ft14,\ft14|-\ft1{2x}}= (2x)^{-1/4} \ft12\Gamma(\ft14)\left[
1-\frac1{32 x}+\ldots\right]\,,
\\  \label{U-large}
U_1^-(x) &= (2x)^{-3/4} \ft12\Gamma(\ft34)F\lr{\ft34,\ft34|-\ft1{2x}} = (2x)^{-3/4} \ft12\Gamma(\ft34)\left[ 1
-\frac{9}{32 x} +\ldots  \right]\,,
\end{align}
where the function $F(a,b|-\ft1{2x})$ is defined in  \re{F-Borel}.

Notice that the expansion coefficients in \re{U-large} grow factorially to higher orders but
the series are Borel summable for $x>0$.
For $x<0$ one has to distinguish the functions $U_n^\pm(x+i\epsilon)$
and $U_n^\pm(x-i\epsilon)$ (with $\epsilon\to 0$) which define analytical
continuation of the function $U_n^\pm(x)$ to the upper and lower edges of the
cut, respectively. In contrast with this, the functions $\I_n(x)$ are
well-defined on the whole real axis. Still, to make use of the relations
\re{I-U-relation} we have to specify the $U-$functions on the cut. As an example,
let us consider $\I_0(-\pi g)$ in the limit $g\to\infty$ and apply
\re{I-U-relation}
\begin{align}\label{cut-choice}
\I_0(-\pi g) & =\frac{2 \sqrt{2}}{\pi}\e^{- \frac{3 i\pi}{4}} \e^{\pi g}\left[
U_0^+(-\pi g+ i\epsilon )+ \e^{-2\pi g} U_0^-(\pi g)\right],
\end{align}
where $\epsilon\to 0$ and we have chosen to define the $U-$functions on the upper
edge of the cut. Written in this form, both terms inside the square brackets are
well-defined separately. Replacing $U_0^\pm$ functions in \re{cut-choice} by
their expressions \re{U-large} in terms of $F-$functions we find
\begin{align}\label{I0-sep}
\I_0(-\pi g) & =\frac{(2\pi g)^{-5/4} \e^{\pi g}}{\Gamma(\ft34)}\left[
F\lr{\ft14,\ft54|\ft1{2\pi g}+i\epsilon} + \Lambda^2 F\lr{-\ft14,\ft34|-\ft1{2\pi
g}}\right]\,,
\end{align}
with $\Lambda^2$   given by
\be \label{phase}
\Lambda^2 = \sigma \frac{\Gamma(\ft34)}{\Gamma(\ft54)}\e^{-2\pi g} (2\pi g)^{1/2}\,,\qqquad
\sigma = \e^{- \frac{3 i\pi}{4}}\,.
\ee
Since the second term in the right-hand side of \re{I0-sep} is exponentially
suppressed at large $g$ we may treat it as a nonperturbative correction. Repeating
the same analysis for $\I_1(-\pi g)$, we obtain from \re{I-U-relation} and
\re{U-large}
\begin{align}\label{I1-sep}
\I_1(-\pi g) & =\frac{(2\pi g)^{-5/4} \e^{\pi g}}{2\Gamma(\ft34)}\left[ 8\pi g
F\lr{\ft14,\ft14|\ft1{2\pi g}+i\epsilon} + \Lambda^2 F\lr{\ft34,\ft34|-\ft1{2\pi
g}}\right]\,,
\end{align}
We would like to stress that the `$+i\epsilon$' prescription in the first term in
\re{I0-sep} and the phase factor $\sigma=\e^{- \frac{3 i\pi}{4}}$ in \re{phase}
follow unambiguously from \re{cut-choice}. Had we defined the $U-$functions on
the lower edge of the cut, we would get the expression for $\I_0(-\pi g)$ with
`$-i\epsilon$' prescription and the phase factor  $\e^{\frac{3 i\pi}{4}}$. The
two expressions are however equivalent since discontinuity of the first term in
\re{I0-sep}  compensates the change of the phase factor in front of the second
term
\be\label{disc-F}
  F\lr{\ft14,\ft54|\ft1{2\pi g}+i\epsilon} -  F\lr{\ft14,\ft54|\ft1{2\pi g}-i\epsilon} = \frac{i\sqrt{2} {\Lambda^2}}{\sigma}
   F\lr{-\ft14,\ft34|-\ft1{2\pi g}}\,.
\ee

If one neglected `$+i\epsilon$' prescription in \re{cut-choice} and formally
expanded the first term in \re{I0-sep}  in powers of $1/g$, this would lead to
non-Borel summable series. This series suffers from Borel ambiguity which are
exponentially small for large $g$ and produce the contribution of the same order
as the second term in the right-hand side of \re{I0-sep}. The relation
\re{I0-sep} suggests how to give a meaning to this series. Namely, one should
first resum the series for negative $g$ where it is Borel summable and, then,
analytically continue it to the upper edge of the cut at positive $g$.

\section{Expression for the mass gap}\label{App-m}

In this appendix we derive the expression for the mass gap \re{m=U}. To this end, we
replace $\Gamma(4\pi g it)$ in  \re{m=int1} by its expression \re{G-gen1} and perform integration 
over $t$ in the right-hand side of  \re{m=int1}. We recall that, in the relation  \re{G-gen1},  $V_{0,1}(4\pi gt)$ are entire functions
of $t$, while $f_{0,1}(4\pi gt)$ are meromorphic functions defined in \re{ff1}. It is convenient
to decompose ${\Gamma(4\pi g it)}/(t+\ft14)$ into a sum of simple poles as
\begin{align}\label{G-sum}
\frac{\Gamma(4\pi g it)}{t+\ft14} &= \sum_{k=0,1}f_k(-\pi g) \frac{V_k(4\pi gt)}{t+\ft14}
+\sum_{k=0,1}\frac{f_k(4\pi gt)-f_k(-\pi g)}{t+\ft14}V_k(4\pi gt) \,,
\end{align}
where the second term is regular at $t=-1/4$.
Substituting this relation into \re{m=int1} and replacing $f_k(4\pi gt)$ by their expressions \re{ff1}, we encounter the following integral
\begin{align}\label{R-fun}
R_k(4\pi g s)=\Re\left[  \int_0^{-i\infty}  {dt\,
\e^{-4\pi g t-i\pi/4}} \frac{V_k(4\pi g t)}{t-s} \right] = \Re\left[  \int_0^{-i\infty}  {dt\,
\e^{-t-i\pi/4}} \frac{V_k(t)}{t-4\pi g s} \right] \,.
\end{align}
Then, the integral in \re{m=int1} can be expressed in terms of $R-$function as
\begin{align} \notag
\Re\bigg[  \int_0^{-i\infty}  &{dt\,\e^{-4\pi g t-i\pi/4}} \frac{\Gamma(4\pi g it)}{t+\ft14} \bigg]
 = f_0(-\pi g) R_0(-\pi g) + f_1(-\pi g) R_1(-\pi g) 
\\ \notag
&-\sum_{n\ge 1} \frac{n c_+(n,g)}{n+\ft14}\left[U_1^+(4\pi ng)  {R_0(4\pi gn)} + U_0^+(4\pi ng)  {R_1(4\pi gn)} \right]
\\ \label{big}
&+\sum_{n\ge 1} \frac{n c_-(n,g)}{n-\ft14}\left[U_1^-(4\pi ng)  {R_0(-4\pi gn)} - U_0^-(4\pi ng){R_1(-4\pi gn)} \right]\,,
\end{align}
where the last two lines correspond to the second sum in the right-hand side of \re{G-sum} and we
took into account that the coefficients $c_\pm(n,g)$ are real.

Let us evaluate the integral \re{R-fun} and choose for simplicity $R_0(s)$. We have to distinguish
two cases: $s>0$ and $s<0$. For $s>0$ we have
\begin{align} \notag
R_0(s)=& 
 - \Re\left[\e^{-i\pi/4} \int^{1}_{-\infty} dv\,\e^{-(1-v)s} \int_0^{-i\infty}  {dt\,
\e^{-vt}}\,V_0(t)\right] 
\\ \label{R0}
&=  \frac{\sqrt{2}}{\pi}\Re\bigg[\e^{-i\pi/4} \int^{1}_{-\infty} dv\,\e^{-(1-v)s}  \int_{-1}^{1}du\, \frac{(1+u)^{1/4}(1-u)^{-1/4}}{u-v-i\epsilon}\bigg] \,,
\end{align}
where in the second relation we replaced $V_0(t)$ by its integral representation \re{I-U}. Integration over $u$ can be carried out with a help of identity
\begin{align}
\frac1{\sqrt{2}\pi}\int_{-1}^{1}du\,\frac{(1+u)^{1/4-k}(1-u)^{-1/4}}{u-v-i\epsilon} =\delta_{k,0} - (v+1)^{-k}\times  \left\{\begin{array}{ll} \lr{\frac{v+1}{v-1}}^{1/4}\,, & v^2>1  \\[3mm] \e^{-i\pi/4}  \lr{\frac{1+v}{1-v}}^{1/4} \,, & v^2<1    \end{array}\right.
\end{align}
In this way, we obtain from \re{R0}  
\begin{align}\label{R0-pos}
R_0(s)\stackrel{s>0}{=} \sqrt{2}\left[ \frac1s - \int_{-\infty}^{-1} dv\,\e^{-(1-v)s}\lr{\frac{v+1}{v-1}}^{1/4} \right] = \sqrt{2}\left[
\frac1{s}-2 \e^{-2s} \, U_0^+(s)\right]\,,
\end{align} 
with the function $U_0^+(s)$ defined in \re{I-U}. In the similar manner, for $s<0$ we get
\begin{align}\label{R0-neg}
R_0(s)\stackrel{s<0}{=}  \sqrt{2}\left[
 \frac1{s}+2  U_0^-(-s)\right]\,,
\end{align} 
together with
\begin{align}\label{R1}
R_1(s) = 2\sqrt{2}\left[ \theta(-s)U_1^-(-s)+\theta(s) \e^{-2s} U_1^+(s)\right]\,.
\end{align}
Then, we substitute the relations \re{R0-pos}, \re{R0-neg} and \re{R1} into  \re{big} and find
\begin{align} \notag
& \Re\bigg[  \int_0^{-i\infty}   {dt\,\e^{-4\pi g t-i\pi/4}} \frac{\Gamma(4\pi g it)}{t+\ft14} \bigg]
\\ \label{RR}
& = 2\sqrt{2} f_0(-\pi g) \left[ U^-_0(\pi g) - \frac1{2\pi g}\right] + 2\sqrt{2} f_1(-\pi g) U^-_1(\pi g) 
+\frac{\sqrt{2}}{\pi g}\left[f_0(-\pi g)+1 \right]\,,
\end{align}
where the last term in the right-hand side corresponds to the last two lines in \re{big} (see Eq.~\re{ff}).
Substitution of \re{RR} into \re{m=int1} yields the expression for the mass scale \re{m=U}.


\begin{thebibliography}{99}

\bibitem{Mal97}
J.~M.~Maldacena,
  ``The large N limit of superconformal field theories and supergravity,''
  Adv.\ Theor.\ Math.\ Phys.\  {\bf 2} (1998) 231
  [arXiv:hep-th/9711200];
\\
S.~S.~Gubser, I.~R.~Klebanov and A.~M.~Polyakov,
  ``Gauge theory correlators from non-critical string theory,''
  Phys.\ Lett.\  B {\bf 428} (1998) 105
  [arXiv:hep-th/9802109];
\\
E.~Witten,
  ``Anti-de Sitter space and holography,''
  Adv.\ Theor.\ Math.\ Phys.\  {\bf 2} (1998) 253
  [arXiv:hep-th/9802150].

\bibitem{GKP} S.~S.~Gubser, I.~R.~Klebanov and A.~M.~Polyakov,
``A semi-classical limit of the gauge/string correspondence,'' Nucl.\ Phys.\  B {\bf 636} (2002) 99
[arXiv:hep-th/0204051].

\bibitem{FT}
 S.~Frolov and A.~A.~Tseytlin,
  ``Semiclassical quantization of rotating superstring in $\rm AdS(5) \times S(5)$,''
  J. High Ener. Phys. {\bf 0206} (2002) 007
  [arXiv:hep-th/0204226].
  
\bibitem{BKM03}
  V.~M.~Braun, G.~P.~Korchemsky and D.~Mueller,
  ``The uses of conformal symmetry in QCD,''
  Prog.\ Part.\ Nucl.\ Phys.\  {\bf 51} (2003) 311
  [arXiv:hep-ph/0306057].

\bibitem{BGK06}
A.V.~Belitsky, A.S.~Gorsky and G.P.~Korchemsky, ``Logarithmic scaling in gauge/string
correspondence,'' Nucl.\ Phys.\  B {\bf 748} (2006) 24
 [arXiv:hep-th/0601112].

\bibitem{AM07}
L.~F.~Alday and J.~M.~Maldacena, ``Comments on operators with large spin,''
 J. High Ener. Phys. {\bf 0711} (2007) 019
 [arXiv:0708.0672 [hep-th]].

\bibitem{FRS}
  L.~Freyhult, A.~Rej and M.~Staudacher,
  ``A Generalized Scaling Function for AdS/CFT,''
  arXiv:0712.2743 [hep-th].

\bibitem{FTT06}
S.~Frolov, A.~Tirziu and A.~A.~Tseytlin, ``Logarithmic corrections to higher twist scaling at strong
coupling from AdS/CFT,''
  Nucl.\ Phys.\  B {\bf 766} (2007) 232
  [arXiv:hep-th/0611269].

\bibitem{CK07}
P.~Y.~Casteill and C.~Kristjansen,
  ``The Strong Coupling Limit of the Scaling Function from the Quantum String
  Bethe Ansatz,''
  Nucl.\ Phys.\  B {\bf 785} (2007) 1
  [arXiv:0705.0890 [hep-th]];
\\
A.~V.~Belitsky,
  ``Strong coupling expansion of Baxter equation in N=4 SYM,''
  Phys.\ Lett.\  B {\bf 659} (2008) 732
  [arXiv:0710.2294 [hep-th]].

\bibitem{KR87}
G.~P.~Korchemsky and A.~V.~Radyushkin, ``Renormalization of the Wilson Loops
Beyond the Leading Order,'' Nucl.\ Phys.\  B {\bf 283} (1987) 342;
\\
G.P.~Korchemsky, ``Asymptotics of the Altarelli-Parisi-Lipatov Evolution Kernels
of Parton Distributions,'' Mod.\ Phys.\ Lett.\  A {\bf 4} (1989) 1257.

\bibitem{BGK03}
A.~V.~Belitsky, A.~S.~Gorsky and G.~P.~Korchemsky, ``Gauge/string duality for QCD
conformal operators,'' Nucl.\ Phys.\  B {\bf 667} (2003) 3
[arXiv:hep-th/0304028].

\bibitem{P80}
A.~M.~Polyakov,
  ``Gauge Fields As Rings Of Glue,''
  Nucl.\ Phys.\  B {\bf 164} (1980) 171.

\bibitem{Brandt:1981kf}
  V.~S.~Dotsenko and S.~N.~Vergeles,
 ``Renormalizability Of Phase Factors In The Nonabelian Gauge Theory,''
  Nucl.\ Phys.\  B {\bf 169} (1980) 527;
  \\
  R.~A.~Brandt, F.~Neri and M.~a.~Sato,
 ``Renormalization Of Loop Functions For All Loops,''
  Phys.\ Rev.\  D {\bf 24} (1981) 879;
  \\
  H.~Dorn,
  ``Renormalization of path ordered phase factors and related hadron operators in gauge field
  theories,''
  Fortsch.\ Phys.\  {\bf 34} (1986) 11.

\bibitem{KK92}
I.~A.~Korchemskaya and G.~P.~Korchemsky,
  ``On lightlike Wilson loops,''
  Phys.\ Lett.\  B {\bf 287} (1992) 169;
\\ A.~Bassetto, I.~A.~Korchemskaya, G.~P.~Korchemsky and G.~Nardelli,
  ``Gauge invariance and anomalous dimensions of a light cone Wilson loop in
  lightlike axial gauge,''
  Nucl.\ Phys.\  B {\bf 408} (1993) 62
  [arXiv:hep-ph/9303314].


\bibitem{Korchemsky:1985xj}
 S.~V.~Ivanov, G.~P.~Korchemsky and A.~V.~Radyushkin,
  ``Infrared Asymptotics Of Perturbative QCD: Contour Gauges,''
  Yad.\ Fiz.\  {\bf 44} (1986) 230
  [Sov.\ J.\ Nucl.\ Phys.\  {\bf 44} (1986) 145];
  \\
  G.~P.~Korchemsky and A.~V.~Radyushkin,
  ``Loop Space Formalism And Renormalization Group For The Infrared Asymptotics
  Of QCD,''
  Phys.\ Lett.\  B {\bf 171} (1986) 459.

\bibitem{Korchemsky:1988hd}
  G.~P.~Korchemsky,
  ``Double logarithmic asymptotics in QCD,''
  Phys.\ Lett.\  B {\bf 217} (1989) 330;
  ``Sudakov Form-Factor In QCD,''
  Phys.\ Lett.\  B {\bf 220} (1989) 629.

\bibitem{Korchemskaya:1996je}
  I.~A.~Korchemskaya and G.~P.~Korchemsky,
  ``Evolution equation for gluon Regge trajectory,''
  Phys.\ Lett.\  B {\bf 387} (1996) 346
  [arXiv:hep-ph/9607229];
  ``High-energy scattering in QCD and cross singularities of Wilson loops,''
  Nucl.\ Phys.\  B {\bf 437} (1995) 127
  [arXiv:hep-ph/9409446].


\bibitem{K95}
  G.~P.~Korchemsky,  ``Bethe Ansatz For QCD Pomeron,''
  Nucl.\ Phys.\  B {\bf 443} (1995) 255
  [arXiv:hep-ph/9501232];
%
  ``Quasiclassical QCD pomeron,''
  Nucl.\ Phys.\  B {\bf 462} (1996) 333
  [arXiv:hep-th/9508025].

\bibitem{BES}
B.~Eden and M.~Staudacher,
 ``Integrability and transcendentality,''
J.\ Stat.\ Mech.\  {\bf 0611} (2006) P014
  [arXiv:hep-th/0603157];
 \\
N.~Beisert, B.~Eden and M.~Staudacher,  ``Transcendentality and crossing,'' J.\
Stat.\ Mech.\  {\bf 0701} (2007) P021  [arXiv:hep-th/0610251].

\bibitem{cusp-4loop}
  Z.~Bern, M.~Czakon, L.~J.~Dixon, D.~A.~Kosower and V.~A.~Smirnov,
  ``The Four-Loop Planar Amplitude and Cusp Anomalous Dimension in Maximally
  Supersymmetric Yang-Mills Theory,''
  Phys.\ Rev.\  D {\bf 75} (2007) 085010
  [arXiv:hep-th/0610248];
\\ F.~Cachazo, M.~Spradlin and A.~Volovich,
  ``Four-Loop Cusp Anomalous Dimension From Obstructions,''
  Phys.\ Rev.\  D {\bf 75} (2007) 105011
  [arXiv:hep-th/0612309].

\bibitem{Benna06}
M.~K.~Benna, S.~Benvenuti, I.~R.~Klebanov and A.~Scardicchio, ``A test of the
AdS/CFT correspondence using high-spin operators,'' Phys.\ Rev.\ Lett.\  {\bf 98}
(2007) 131603  [arXiv:hep-th/0611135].

\bibitem{Kotikov:2006ts}
  A.~V.~Kotikov and L.~N.~Lipatov,
  Nucl.\ Phys.\  B {\bf 769} (2007) 217
  [arXiv:hep-th/0611204].

\bibitem{Alday07}
L.~F.~Alday, G.~Arutyunov, M.~K.~Benna, B.~Eden and I.~R.~Klebanov, ``On the
strong coupling scaling dimension of high spin operators,'' J. High Ener. Phys.
{\bf 0704} (2007) 082 [arXiv:hep-th/0702028].

\bibitem{KSV07}
I.~Kostov, D.~Serban and D.~Volin,
  ``Strong coupling limit of Bethe ansatz equations,''
  Nucl.\ Phys.\  B {\bf 789} (2008) 413
  [arXiv:hep-th/0703031];
  ``Functional BES equation,'' arXiv:0801.2542 [hep-th];
\\
M.~Beccaria, G.~F.~De Angelis and V.~Forini,
 ``The scaling function at strong coupling from the quantum string Bethe
 equations,''
  J. High Ener. Phys.
{\bf 0704} (2007) 066.
  [arXiv:hep-th/0703131].

\bibitem{BKK07}
  B.~Basso, G.~P.~Korchemsky and J.~Kotanski,
  ``Cusp anomalous dimension in maximally supersymmetric Yang-Mills theory at
  strong coupling,''
  Phys.\ Rev.\ Lett.\  {\bf 100} (2008) 091601
  [arXiv:0708.3933 [hep-th]].

\bibitem{KSV08}
I.~Kostov, D.~Serban and D.~Volin,
  ``Functional BES equation,'' arXiv:0801.2542 [hep-th].

\bibitem{Zinn-Justin}
  J.~Zinn-Justin,
  ``Quantum field theory and critical phenomena,''
  Int.\ Ser.\ Monogr.\ Phys.\  {\bf 113} (2002) 1;
\\
  J.~C.~Le Guillou and J.~Zinn-Justin,
  ``Large order behavior of perturbation theory,''
{\it  Amsterdam, Netherlands: North-Holland (1990)} .

\bibitem{Bender}
C.~M.~Bender and S.~A.~Orszag, ``Advanced Mathematical Methods for Scientists and
Engineers'', {\it   McGraw-Hill, 1978}.

\bibitem{Roiban:2007dq}
  R.~Roiban and A.~A.~Tseytlin,
  ``Strong-coupling expansion of cusp anomaly from quantum superstring,''
  JHEP {\bf 0711} (2007) 016
  [arXiv:0709.0681 [hep-th]];
  \\
  R.~Roiban, A.~Tirziu and A.~A.~Tseytlin,
  ``Two-loop world-sheet corrections in $AdS_5 \times S^5$ superstring,''
  JHEP {\bf 0707} (2007) 056
  [arXiv:0704.3638 [hep-th]].

\bibitem{G08}
  N.~Gromov,
  ``Generalized Scaling Function at Strong Coupling,''
  JHEP {\bf 0811} (2008) 085
  [arXiv:0805.4615 [hep-th]].

\bibitem{ZZ78}
  A.~B.~Zamolodchikov and Al.~B.~Zamolodchikov,
``Relativistic Factorized S Matrix In Two-Dimensions Having O(N) Isotopic
  Symmetry,''
  Nucl.\ Phys.\  B {\bf 133} (1978) 525
  [JETP Lett.\  {\bf 26} (1977) 457];
  ``Factorized S-matrices in two dimensions as the exact solutions of  certain
  relativistic quantum field models,''
  Annals Phys.\  {\bf 120} (1979) 253.

\bibitem{PW83}
  A.~M.~Polyakov and P.~B.~Wiegmann,
  ``Theory of nonabelian Goldstone bosons in two dimensions,''
  Phys.\ Lett.\  B {\bf 131} (1983) 121;
\\
  P.~B.~Wiegmann,
  ``Exact Solution Of The O(3) Nonlinear Sigma Model,''
  Phys.\ Lett.\  B {\bf 152} (1985) 209.

\bibitem{FR85}
  L.~D.~Faddeev and N.~Y.~Reshetikhin,
  ``Integrability Of The Principal Chiral Field Model In (1+1)-Dimension,''
  Annals Phys.\  {\bf 167} (1986) 227.

\bibitem{HMN90}
  P.~Hasenfratz, M.~Maggiore and F.~Niedermayer,
  ``The Exact mass gap of the O(3) and O(4) nonlinear sigma models in d = 2,''
  Phys.\ Lett.\  B {\bf 245} (1990) 522;
\\
  P.~Hasenfratz and F.~Niedermayer,
  ``The Exact mass gap of the O(N) sigma model for arbitrary N is $>= 3$ in d =
  2,''
  Phys.\ Lett.\  B {\bf 245} (1990) 529.

\bibitem{BK08}
B.~Basso and G.~P.~Korchemsky, ``Embedding nonlinear O(6) sigma model into N=4 super-Yang-Mills
theory,'' arXiv:0805.4194 [hep-th].

\bibitem{FGR08a}
 D.~Fioravanti, P.~Grinza and M.~Rossi, ``Strong coupling for planar ${\cal N}=4$ SYM theory: an
 all-order result,'' arXiv:0804.2893 [hep-th];
\\
  D.~Fioravanti, P.~Grinza and M.~Rossi,
  ``The generalised scaling function: a note,''
  arXiv:0805.4407 [hep-th];
\\
  F.~Buccheri and D.~Fioravanti,
  ``The integrable O(6) model and the correspondence: checks and predictions,''
  arXiv:0805.4410 [hep-th];
\\
  D.~Fioravanti, P.~Grinza and M.~Rossi,
  ``The generalised scaling function: a systematic study,''
  arXiv:0808.1886 [hep-th].

\bibitem{RT07}
  R.~Roiban and A.~A.~Tseytlin,
  ``Spinning superstrings at two loops: strong-coupling corrections to
  dimensions of large-twist SYM operators,''
  Phys.\ Rev.\  D {\bf 77} (2008) 066006
  [arXiv:0712.2479 [hep-th]].

\bibitem{BBBKP08}
  Z.~Bajnok, J.~Balog, B.~Basso, G.~P.~Korchemsky and L.~Palla,
  ``Scaling function in AdS/CFT from the O(6) sigma model,''
  arXiv:0809.4952 [hep-th].

\bibitem{V08}
  D.~Volin,
  ``The 2-loop generalized scaling function from the BES/FRS equation,''
  arXiv:0812.4407 [hep-th].
  
\bibitem{Dorey:1998yh}
  N.~Dorey,
  ``The BPS spectra of two-dimensional supersymmetric gauge theories with
  twisted mass terms,''
  JHEP {\bf 9811} (1998) 005
  [arXiv:hep-th/9806056].

\bibitem{Shifman:2004dr}
  M.~Shifman and A.~Yung,
  ``Non-Abelian string junctions as confined monopoles,''
  Phys.\ Rev.\  D {\bf 70} (2004) 045004
  [arXiv:hep-th/0403149].

\bibitem{BMN02}
  D.~E.~Berenstein, J.~M.~Maldacena and H.~S.~Nastase,
  ``Strings in flat space and pp waves from N = 4 super Yang Mills,''
  JHEP {\bf 0204} (2002) 013
  [arXiv:hep-th/0202021].

\bibitem{Frolov:2003}
  S.~Frolov and A.~A.~Tseytlin,
  ``Multi-spin string solutions in AdS(5) x S**5,''
  Nucl.\ Phys.\  B {\bf 668} (2003) 77
  [arXiv:hep-th/0304255];
\\
  N.~Beisert, J.~A.~Minahan, M.~Staudacher and K.~Zarembo,
  ``Stringing spins and spinning strings,''
  JHEP {\bf 0309} (2003) 010
  [arXiv:hep-th/0306139];
\\
  S.~Frolov and A.~A.~Tseytlin,
  ``Rotating string solutions: AdS/CFT duality in non-supersymmetric
  sectors,''
  Phys.\ Lett.\  B {\bf 570} (2003) 96
  [arXiv:hep-th/0306143].

 \bibitem{BT05}
  N.~Beisert and A.~A.~Tseytlin,
  ``On quantum corrections to spinning strings and Bethe equations,''
  Phys.\ Lett.\  B {\bf 629} (2005) 102
  [arXiv:hep-th/0509084];
\\  
  J.~A.~Minahan, A.~Tirziu and A.~A.~Tseytlin,
  ``1/J**2 corrections to BMN energies from the quantum long range
  Landau-Lifshitz model,''
  JHEP {\bf 0511} (2005) 031
  [arXiv:hep-th/0510080].

\bibitem{K08}
  J.~Kotanski,
  ``Cusp anomalous dimension in maximally supersymmetric Yang-Mills theory,''
  arXiv:0811.2667 [hep-th].

\bibitem{Mikhlin}
S.~G.~Mikhlin, {\it Linear Integral Equations}, New York: Gordon \& Breach, 1961.

\bibitem{WW}
E.~T.~ Whittaker and G.~N.~Watson, ``A Course of Modern Analysis'', {\it
Cambridge University Press 1927 - 4th Edition, 1980}.

\bibitem{David}
  F.~David,
  ``On The Ambiguity Of Composite Operators, Ir Renormalons And The Status Of
  The Operator Product Expansion,''
  Nucl.\ Phys.\  B {\bf 234} (1984) 237;
  ``Nonperturbative Effects And Infrared Renormalons Within The 1/N Expansion
  Of The O(N) Nonlinear Sigma Model,''
  Nucl.\ Phys.\  B {\bf 209} (1982) 433.

\bibitem{NSVZ84}
  V.~A.~Novikov, M.~A.~Shifman, A.~I.~Vainshtein and V.~I.~Zakharov,
  ``Two-Dimensional Sigma Models: Modeling Nonperturbative Effects Of Quantum
  Chromodynamics,''
  Phys.\ Rept.\  {\bf 116} (1984) 103
  [Sov.\ J.\ Part.\ Nucl.\  {\bf 17} (1986\ FECAA,17,472-545.1986) 204.1986\ FECAA,17,472].

\bibitem{Broadhurst:1993ib}
  D.~J.~Broadhurst,
  ``Summation of an infinite series of ladder diagrams,''
  Phys.\ Lett.\  B {\bf 307} (1993) 132.



\end{thebibliography}
\end{document}